\documentclass[reprint]{revtex4-2}
\usepackage{braket,amsmath,amssymb,graphicx,color,ulem}
\usepackage[hidelinks]{hyperref}
\allowdisplaybreaks
\newcommand{\alert}[1]{#1}

\begin{document}
\title{Holographic entanglement renormalisation for fermionic quantum matter}
\author{Abhirup Mukherjee}
\email{am18ip014@iiserkol.ac.in\\
(Author to whom any correspondence should be addressed.)}
\affiliation{Department of Physical Sciences, Indian Institute of Science Education and Research Kolkata, W.B. 741246, India}
\author{Siddhartha Patra}
\email{sp14ip022@iiserkol.ac.in }
\affiliation{Department of Physical Sciences, Indian Institute of Science Education and Research Kolkata, W.B. 741246, India}
\author{Siddhartha Lal}
\email{slal@iiserkol.ac.in}
\affiliation{Department of Physical Sciences, Indian Institute of Science Education and Research Kolkata, W.B. 741246, India}

\begin{abstract}
	We demonstrate the emergence of a holographic dimension in a system of 2D non-interacting Dirac fermions placed on a torus, by studying the scaling of multipartite entanglement measures under a sequence of renormalisation group (RG) transformations \alert{applied in momentum space}. Geometric measures defined in this emergent space can be related to the RG beta function of the spectral gap, hence establishing a holographic connection between the spatial geometry of the emergent spatial dimension and the entanglement properties of the boundary quantum theory. We prove, analytically, that changing the boundedness of the holographic space involves a topological transition accompanied by a critical Fermi surface in the boundary theory. We go on to show that this results in the formation of a quantum wormhole geometry that connects the UV and the IR of the emergent dimension. The additional conformal symmetry at the transition also supports a relation between the emergent metric and the stress-energy tensor. In the presence of an Aharonov-Bohm flux, the entanglement gains a geometry-independent piece which is shown to be topological, sensitive to changes in boundary conditions, and related to the Luttinger volume of the system. Upon the insertion of a strong transverse magnetic field, we show that the Luttinger volume is linked to the Chern number of the occupied single-particle Landau levels.
\end{abstract}
\maketitle

\section{Introduction}

The present millennium has seen a rapid surge in the use of quantum entanglement towards the study of quantum and condensed matter systems~\cite{terhal_physicstoday_2003,horodecki2009quantum,eisert_entanglement_2010,laflorencie2016quantum,zeng2019,Erhard2020}. By tracking quantum correlations among particles across both small and large distances, entanglement measures offer valuable insights into the nature of the ground state and low-energy excitations. This, for example, allows us to distinguish strongly-correlated topologically ordered states characterised by long-ranged entanglement (in the form of a sub-leading topological term in their entanglement entropy~\cite{wen_1990,Wen_fqhe_1990,wen1995topological,niu1985quantized}) from weakly-interacting topologically trivial phases that have short-ranged area-law entanglement~\cite{das_shankaranarayanan_2006,cramer_eisert_2006,eisert_entanglement_2010}.

More pertinent to the present work is the entanglement of non-interacting relativistic fermionic systems. Such phases are known to emerge in several condensed matter systems~\cite{vafek_2014_diracreview}, appearing at the surfaces of three-dimensional topological insulators~\cite{bernevig_2013,cayssol2013,moessner_moore_2021,Rachel_2018}, quasi-2D organic materials~\cite{wang_2015,Wu_2016,ni_2022_dirac} and in artificial quantum simulators such as cold atom gases, molecular and microwave crystals~\cite{zhu_2007_coldatoms,Lepori_2010,Boada_2011,Kuno2018,Gomes2012,Yuan2016,bellec_2013}.
Massless critical systems (or conformal field theories (CFTs)) differ from gapped systems due to the fact that their entanglement satisfies a modified area law; in one spatial dimension, the entanglement entropy ($S$) of a single subsystem of length \(l\) scales as \(S \sim \frac{c}{3}\ln l\)~\cite{vidal2003entanglement,latorre_extanglement_2003,lattore_pra_2005,calabrese2004,wolf2008area}, where \(c\) is the conformal central charge of the CFT.
This violation arises from the non-local nature of the gapless quantum fluctuations that lie proximate to the $d-1$-dimensional Fermi surface of a fermionic system in $d$ spatial dimensions. For massive fermions, an entropic \(c-\)function (defined as the spatial derivative of the entanglement entropy) can be expressed in the form of differential equations that need to be solved numerically~\cite{Casini_2005,Cardy2008,Casini_2009_multi,doyon_2009}.

Unlike topologically ordered phases, there is no geometry-independent topological term in the entanglement entropy of free fermionic systems~\cite{Chen_2017,bueno_2017}. It is, however, possible to generate such a term even here by placing the system on a multiply connected manifold such as a cylinder, and inserting a magnetic flux through the cylinder~\cite{Metlitski_2009,Arias_2015,Whitsitt_2017,Chen_2017,zhu_kagome_2018}.
It has been shown that if there are \(\phi\) number of electronic flux quanta piercing the system, an additional term is generated in the entanglement entropy of a subsystem of massless Dirac fermions of sufficiently large length (as defined in Ref.~\cite{Chen_2017}) with the form $\frac{1}{6}\ln \big|2\sin \frac{\phi}{2}\big|$~\cite{Chen_2017,Arias_2015}. Further, there is evidence from numerical computations on lattice models of interacting electrons that such a term is a signature of universality~\cite{zhu_kagome_2018,Hu_triangular_2019,crepel_2021}.

The notion of entanglement also plays a key role in the context of holographic duality, which posits that there exists a duality relation between a quantum field theory in \(d-\)spatial dimensions and a gravity theory in \(d+1-\)spatial dimensions~\cite{hooft_1974,polyakov_1978,polyakov_strings,gross_1993,bousso_2002,zaanen_liu_sun_schalm_2015,hartnoll_lucas_sachdev2018}. It is widely believed that the additional dimension can be visualised as a renormalisation group flow that starts from the (almost) critical field theory on the boundary and extends into the bulk~\cite{akhmedov_1998,alvarez_1999,freedman_1999,porrati_1999,kachru_2000,girardello_2000,de2000,
de_2001,Johnson_2001,Yamaguchi_2003}. The most popular realisation of the holographic principle is the AdS/CFT correspondence that links a conformal field theory with a theory of quantum gravity through a strong-weak duality relation~\cite{bekenstein_1973,Hawking_1974,Hawking_1975,maldacena1999large,gubser_1998,witten1998anti,
ryu2006,ryu2006aspects}.
This has led to new insights in various areas of high energy and condensed matter physics, such as the viscosity of the quark-gluon plasma~\cite{kovtun_2005}, the response functions of the Bose-Hubbard model~\cite{damle_1997,hartnoll_2007}, and the phenomenology of non-Fermi liquids~\cite{lee_2009_nonfermi,lio_2011,faulkner_2011,mihailo_2009}, holographic superconductors~\cite{gubser_2010,gauntlett_2009,hartnoll_2008_building,Hartnoll_2008,Cai2015,Wang2020}, striped phases~\cite{Donos2011,Donos2013,Cremonini2018} and holographic Fermi liquids~\cite{kachru_2008,cubrovic_2011,senthil_2003,sachdev_2011,Hartnoll_2011_stellar}.

Conversely, the holographic principle also implies that studying the boundary conformal theory can provide insights into the nature of the bulk quantum gravity theory. This requires constructing the emergent dimension from first principles by investigating the RG flow of the (often strongly coupled) quantum field theory, and has proved to be considerably more difficult.
Nevertheless, there exist some examples of constructive attempts towards a bulk gravity theory, including ~\cite{matsueda_2012_kondo,czech_2018,Lee2012,lee2014,vidal2007,vidal2008,swingle2012a,evenbly_2014,anirbanurg1,anirbanurg2}.
These include the momentum-shell renormalisation approach of Refs.\cite{lee2010,lee_2010_lectures,Lee2012,lee2014}, the multi-scale entanglement renormalisation ansatz (MERA) approach of Refs.\cite{vidal2007,vidal2008,swingle2012a,evenbly_2014,milsted_2018}, the exact holographic mapping (EHM)~\cite{qi2013exact,lee2016}, the Wilsonian RG-based approach of Refs.~\cite{kim2019,kim_2017_supeconductor,kim_2016} and the unitary renormalisation group (URG) approach~\cite{anirbanurg1,anirbanurg2,Mukherjee_mott_merg}.

Questions on entanglement and holography become particularly interesting in systems of critical fermions. As mentioned earlier, such systems differ from bosonic and gapped systems through the presence of longer-ranged entanglement that follows a modified area-law~\cite{wolf2008area,swingle2010} as well as topological features arising from the incompressibility of the Fermi volume~\cite{oshikawa2000topological,seki2017topological}. Fermi surface gapping phase transitions in such systems are often driven by changes in topological quantum numbers and the nature of the inter-particle entanglement~\cite{anirbanurg1,anirbanurg2,Gegenwart2008}. Constructing a framework for fermionic criticality is therefore an active area of research with consequences for important systems such as unconventional superconductors, strange metals and heavy fermions~\cite{anirbanmott1,anirbanmott2,anirban_mott_2022,Custers2003}. Still less is known about the holographic implications of such systems. We extend this body of work by providing an explicit demonstration of the holographic principle in the form of a simple and tractable exact holographic mapping (EHM) for a free fermion system with and without a mass gap. Our program yields analytic relations between the RG flow parameters and their dual bulk quantities such as distances and curvature. This also allows us to study the consequences of a critical Fermi surface (lying precisely at the quantum phase transition) on the nature of the holographic space.

We first briefly describe the strategy employed in the present work. (i)~We perform a set of scaling transformations \(T_j\) (corresponding to a step index \(j\)) on the set of \(k-\)states in the Hilbert space of our system of fermions, and we show that the renormalisation of entanglement measures along the sequence of Hamiltonians \(T_j^\dagger H_j T_j\) constitutes an emergent additional spatial dimension. (ii)~By employing the mutual information between multiple subsets of the Hilbert space, we define a measure of distance that is used to obtain holographic relations between boundary and bulk quantities. Additional results are obtained upon further analysis of the entanglement and these holographic relations.
The approach presented in this work is similar in spirit to other renormalisation group approaches that apply decoupling transformations on the Hamiltonian, such as the continuous unitary transformation (CUT) RG~\cite{wegner1994flow,glazekwilson1993,glazekwilson1994,savitz2017stable}, the strong-disorder RG~\cite{rademaker2016explicit,monthus2016flow} and the spectrum-bifurcation RG~\cite{you2016}. As the system we consider is non-interacting, the scaling transformations implemented in this work do not lead to the decoupling of states, and only result in the coarse-graining of the system in momentum space, through non-local transformations.

The manuscript is structured in the following manner. Sec.~\ref{prelims} describes the system we work with and defines certain physical ideas and measures of entanglement that are employed in later sections. In Sec.~\ref{hierarchy}, we define the scaling transformations mentioned earlier, and discuss its implications on the scaling of entanglement in the system under study.
In Sections \ref{EHM},\ref{holography} and \ref{curvature}, we probe the holographic nature of the entanglement, unveiling an emergent spatial dimension in the process and obtain several characteristics of this emergent geometry.
Section \ref{fermi_critical} discusses the consequences of a critical Fermi surface on the holographic space, and how the emergent curvature can be linked to a convergence parameter associated with the RG flows.
In Section ~\ref{topologicalContent}, we probe the topological aspects of the entanglement, and discuss its implications for the system at hand.
We conclude with a discussion of our results in Sec.~\ref{conclusions} and point out some open questions. 

\subsection*{Summary of our main results}
For the convenience of the reader, we present below a summary of the main results obtained by us.

1.~Quantum information measures (such as entanglement entropy and multipartite information) reveal a hierarchical structure in the nature of entanglement within the system. We relate geometric parameters like distance and curvature of the holographic dimension to an RG beta function of the mass gap, providing thereby an explicit analytic manifestation of the holographic principle. We show that a change in the in the boundedness of the space corresponds to a change in certain topological winding numbers.

2.~We go on to argue that this topological transition and the underlying critical Fermi surface coincides with the formation of a quantum wormhole geometry that connects the UV and the IR of the emergent dimension. The additional (conformal) symmetry at the transition links the emergent metric and the stress-energy tensor.

3.~We show that the geometry-independent part of the entanglement is invariant under appropriate scaling transformations for a system of fixed number density, and is related to the Luttinger volume of the gapless fermionic system. In the presence of a strong transverse magnetic field, the topological Luttinger volume in the metallic state transforms into the Chern number \(\mathcal{C}\) of the insulating integer quantum Hall state.

\begin{figure}
	\centering
	\includegraphics[width=0.45\textwidth]{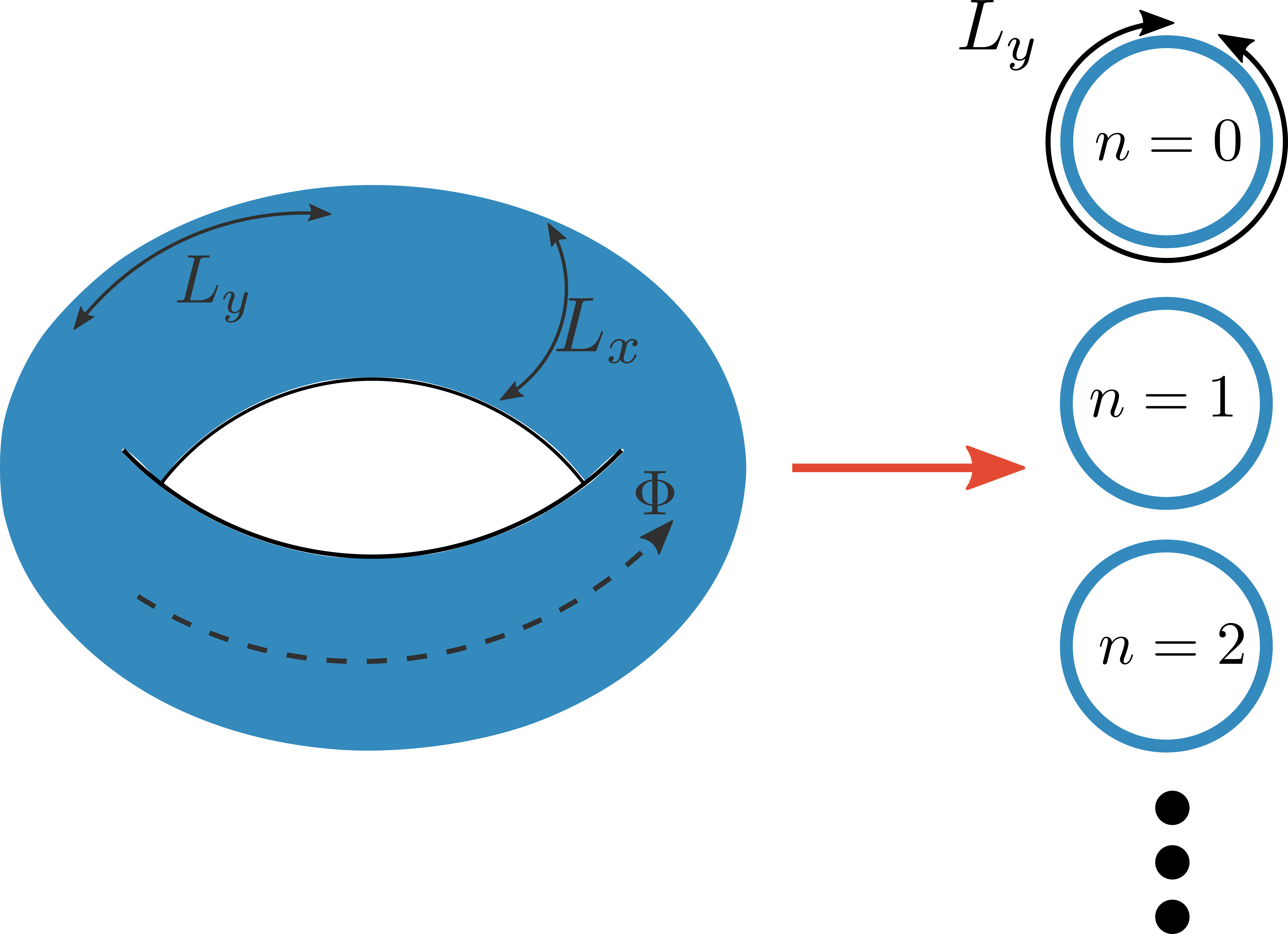}
	\caption{The model on the left consists of a system of non-interacting electrons placed on a torus with a flux threading through it. The model on the right consists of a sum of \(1+1-\)dimensional systems, each consisting of massive electrons placed on a ring; the \(1+1-\)D systems are decoupled from each other. The mass arises from the flux. These two models can be shown to be equivalent to each other~\cite{chung_2000,Arias_2015,Chen_2017,Murciano_2020}.}
	\label{dim-red}
\end{figure}
\section{Preliminaries and definitions}
\label{prelims}
\subsection{The system}
In this work, we focus on the case of Dirac fermions in two spatial dimensions described by the Lagrangian (in natural units \(\hbar=c=1\)) \(\mathcal{L} = \int \mathrm{d}x~\mathrm{d}y~\overline\psi(x,y,t)\left(i\gamma^\mu\partial_\mu - m\right)\psi(x,y,t)\). The system is placed on a \(2-\)torus of dimensions \(L_x\times L_y\) with periodic boundary conditions, such that \(\psi(x,y,t) = \psi(x+L_x,y,t) = \psi(x,y+L_y,t)\). Entanglement measures will be calculated in the presence of a gauge field \(A\) that transforms the Lagrangian by shifting the momenta: \(i\partial_\mu \to i\partial_\mu + eA\). We choose the vector potential $A$ such that, in any particular setup, it couples to just one component of the momentum (say, the \(x-\)component). This results in \(\phi=eAL_x/2\pi\) units of flux quantum threading the torus in the \(y-\)direction. One of our goals is to express the Luttinger volume \(V_L\) in terms of multi-partite measures of entanglement \(\left( I^g_{\left\{ \mathcal{A}_i \right\} }(\phi) \right)\) between multiple subsystems \(\left\{\mathcal{A}_i, i=1,2,\ldots,g\right\}\)~\cite{siddhartha_TEE}, in the presence of the flux \(\phi\).
The precise definition of \(I^g\) and the method of choosing the subsystems will be given later.
In order to calculate \(I^g_{\left\{\mathcal{A}_i\right\}}\), however, we find it convenient to take the thin torus limit: \(\left( L_y/L_x \right) \to \infty\), keeping \(l/L_y\) fixed. Importantly, this operation does not affect quantities such as \(V_L\) as long as we keep the number density unchanged.

In order to calculate entanglement measures, we will express the \(2+1-\)dimensional system as a sum of an infinite number of massive \(1+1-\)dimensional systems
~\cite{chung_2000,Arias_2015,Chen_2017,Murciano_2020}. In the presence of a gauge field \(\vec A = A \hat x\), the complete Lagrangian (in natural units) is
\begin{equation}\begin{aligned}
	\label{lagrangian}
	\mathcal{L} = \int \mathrm{d}x~\mathrm{d}y~\overline\psi(x,y)\left[\gamma^\mu\left(i\partial_\mu + eA_\mu\right) -m \right]\psi(x,y)~,
\end{aligned}\end{equation}
where \(A_\mu = A \delta_{\mu,x}\) and \(\phi = eAL_x/2\pi\). The integral ranges over the surface area of the torus, and the time dependence of the fields has been suppressed. Due to the periodic boundary conditions (PBCs) in the \(x-\)direction, the momenta \(k_x\) are quantised:~\( k_x^n = \frac{2\pi n}{L_x}, n \in \mathbb{Z}\). We then expand the fields \(\psi(x,y)\) in these momenta: 
\begin{equation}\begin{aligned}
\psi(x,y) = \sum_{n=-\infty}^{\infty} e^{ik_x^n x} \psi(k_x^n,y)~.
\end{aligned}\end{equation}
In terms of these dual fields, the Lagrangian can be shown to be equivalent to a tower of massive \(1+1-\)dimensional fermions (see Fig.~\ref{dim-red}). Details are provided in \alert{Appendix~\ref{app-red}}:
\begin{equation}\begin{aligned}
\mathcal{L} = \sum_{k_x^n} \int \mathrm{d}y~\overline\psi(k_x^n,y)\left(i\gamma^\mu \partial_\mu - M(n,\phi)\right) \psi(k_x^n,y)~,
\end{aligned}\end{equation}
with an effective mass $M_{n,\phi}$ given by
\begin{equation}\begin{aligned}
	M_{n,\phi} = \sqrt{m^2 + \left(k^n_x + eA\right)^2} = \sqrt{m^2 + \frac{4 \pi^2}{L_x^2}\left(n+\phi\right)^2}~.
\end{aligned}\end{equation}
This decoupling ensures that the various modes \(n\in\mathbb{Z}\) are disentangled from one another, such that the total density matrix \(\rho_{2D}\) of the \(2+1-\)dimensional system can be written as a product of the density matrices \(\rho_{1D}^n\) for each of the modes:
\begin{equation}\begin{aligned}
	\label{prod_rhod}
	\rho_\text{2D} = \prod_{n=-\infty}^\infty \rho^n_\text{1D}~.
\end{aligned}\end{equation}
We will focus primarily on the case of massless \(2+1-\)dimensional electrons by setting \(m=0\), such that the effective mass is given by~\(M_{n,\phi} = \frac{2\pi}{L_x}|n+\phi|\). Certain results will, however, be generalised to the case of massive fermions ($m\neq 0$). While the massless case corresponds to the gapless excitations of a Fermi liquid or those at the surface of a topological insulator, the massive case corresponds to gapped or insulating states of matter, e.g., the Bogoliubov-de Gennes quasiparticles above an \(s-\)wave superconducting gap. Massive Dirac fermions have also been observed experimentally in several materials like ultrathin films~\cite{lu_2010}, topological insulating crystals~\cite{zhao_2021_massive} and van der Waals heterostructures~\cite{hunt_2013}. They are also emergent at low-energies, for instance, in fermionic nonlinear \(\sigma-\)models~\cite{abanov2000}, and can also be realised near the IR fixed point of the kagome Heisenberg quantum antiferromagnets placed in an external magnetic field along the \(z-\)direction~\cite{santanukagome}.

\begin{figure*}
	\centering
	\includegraphics[width=0.45\textwidth]{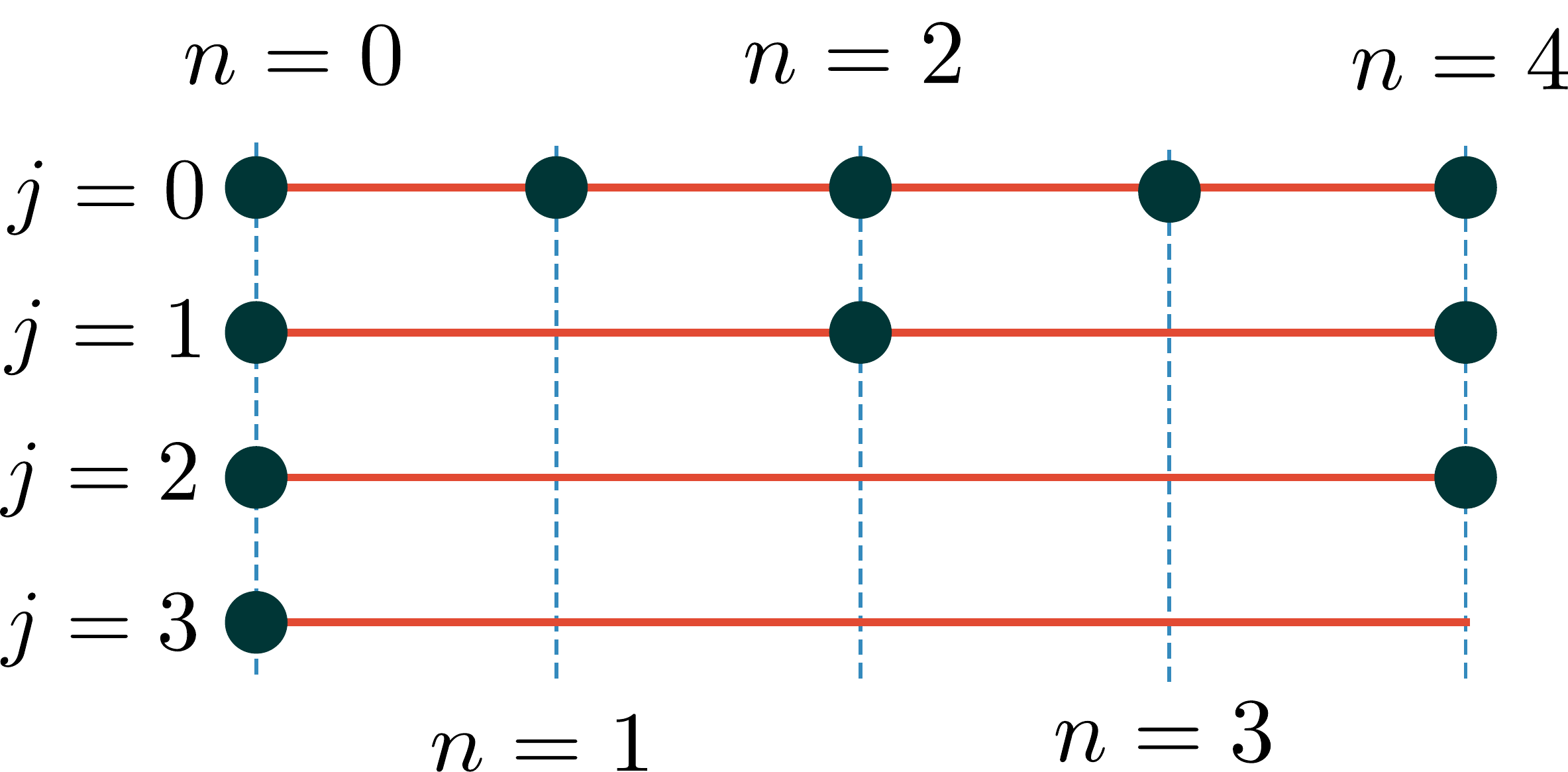}
	\includegraphics[width=0.45\textwidth]{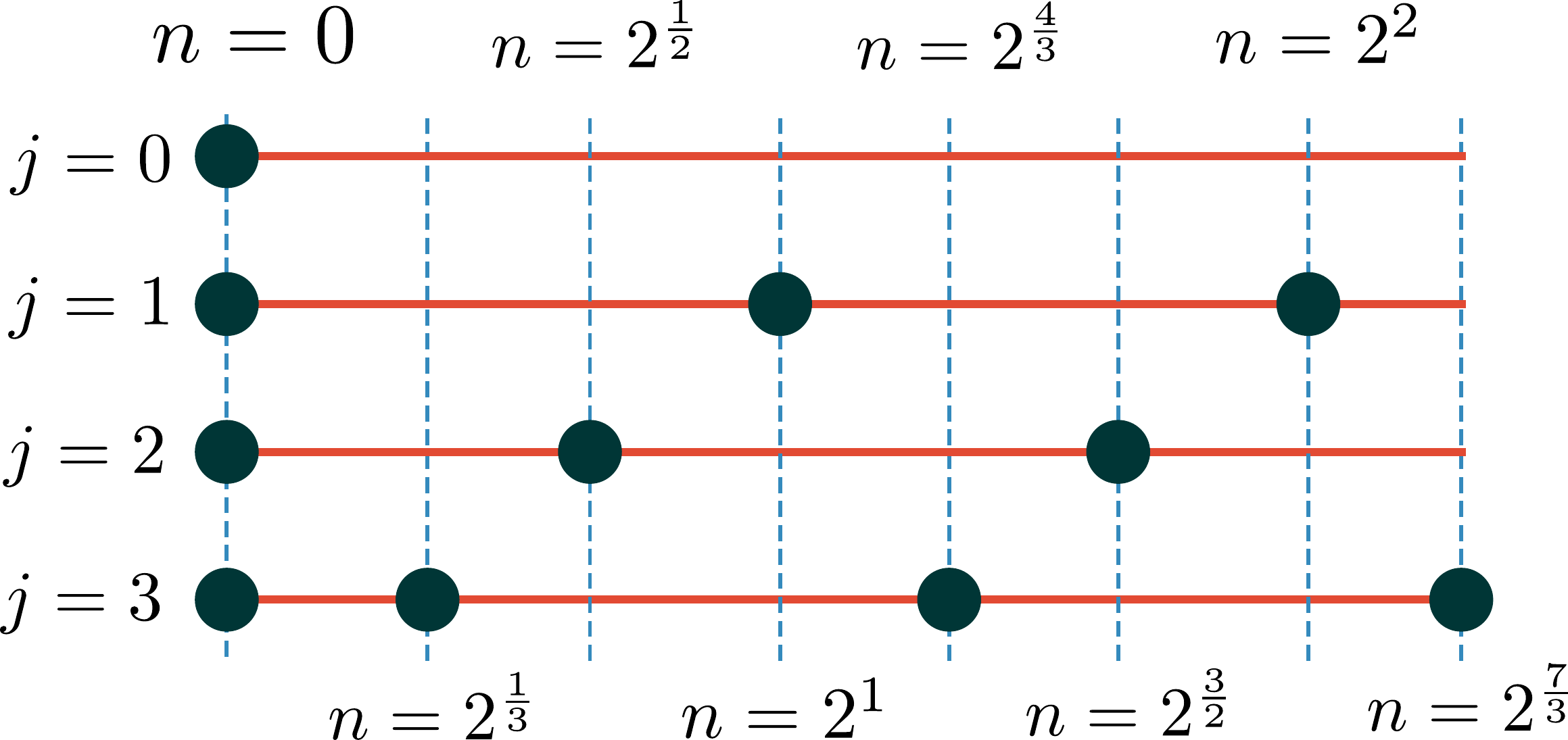}
	\caption{Visual depiction of the subsets \(\mathcal{A}_z(j)\) employed by us in this work, for \(z=1\) and \(z=-1\). As \(j\) is changed by \(1\), the spacing between the modes in the subset increases by a factor of \(2^{j^z}\). {\it Left:} \(z = 1\). The initial subset \(\mathcal{A}_z(0)\) has \(k-\)states at the modes \(n=0,1,2,3,4\). Upon changing \(j\) by 1, the new subset obtained has double the spacing between the modes \((2^{j^z} = 1/2^j)\) of the previous number of modes. That is, if \(j=0\) has \({0,1,2,3,4}\), then \(j=1\) has \({0,2,4}\), \(j=2\) has \({0,4}\) and \(j=3\) has \({0}\).
		{\it Left:} \(z = -1\). In this case, the number of modes get double at each state (\(2^{-i^z} = 2^i\)). We start with \(k-\)modes present only at \(n=0\), but the scaling transformations lead to \(j=1\) having the modes \({0,4}\), \(j=2\) having \({0,2,4}\), and so on.
}
	\label{fine-graining}
\end{figure*}
\subsection{Entanglement measures}

We will work with two measures of entanglement discussed below.
Given the density matrix \(\rho = \ket{\Psi}\bra{\Psi}\) for the ground state \(\ket{\Psi}\) of the complete system, the entanglement entropy (EE) of a set of subsystems \(\left\{ \mathcal{A}_i, i=1,2,\ldots,g \right\}\) with the rest of the system is given by:
\begin{equation}\begin{aligned}
	S_{\left\{ \mathcal{A}_i\right\}} = - \text{Tr}\left[\rho_{\left\{ \mathcal{A}_i\right\}} \ln \rho_{\left\{ \mathcal{A}_i\right\}}\right];\qquad \rho_{\left\{ \mathcal{A}_i\right\}} = \text{Tr}_{\left\{ \mathcal{A}_i\right\}}\left[\rho\right] ~.
\end{aligned}\end{equation}
\(\text{Tr}\left[\cdot\right] \) is the trace operation over all degrees of freedom, while \(\text{Tr}_{\left\{ \mathcal{A}_i\right\}}\left[\cdot\right]\) is the partial trace over only the states corresponding to the set of subsystems \({\left\{ \mathcal{A}_i\right\}}\).
\(\rho_{\left\{ \mathcal{A}_i\right\}}\) is referred to as the reduced density matrix for the rest of the system (complement of the set \(\left\{ \mathcal{A}_i\right\}\)).
Using EE, one can define the mutual information \(I(\left\{ \mathcal{A}_i\right\}:\left\{ \mathcal{B}_i\right\})\) between two subsystems \(A\) and \(B\):
\begin{equation}\begin{aligned}
	\label{mut-inf}
	I_2\left(\left\{ \mathcal{A}_i\right\}:\left\{ \mathcal{B}_i\right\}\right) = S(\left\{ \mathcal{A}_i\right\}) + S(\left\{ \mathcal{B}_i\right\}) - S(\left\{ \mathcal{A}_i\right\}\cup \left\{ \mathcal{B}_i\right\})~,
\end{aligned}\end{equation}
where \(S(\left\{ \mathcal{A}_i\right\}\cup \left\{ \mathcal{B}_i\right\})\) is the EE of \(\left\{ \mathcal{A}_i\right\}\cup \left\{ \mathcal{B}_i\right\}\) with the rest.
Higher order measures of information among $N$ subsystems can also be defined similarly~(see, e.g., \cite{siddhartha_TEE}). Such multi-partite measures of quantum information convey the degree of correlations present within the system - the presence of \(N-\)particle correlations ensures all information measures up to the \(N-\)partite information will be non-zero.

One final object of interest in this context is the entanglement Hamiltonian \(\mathcal{H}(\left\{ \mathcal{A}_i\right\})\) for a subsystem \(\left\{ \mathcal{A}_i\right\}\), defined through the relation~\cite{li2008entanglement,CalabreseLefevre2008}
\begin{equation}\begin{aligned}\label{ent_Ham}
	\rho_{\left\{ \mathcal{A}_i\right\}} = \frac{e^{-\mathcal{H}(\left\{ \mathcal{A}_i\right\})}}{\text{Tr}\left( e^{-\mathcal{H}(\left\{ \mathcal{A}_i\right\})} \right) }~.
\end{aligned}\end{equation}
The entanglement Hamiltonian is essentially the spectrum of the reduced density matrix, and contains additional information on the nature of correlations beyond the entanglement entropy. For the system we are considering (described in the previous subsection), the total reduced density matrix can be decomposed into decoupled 1D modes (eq.~\eqref{prod_rhod}). This implies that the total entanglement Hamiltonian is simply a sum of the individual Hamiltonians for the 1D modes:
\begin{equation}\begin{aligned}\label{ent_ham_2d}
	\rho_\text{2D} = \prod_{n=-\infty}^\infty \rho^n_\text{1D} \sim \prod_{n=-\infty}^\infty e^{-\mathcal{H}^n_\text{1D}(\left\{ \mathcal{A}_i\right\})} = e^{-\mathcal{H}_\text{2D}(\left\{ \mathcal{A}_i\right\})}~,
\end{aligned}\end{equation}
where \(\mathcal{H}_\text{2D} = \sum_n\mathcal{H}^n_\text{1D}\).

\subsection{Luttinger volume}
The Dirac electrons considered throughout this work should be understood as the low-energy effective description of some microscopic lattice model of interacting electrons that has a well-defined Fermi surface at $T=0$. The Luttinger volume \(V_L\) is defined as the number of \(\vec k-\)space points inside the Fermi surface~\cite{luttinger1960fermi,luttinger1960ground} :
\begin{equation}\label{Luttvol}\begin{aligned}
	V_L = \int_{G_{\vec k}(\omega=0) > 0} \frac{1}{\left( 2\pi \right) ^d} d\vec k = \int_0^{k_F}\frac{1}{\left( 2\pi \right) ^d} d\vec k~.
\end{aligned}\end{equation}
The integral (referred to as Luttinger's integral) runs over all \(\vec k-\)space points where the \(\omega=0\) Greens function is positive, and the Fermi momentum \(k_F\) is defined as the set of points in \(\vec k-\)space where \(G_{\vec k}(\omega=0)\) changes sign. Luttinger's integral can be shown to be topological in nature~\cite{seki2017topological,oshikawa2000topological}, promoting Luttinger's volume to the status of a topological invariant that characterises the system. For systems with long-lived excitations close to the Fermi surface, \(V_L\) is equal to the average number density of electrons in the system, leading to Luttinger's theorem~\cite{luttinger1960fermi,luttinger1960ground,Heath_2020}.  For systems with gapped excitations in the momentum window \(k \leq k^*\) above the ground state, the violation \(\Delta V_L\) of Luttinger's theorem can be quantified through a count of the bound states~\cite{anirbanurg1}:
\begin{equation}\begin{aligned}\label{LT-violation}
	\Delta V_L = \int_0^{k_F} \frac{1}{\left( 2\pi \right) ^d} d\vec k - \int_{k^*}^{k_F} \frac{1}{\left( 2\pi \right) ^d} d\vec k = \int_{0}^{k^*} \frac{1}{\left( 2\pi \right) ^d} d\vec k
\end{aligned}\end{equation}

\section{Entanglement hierarchy in mixed momentum and real space}
\label{hierarchy}

Before beginning our calculations of entanglement measures, we define the subsystem(s) whose entanglement we want to obtain. The subsystems we will use in this work are of the following kind:
\begin{equation}\begin{aligned}
	\mathcal{A}_{z}(i) = \widetilde{A}_{z}(i) \times A_{z}(i)~,
\end{aligned}\end{equation}
where \(i\) takes integral values starting from zero, \(z\) can take all integral values apart from 0 and \(\times\) represents the Cartesian product between the set of real space points \(\widetilde{A}_{z}(i) = \left\{0\leq y \leq l\right\}\) and the set of momentum space points 
\begin{gather}
	A_{z}(i) = \bigg\{k_x^n\bigg\};~ k_x^n = \frac{2\pi}{L_x}n;~\text{max}(\{k_x^n\}) = \frac{2\pi N}{L_x}~;\nonumber\\
	~ n \in \left\{-N, -N + t_{z}(i),\ldots,-t_{z}(i),0,t_{z}(i),N\right\};~ t_{z}(i) = 2^{i^z}~,
\end{gather}
and where \(t_{z}(i) = 2^{i^z}\) is the difference between consecutive values of \(n\).
The subset \(\mathcal{A}_{z}(i)\) is therefore a region in mixed momentum and real spaces, spanned by the variables \(k_x\) and \(y\) respectively. In \(y-\)space, the region is of length \(l\) and bounded by the points \(y=0\) and \(y=l\). In \(k_x-\)space, \(\mathcal{A}_{z}(i)\) comprises the points \(\left\{0,\pm 2^{i^z} \frac{2\pi}{L_x},\pm 2^{i^z} \frac{4\pi}{L_x},\ldots\right\}\) such that \(|k_x| \leq \frac{2\pi N}{L_x}\). For example, for the case of \(z=1\) and \(N\) being a multiple of 4, we have
\begin{widetext}
\begin{equation}\begin{aligned}
	A_{1}(0) = \left[-\frac{2\pi}{L_x}N, -\frac{2\pi}{L_x}(N-1), \ldots, -\frac{2\pi}{L_x}2, -\frac{2\pi}{L_x}1, 0,\frac{2\pi}{L_x}1, \frac{2\pi}{L_x}2, \ldots, \frac{2\pi}{L_x}(N-1), \frac{2\pi}{L_x}N\right] ~,\\
	A_{1}(1) = \left[-\frac{2\pi}{L_x}N, -\frac{2\pi}{L_x}(N-2), \ldots, -\frac{2\pi}{L_x}4, -\frac{2\pi}{L_x}2, 0,\frac{2\pi}{L_x}2, \frac{2\pi}{L_x}4, \ldots, \frac{2\pi}{L_x}(N-2), \frac{2\pi}{L_x}N\right] ~,\\
	A_{1}(2) = \left[-\frac{2\pi}{L_x}N, -\frac{2\pi}{L_x}(N-4), \ldots, -\frac{2\pi}{L_x}8, -\frac{2\pi}{L_x}4, 0,\frac{2\pi}{L_x}4, \frac{2\pi}{L_x}8, \ldots, \frac{2\pi}{L_x}(N-4), \frac{2\pi}{L_x}N\right]~.
\end{aligned}\end{equation}
\end{widetext}
Note that \(N\) represents the edge of the Brillouin zone, and will be sent to infinity when we calculate quantities in the continuum limit. In Fig.~\ref{fine-graining}, we have shown two such families of subsets \(\mathcal{A}_z(0), \mathcal{A}_z(1), \mathcal{A}_z(2), \mathcal{A}_z(3), \ldots\), for \(z=1\) and \(z=-1\).

We note here that the transformations for \(z<0\) correspond to increasing the density (fine-graining) of \(k_x-\)states by adding more momentum states while keeping the bounds fixed. This involves reducing the interval in \(k_x-\)space, and hence equivalent to increasing the system size \(L_x\). This is similar to methods such as numerical renormalisation group treatment of the Kondo problem ~\cite{wilson1975} where an increasing number of states are added to the system at increasingly larger distances from the impurity site, such that the system asymptotically goes to the thermodynamic limit. 
The other case of \(z>0\) is, on the contrary, a decimation process in \(k_x-\)space, and equivalent to reducing the system size. Indeed, it will be shown later that the sequence of Hamiltonians generated in the process (for both positive and negative \(z\)) are related to one other by renormalisation group transformations. 

\begin{figure}
	\includegraphics[width=0.45\textwidth]{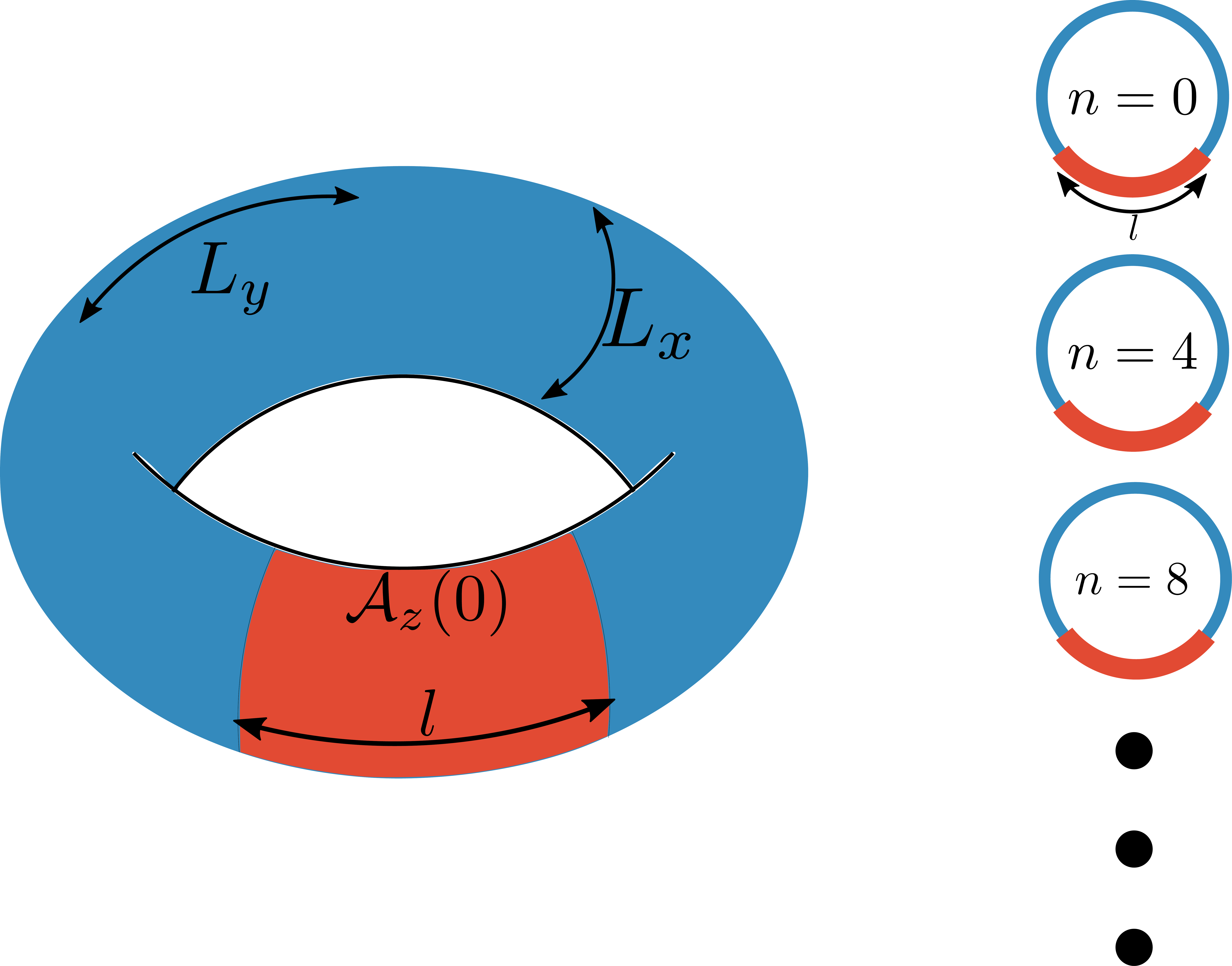}
	\caption{{\it Left:} Depiction of the subset \(\mathcal{A}_z(i)\) for the choice \(i=0\). Setting \(i=0\) leads to unchanged length \(L_x\) along \(x\) and the constraint \(0\leq y \leq l\) in the \(y-\)direction. This creates an annulus (red region) of height \(l\) and circumference \(L_x\).\\
	{\it Right:} Typical \(k-\)modes present in the subset \(\mathcal{A}_z(i)\) for \(z=1\) and \(i=2\). This is because, if \(i=0\) has the modes \(n=0,1,2,3,4,\ldots,\), increasing \(i\) by 2 increases the spacing between the modes to \(2^{2^z}=4\), so that the mode just after \(n=0\) is \(n=4\), and the one following that is \(n=8\), and so on.}
	\label{Am-1}
\end{figure}

\begin{figure}
	\centering
	\includegraphics[width=0.48\textwidth]{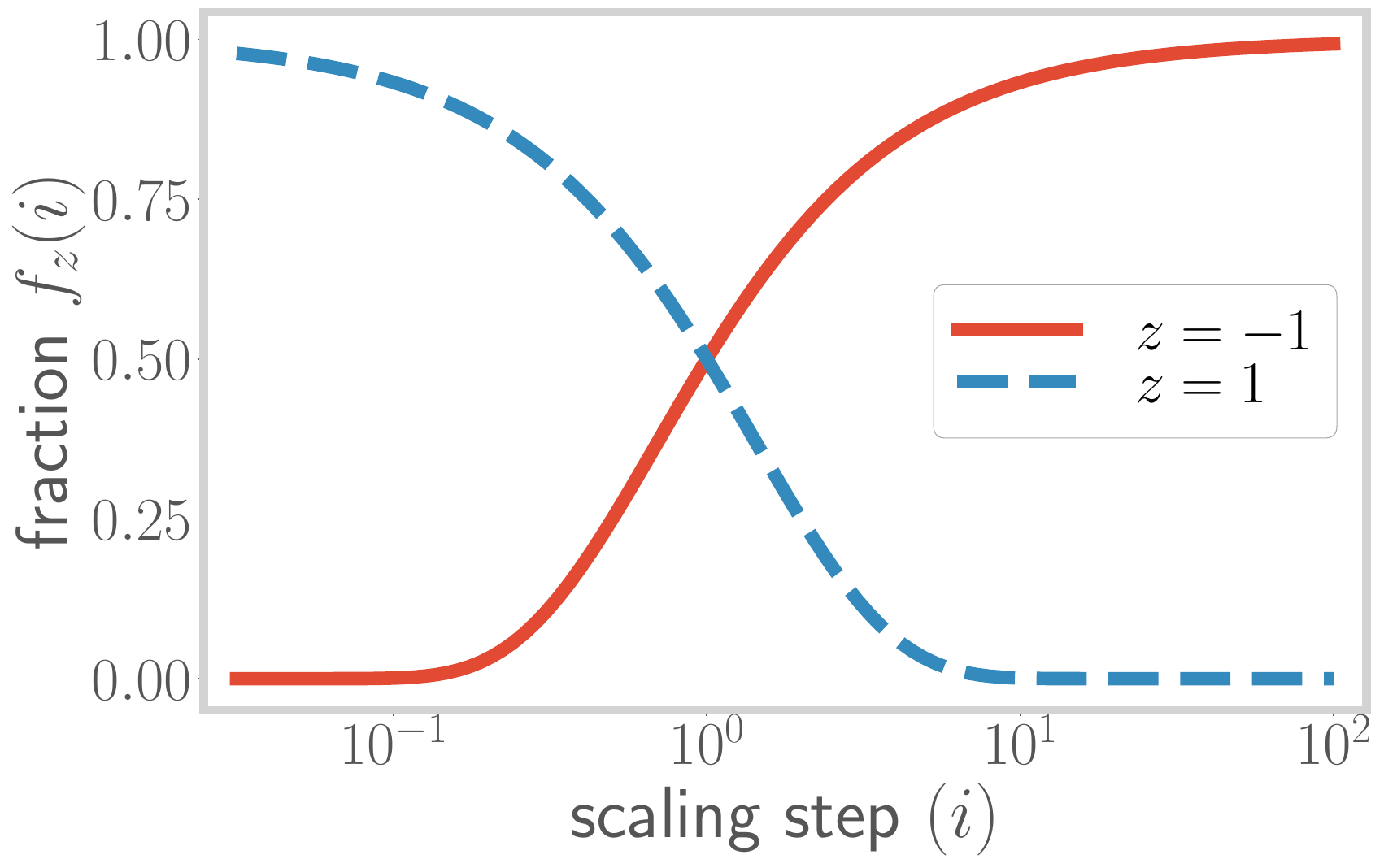}
	\caption{A plot of the fraction of states remaining upon performing \(i\) number of transformations on the starting Hamiltonian. Obtained by plotting the expression \(f = 2^{-i^z}\), for a range of \(i\), and for two values of \(z\) (red and blue). This fraction decreases for \(z>0\) (blue curve), and increases for \(z<0\) (red curve).}
	\label{fraction}
\end{figure}

Within a given set \(\mathcal{A}_{z}(i)\), the distance between adjacent \(k-\)space points is \(\frac{2\pi}{L_x} t_{z}(i)\) where \(t_{z}(i) = 2^{i^z}\). Choosing \(i=0\) keeps the set of \(k_x-\)space points unchanged, such that we recover the entire Lagrangian of eq.~\eqref{lagrangian} in the \(x-\)direction for system size \(L_x\), while the \(y-\)direction is constrained to length \(l\).
This means that \(\mathcal{A}_z(0)\) is just an annulus that wraps around the \(x-\)direction of the torus and extends to a length \(l\) longitudinally in the \(y-\)direction (shown in Fig.~\ref{Am-1}).
Along with the period \(t_{z}(i)\), one can also define the fraction of maximum points present in the set \(\mathcal{A}_{z}(i)\) as
\begin{equation}\begin{aligned}
	f_{z}(i) = 1/t_{z}(i) = 2^{-i^z}~.
\end{aligned}\end{equation}
For example, for \(z>0\), \(f_{z}(i) =2^{-i^z}\) expresses the fact that in going from the set at \(j=0\) to the set at \(j=i\), the number of elements in the set is scaled down by a factor of \(2^{i^z}\) (see Fig.~\ref{fraction}).

We now define the shorthand notation \(S_{z}(i)(\phi) \equiv S_{\mathcal{A}_{z}(i)}(\phi)\) to denote the EE of the subsystem \(\mathcal{A}_{z}(i)\) in the presence of the Aharonov-Bohm flux \(\phi\). In order to calculate \(S_{z}(i)\), we start with \(i=0\). As mentioned earlier, this is essentially the EE between an annulus of length \(l\) and the rest of the \(2-\)torus. Since the density matrices are in product, the total EE is just a sum of the EE arising from each mode:
\begin{equation}\begin{aligned}
	S_{\mathcal{A}_{z}(0)}(\phi) \equiv S_{z}(0)(\phi) = \lim_{N \to \infty}\sum_{n=-N}^N S_{\widetilde{A}_{z}(0)}^n~,
\end{aligned}\end{equation}
where \(\widetilde{A}_{z}(0)\) acts as a projection of the annulus on the \(k_x-y\) plane, and
\(S^n\) indicates that the EE is calculated for the \(n^\text{th}\) mode. Each massive fermionic mode has a correlation length-scale \(\xi_n = 1/M_{n,\phi}\). We choose \(l\) to be much larger than the largest correlation length: \(l \gg \text{max}\left(\left\{\xi_n,\forall~n\right\}\right)\). Using the expression for \(\phi\) and the quantisation of the momenta \(k^n_x\), this condition can be expressed as
\begin{equation}\begin{aligned}
	l \gg \frac{L_x}{2\pi}\times \text{max} \left(\left\{\frac{1}{|n + \phi|}\right\}\right) = \frac{L_x}{2\pi \left\{\phi\right\}}~,
\end{aligned}\end{equation}
where \(\left\{\phi\right\} = \phi - \lfloor \phi \rfloor\) is the fractional part of the flux \(\phi\) in units of the flux quantum. Since we are in the thin torus limit, \(L_x\) must remain finite, and the maximum correlation can diverge only if \(\{\phi\}=0\). We ensure that \(l\) is larger than the correlation lengths by choosing \(\phi\) to be necessarily non-integral, as this guarantees that the correlation lengths remain finite. Then, the EE of the annulus is given by~\cite{Arias_2015,Chen_2017}
\begin{equation}\begin{aligned}
	\label{annulus}
	S_{z}(0)(\phi) = c\left[\alpha \frac{L_x}{\epsilon} - \ln \big|2\sin\left(\pi\phi\right) \big|\right]~,
\end{aligned}\end{equation}
where \(c\) is the conformal central charge and \(\epsilon\) is a UV cutoff; \(c\) is \(1/3\) for a Dirac fermion. The computation of \(S_{z}(0)(\phi)\) uses the EE \(S_{\widetilde{A}_{z}(0)}^n\) of a \(1-\)D chain of fermions with a mass \(M_{n,\phi}\)~\cite{calabrese2004,Casini_2009}: \(S_{\widetilde{A}_{z}(0)}^n = c\ln \frac{L_x}{2\pi\epsilon} - c\ln |n+\phi|\). On summing over \(n\) and regularising the UV cutoff, the first term of eq.\eqref{annulus} produces the leading area-law term of \(S_0(\phi)\)~\cite{Chen_2017}:
\begin{equation}\begin{aligned}
	\sum_{n=-\infty}^\infty c\ln \frac{L_x}{2\pi\epsilon} \to c\alpha \frac{L_x}{\epsilon}~,
\end{aligned}\end{equation}
where \(\alpha\) is a cutoff-dependent non-universal constant.The flux-dependent term in \(S_{\widetilde{A}_{z}(0)}^n\) leads to the sub-leading term of \(S_{z}(0)(\phi)\), and requires a regularisation using the Hurwitz zeta function~\cite{Chen_2017}:
\begin{equation}\begin{aligned}
	\sum_{n=-\infty}^\infty \ln|n+\phi| = \sum_{n=-\infty}^\infty \left[\ln|n+\phi| + \ln|n+\left( 1-\phi \right) |\right] \\
	= -\frac{\partial{\zeta(\phi)}}{\partial{\phi}} - \frac{\partial{\zeta(1-\phi)}}{\partial{\phi}}~,
\end{aligned}\end{equation}
where \(\zeta(\phi)=\sum_{n=-\infty}^\infty \left(n+\phi\right)^{-1} \) is the Hurwitz zeta function.
The derivative converges to \(\frac{\partial{\zeta(\phi)}}{\partial{\phi}} = \ln \Gamma(\phi) - \frac{1}{2}\ln 2\pi\), as long as the zero mode \(n+\phi=0\) is excluded from the summation. The latter condition is guaranteed as we have chosen \(\phi\) to be a non-integer. \(\Gamma\) is the Gamma function, satisfying \(\Gamma(\phi)\Gamma(1 - \phi) = \pi/\sin(\pi\phi)\).
Substituting the derivative gives~\cite{Arias_2015,Chen_2017,crepel_2021}
\begin{equation}\begin{aligned}
	\label{gamma}
	\sum_{n=-\infty}^\infty \ln|n+\phi| = \ln 2\pi -\ln \left[\Gamma(\phi)\Gamma(1-\phi)\right] = \ln \big|2\sin\left( \pi\phi \right) \big|~.
\end{aligned}\end{equation}

For a general \(\mathcal{A}_{z}(i)\), there are certain modifications to the expression in eq.~\eqref{annulus}. For the first term in eq.\eqref{annulus}, we have
\begin{equation}\begin{aligned}
	\label{area-term}
\lim_{N \to \infty} \sum_{n=-N,\atop{-N+t_{z}(i),\ldots}}^N c\ln \frac{L_x}{2\pi\epsilon} = \lim_{N \to \infty} \sum_{n^\prime=-\frac{N}{t_{z}(i)},\atop{-\frac{N}{t_{z}(i)}+1,\ldots}}^{\frac{N}{t_{z}(i)}} c\ln \frac{L_x}{2\pi\epsilon}\\
= t^{-1}_{z}(i) \lim_{N \to \infty} \sum_{n=-N}^N c\ln \frac{L_x}{2\pi\epsilon} \rightarrow c f_{z}(i)\frac{\alpha L_x}{\epsilon}~.
\end{aligned}\end{equation}
Similarly, for the second term, we have
\begin{equation}\begin{aligned}
	\lim_{N \to \infty} &\sum_{n=-N,\atop{-N+t_{z}(i),\ldots}}^N \ln|n+\phi| \\
	&= \lim_{N \to \infty} \sum_{n^\prime=-\frac{N}{t_{z}(i)},\atop{-\frac{N}{t_{z}(i)}+1,\ldots}}^{\frac{N}{t_{z}(i)}} \ln|n^\prime t_{z}(i) + \phi| \\
										&= \lim_{N \to \infty} \sum_{n^\prime=-\frac{N}{t_{z}(i)},\atop{-\frac{N}{t_{z}(i)}+1,\ldots}}^{\frac{N}{t_{z}(i)}} \ln \left(f_{z}(i)|n^\prime + f_{z}(i)\phi|\right) \\
										&= \sum_{n^\prime=-\infty}^{\infty} \ln |n^\prime +  f_{z}(i)\phi|~,
\end{aligned}\end{equation}
where we have assumed that \(t_{z}(i)\) is finite, and dropped the constant part \(\ln f_{z}(i)\) (which can be absorbed into the UV cutoff \(\epsilon\) in eq.~\eqref{area-term}). Following eq.~\eqref{gamma}, this gives
\begin{equation}\begin{aligned}
	\lim_{N \to \infty} \sum_{n=-N,\atop{-N+t_{z}(i),\ldots}}^N \ln|n+\phi| = \ln |2\sin\left(\pi f_{z}(i)\phi\right) |~.
\end{aligned}\end{equation}
Combining both parts, we obtain the expression for the EE that will be employed in what lies ahead:
\begin{equation}\begin{aligned}
	\label{EE-subsystem}
	S_{z}(i)(\phi) = c f_{z}(i) \frac{\alpha L_x}{\epsilon} - c\ln \big|2\sin\left(\pi f_{z}(i)\phi\right) \big|~.
\end{aligned}\end{equation}
It is evident from the expression in eq.~\eqref{EE-subsystem} that, as modes are added into the subsystem, the \(L_x-\)dependent geometric part of the EE increases, leading to a {\it hierarchical structure} in the entanglement. A similar result was previously shown to hold for geometric measures of entanglement of quantum-mechanical systems with eigenstates such as the \(W\) and \(GHZ\) states and their symmetric superpositions~\cite{blasone_2008}. The variation of the geometric part with the subsystem index \(j\) is shown in Fig.~\ref{fraction}.

The above-mentioned hierarchy in entanglement measures leads to an interesting property for the entanglement: the combined entanglement entropy of any number of systems \(\left\{ \mathcal{A}_{z}(i) \right\} \) is equal to the entanglement entropy of the most dense system among them. For \(z > 0\), the sets become smaller as the RG index increases, while for \(z < 0\), the sets become larger as the RG index increases. Thus, we may write
\begin{equation}\begin{aligned}\label{ent-union}
	S_{\left\{ \mathcal{A}_{z}(i) \right\}} = \theta(z) S_{\text{min}\left\{ i \right\} ,z} + \theta(-z) S_{\text{max}\left\{ i \right\},z},~.
\end{aligned}\end{equation}
The mutual information \(I^2(i,j)\) between the systems at two RG steps \(i\) and \(j\) can then be computed:
\begin{equation}\begin{aligned}\label{mutinfo}
	I^2(i,j) &\equiv S_{z}(i) + S_{j,z} - S_{i \cup j,z} &\\
		 &= \theta(-z) S_{\text{min}(i,j),z} + \theta(z) S_{\text{max}(i,j),z}~.
\end{aligned}\end{equation}

In light of this subsystem-based hierarchical structure, we can also consider the \(g-\)partite information \(I^g_{\left\{ \mathcal{A}_{z}(i) \right\}} \) among \(g\) sets \(\left\{\mathcal{A}_{z}(i)\right\}\). $I^g$ is defined as~\cite{siddhartha_TEE}
\begin{equation}\begin{aligned}
	\label{Ig_def}
	I^g_{\left\{ \mathcal{A}_{z}(i) \right\}} = \sum_{r=1}^g \left( -1 \right)^{r+1} \sum_{\beta \in \mathcal{P}\left(\left\{ \mathcal{A}_{z}(i) \right\}\right)\atop{|\beta|=r}} S_\beta~,
\end{aligned}\end{equation}
where \(\mathcal{P}\left(\left\{ \mathcal{A}_{z}(i) \right\}\right)\) is the power set of \(\left\{ \mathcal{A}_{z}(i) \right\}\), and \(|\beta|=r\) indicates that the sum is carried out only over those sets \(\beta\) that have \(r\) number of elements in them. For example, for \(g=3\), we have
\begin{equation}\begin{aligned}
	I^3 =& S_{\mathcal{A}_{z}(0)} + S_{\mathcal{A}_{z}(1)} + S_{\mathcal{A}_{z}(2)} - S_{\mathcal{A}_{z}(0)\cup \mathcal{A}_{z}(1)} \\
	     &- S_{\mathcal{A}_{z}(1)\cup \mathcal{A}_{z}(2)} - S_{\mathcal{A}_{z}(2)\cup \mathcal{A}_{z}(0)} + S_{\mathcal{A}_{z}(0)\cup \mathcal{A}_{z}(1) \cup \mathcal{A}_{z}(2)}~.
\end{aligned}\end{equation}
For our choice of subsystems, the \(g-\)partite information can be shown to be equal to the EE of the {\it intersection set} of the concerned \(g-\)sets (details in \alert{Appendix~\ref{g-partite info}}). For the case of \(z > 0\), this takes the form
\begin{equation}\begin{aligned}
	\label{manyparty-info}
	I^g_{\left\{ \mathcal{A}_{z}(i) \right\}}(\phi) = S_{g}(\phi) = c f_{g,z}\alpha \frac{L_x}{\epsilon} - c\ln \big|2\sin\left(\pi f_{g,z} \phi\right) \big|~.
\end{aligned}\end{equation}
Since the \(g-\)partite information is simply the EE of the \(g^\text{th}\) subsystem, the hierarchy in the EE can thus be seen to lead to multiple levels of correlation in the system, corresponding to an increasing number of parties entering the set.

Finally, we note that the geometric (\(L_x-\)dependent) part of the EE in eq.~\eqref{EE-subsystem}, as well as any of the information measures $I^g$ (eq.~\eqref{manyparty-info}), has the same form even if we start with a non-zero mass \(m > 0\)~\cite{Chen_2017}. This ensures that the above observations regarding the hierarchical structure of the entanglement hold even for gapped electronic systems.

\section{EHM, the Ryu-Takayanagi bound and holographic inequalities}
\label{EHM}
As we will now show, the approach laid out above to the creation of subsystems amounts essentially to performing a sequence of renormalisation group (RG) transformations on the momentum modes of the Hamiltonian in the \(x-\) direction. We consider here the general case of massive Dirac electrons \(M_{n,\phi} = \sqrt{m^2 + \frac{\pi^2}{L_x^2}\left( n + \phi \right)^2}\). Further, we define the superset \(\mathcal{A}^{(0)}_z\) of \(k_x-\)states: \(\mathcal{A}^{(0)}_z = \bigcup_{i=0}^\infty \mathcal{A}_{z}(i)\), such that all \(k_x-\)states that can appear at any point of the sequence \(\left\{ \mathcal{A}_{z}(i); i=0,1,2,\ldots \right\} \) are present within it. This superset can now be used to define a parent Hamiltonian \({H}(0)\) that consists of all the modes in the \(x-\)direction:
\begin{equation}\begin{aligned}
	{H}(0) = \sum_{k_x \in \mathcal{A}^{(0)}_{z}}{H}(k_x)~,
\end{aligned}\end{equation}
where \({H}(k_x)\) is the Hamiltonian for a massive Dirac fermion that extends over a length \(L_y\) in the single spatial dimension, and having mass \(M_{n,\phi}\). The Hamiltonian at any given step of the RG is then obtained by projecting onto appropriate sets of momenta.
The set of momentum at step \(j\) is given by \(\mathcal{A}_{j,z} = \frac{2\pi}{L_x}\times\left\{0, \pm t_{z}(j), \pm, 2t_{z}(j), \ldots \right\} \) with \(t_{z}(j) = 2^{j^z}\).
The projector onto this set is given by
\begin{equation}\begin{aligned}
	P_{j,z} = \sum_{k_x^1=0,1}\sum_{k_x^2=0,1}\ldots\sum_{k_x^N=0,1} \ket{\hat n_{k_x^1} \hat n_{k_x^2}\ldots}\bra{\hat n_{k_x^1} \hat n_{k_x^2}\ldots},\\
	k_x^j \in \mathcal{A}_{j,z}~,
\end{aligned}\end{equation}
and the Hamiltonian at the corresponding step is then obtained as
\begin{equation}\begin{aligned}
	\label{mapping}
	{H}_{z}(i) = P_{z}(i) {H}(0) P_{z}(i)~.
\end{aligned}\end{equation}
\begin{figure}
	\centering
	\includegraphics[width=0.45\textwidth]{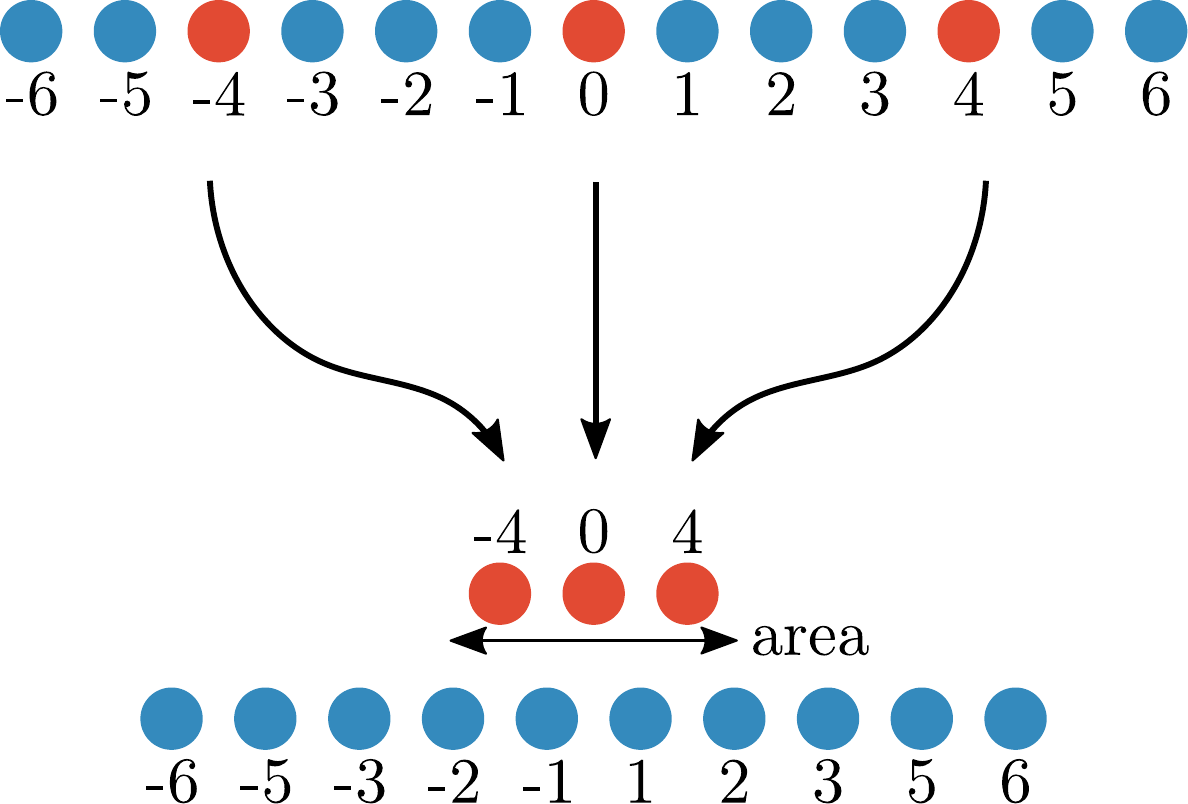}
	\includegraphics[width=0.45\textwidth]{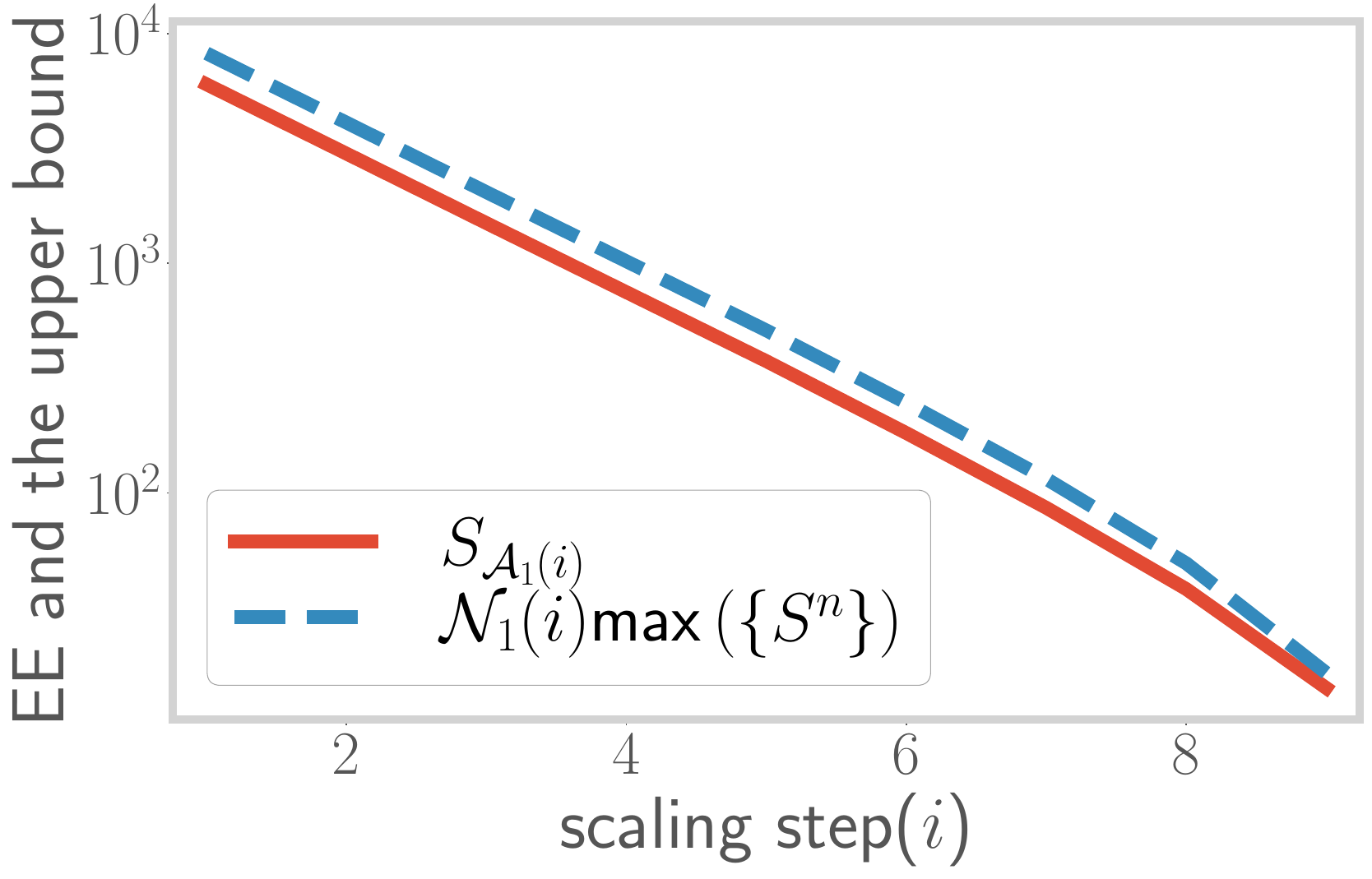}
	\caption{{\it Top:} Modes forming the subsystem \(\mathcal{A}_{z}(i)\) (red circles) form the boundary between \(\mathcal{A}_{z}(i)\) and its complement (blue circles).
	{\it Bottom:} Scaling of the subsystem entanglement entropy (red), and its upper bound (blue), as the RG transformations are performed (denoted by increasing index $i$). The bound is saturated at large \(i\).}
	\label{area}
\end{figure}

In this formulation of the RG, the index \(i\) represents the RG step, while the parameter \(z\) indicates the {\it strength of the scaling} mechanism: for \(z>0 \), a higher value of \(z\) means that the states of the Hamiltonian become rarer as the index \(i\) increases, while for \(z<0\), the states become denser with increasing \(i\). Importantly, we will see subsequently that the parameter $z$ corresponds also to the anomalous dimension of the coupling that leads to a spectral gap in the electronic dispersion, dictating its growth or decay under RG evolution (eq.\eqref{RGbetafunc}).

Since this is a free theory, the \(k-\)modes are decoupled, and the projection operation preserves the spectrum of the remaining degrees of freedom. The renormalised Hamiltonian (eq.~\eqref{mapping}) can then be said to constitute an exact holographic mapping (EHM)~\cite{qi2013exact,lee2016} in the direction of the RG transformations. It is also pertinent to note that the sequence of transformations in eq.~\eqref{mapping} can be thought of as a simple and specific example of a unitary renormalisation group (URG) procedure~\cite{anirbanurg1,anirbanurg2} that first disentangles and then projects out a certain set of degrees of freedom. Because the operation is unitary, the URG preserves the spectrum in the process.
In the present case, the disentanglement operation is unity. The URG has already been shown to lead to a holographic renormalisation of the Hilbert space geometry which can be seen through the evolution of the Fubini-Study metric under the URG transformations~\cite{anirbanurg1}. Our approach differs from MERA implementations like those in refs.~\cite{evenbly2009,evenbly_2016} in the fact that while those are involve a real space renormalisation scheme, we perform our scaling transformations in momentum space.

More evidence for the holographic nature of the construction comes from the fact that the entanglement entropy for both the massless and massive cases satisfy the bound provided by the Ryu-Takayanagi proposal that relates the entanglement of a conformal field theory (CFT) with the geometry of its gravity dual~\cite{ryu2006,ryu2006aspects}:
\begin{equation}\begin{aligned}
	\label{bound}
	S_{\mathcal{A}_{z}(i)} \leq \mathcal{N}_{z}(i) \text{max}\left(\left\{ S^n \right\}\right) ~.
\end{aligned}\end{equation}
In the above constraint, \(\text{max}\left(\left\{ S^n \right\}\right)\) is the maximum single entropy of all the modes \(n\) residing in \(\mathcal{A}_{z}(i)\), and \(\mathcal{N}\) is the area of the minimal surface that separates the subsystem \(\mathcal{A}_{z}(i)\) from the rest.
In order to see how this is satisfied in the presence case, we note that since the subsystem \(\mathcal{A}_{z}(i)\) is interspersed among the rest of the system (see \alert{top panel} of Fig.~\ref{area}), the minimal surface is as large as the subsystem itself.
The area of this surface is therefore given by the number of modes in \(\mathcal{A}_{z}(i)\).
Using this information, the right hand side of eq.~\eqref{bound} can be written as
\begin{equation}\begin{aligned}
	\label{bound1}
	\mathcal{N}_{z}(i) \text{max}\left(\left\{ S^n \right\}\right) = \text{max}\left(\left\{ S^n \right\}\right) N f_{z}(i) = \text{max}\left(\left\{ S^n \right\}\right) \sum_{n \in \mathcal{A}_{z}(i)}~.
\end{aligned}\end{equation}
Following eq.~\eqref{prod_rhod}, the left hand side of eq.~\eqref{bound} can, on the other hand, be written as
\begin{equation}\begin{aligned}
	\label{bound2}
	S_{\mathcal{A}_{z}(i)} = \sum_{n^\prime \in \mathcal{A}_{z}(i)} S^{n^\prime} \leq \sum_{n^\prime \in \mathcal{A}_{z}(i)} \text{max}\left(\left\{S^{n}\right\}\right)~.
\end{aligned}\end{equation}
Comparing eqs.~\eqref{bound1} and \eqref{bound2}, we recover the Ryu-Takayanagi condition of eq.~\eqref{bound}. The scaling of both the left and right hand sides of eq.~\eqref{bound} has been shown in the \alert{bottom panel} of Fig.~\ref{area}.

Entanglement measures like entanglement entropy and mutual information (associated with disjoin intervals) have been shown to satisfy a number of inequalities in quantum field theories with gravity duals. We now consider some of them here, and show that the theory considered by us here also satisfies these inequalities. One such constraint is the monogamy of mutual information, which says that for three disjoint regions \(A_1, A_2\) and \(A_3\), we must have \(I_2(A_1:A_2) + I_2(A_1:A_3) \leq I_2(A_1:A_2\cup A_3)\), where \(I_2(A_1:A_2) = S(A_1) + S(A_2) - S(A_1\cup A_2)\)~\cite{haydenEntanglement2013}. For this, we consider three disjoint regions, all of the kind depicted in the \alert{top panel} of \alert{Fig}.~\ref{Am-1}, but at different positions \(y_i \in \left[l_i, l_i + L\right],i=1,2,3\). We assume that the distance \(l_i - l_{i+1}\) between adjacent regions is much larger than the correlation length. Calabrese and Cardy have shown that the total entanglement entropy of the union of such disjoint regions is obtained by multiplying the entanglement entropy of any one region with the number of disjoin intervals. Putting all this together, we get
\begin{equation}\begin{aligned}\label{identities}
	S(A_1) &= S(A_2) = S(A_3),\\
	S(A_1 \cup A_2) &= S(A_1 \cup A_3) = 2S(A_1),\\
	S(A_1 \cup A_2 \cup A_3) &= 3S(A_1)~,
\end{aligned}\end{equation}
leading to the above inequality being satisfied:
\begin{equation}\begin{aligned}
	I_2(A_1:A_2) + I_2(A_1:A_3) - I_2(A_1:A_2\cup A_3) = 0~.
\end{aligned}\end{equation}

We next consider a cyclic family of inequalities in terms of conditional entropies, introduced and proved in Ref.~\cite{Bao2015}: \(\sum_{i=1}^{2k+1} S(A_i|A_{i+1}\ldots A_{i+k}) \geq S(A_1\ldots A_{2k+1})\), where \(S(A_i|A_j) = S(A_i\cup A_j) - S(A_j)\). Using the results of eq.~\eqref{identities}, we get \(\sum_{i=1}^{2k+1}S(A_i|A_{i+1}\ldots A_{i+k}) = \sum_{i=1}^{2k+1}S(A_i) = (2k+1)S(A_1)\) and \(S(A_1\ldots A_{2k+1}) = (2k+1)S(A_i)\), leading to \(\sum_{i=1}^{2k+1} S(A_i|A_{i+1}\ldots A_{i+k}) = S(A_1\ldots A_{2k+1})\), which also satisfies the inequality. In both cases, we find that the inequality is satisfied by saturating the bound; this is a consequence of the fact that the intervals are separated by distances larger than the correlation length and the total entanglement entropies are simply a sum of the individual entropies.

\section{Holographic geometry of the RG flow}
\label{holography}
\subsection{RG evolution as the emergent dimension}
To obtain a more specific description of the RG flow as the emergence of a geometric space, we introduce a measure of distance. For this, we will employ another bipartite measure of entanglement - the mutual information \(I^2\), as defined in eq.~\eqref{mut-inf}, due to its non-negative nature. We focus on the massless case \(m=0\) for the time-being. Inspired by Refs.~\cite{van2010building,lee2016,hyatt2017,anirban_mott_2022}, we adopt the following definition for distance:
\begin{equation}\begin{aligned}
	\label{info-dist}
	d_z(i,j) \equiv \ln \frac{I^2_\text{max}}{I^2_z(i:j)}~,
\end{aligned}\end{equation}
where \(I^2_z(i:j) \equiv I^2\left( \mathcal{A}_{z}(i), \mathcal{A}_{j,z} \right) \) is the mutual information between the two mentioned sets, and \(I^2_\text{max}\) is the mutual information between two maximally entangled systems. Accordingly, \(d_z(j)\) is zero (i.e., the mutual information is maximum) at \(j=0\) if \(z>0\) and \(j=\infty\) if \(z<0\), and this distance increases (decreases) with the decrease (increase) in the mutual information as the scaling transformations are performed.

We compute the distance between the points \(i\) and \(j\) to obtain the scaling of the distance $d_z(j)$. We assume \( i < j\) without loss of generality. At each step, we tune the flux $\phi$ so as to remove the flux-dependent part from the EE. Following eq.~\eqref{mutinfo}, the mutual information is of the form \(\theta(-z) S_i + \theta(z)S_j\), such that the distance is given by 
\begin{equation}\begin{aligned}
	\label{info-dist2}
	d_z(i,j) = \theta(-z) \ln \frac{S_\text{max}}{S_{z}(i)} + \theta(z) \ln \frac{S_\text{max}}{S_{j,z}} \\
	= \theta(z)j^z\ln 2 + \theta(-z) i^z\ln 2~.
\end{aligned}\end{equation}
\begin{figure}
	\centering
	\includegraphics[width=0.48\textwidth]{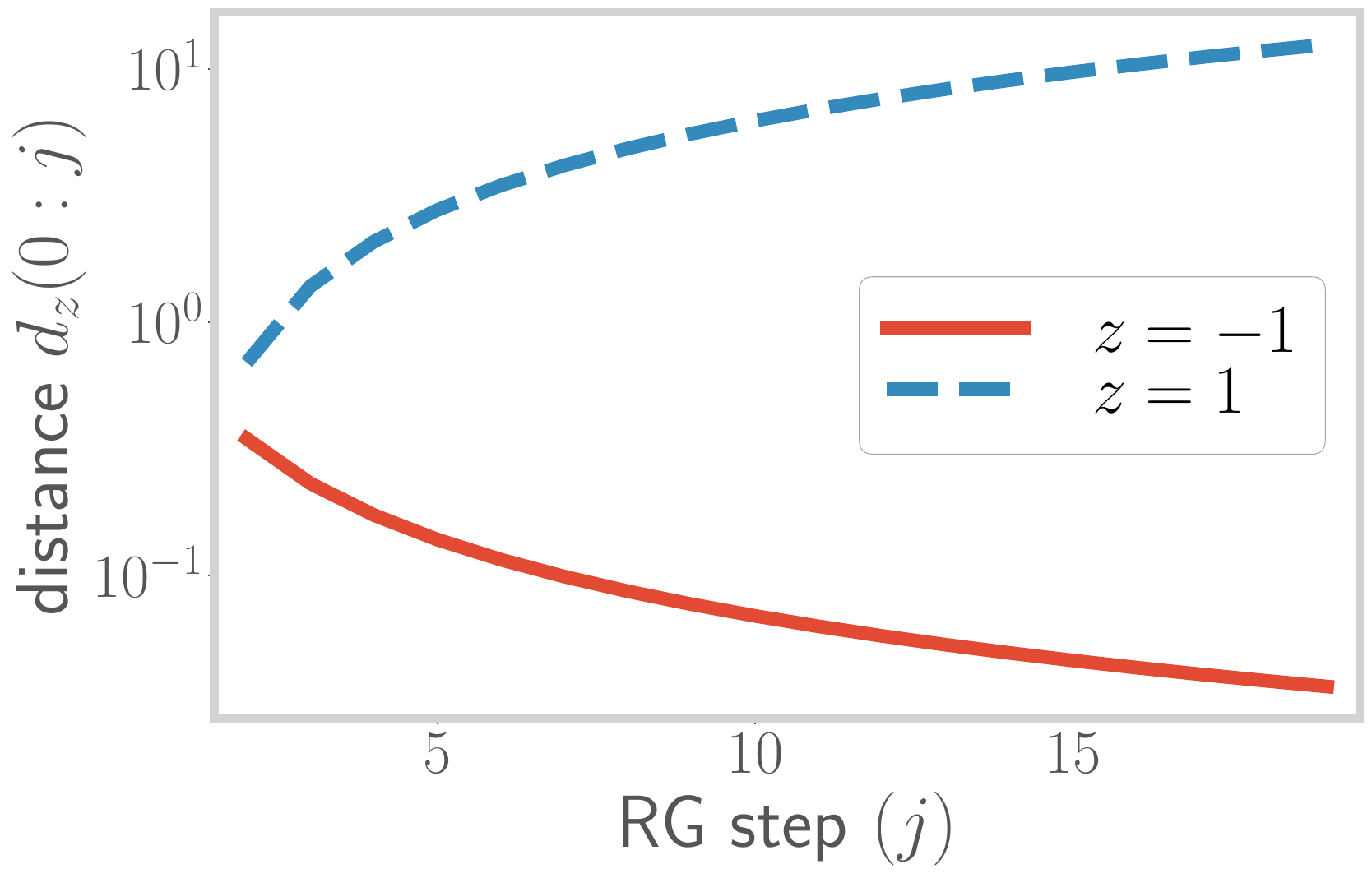}
	\includegraphics[width=0.48\textwidth]{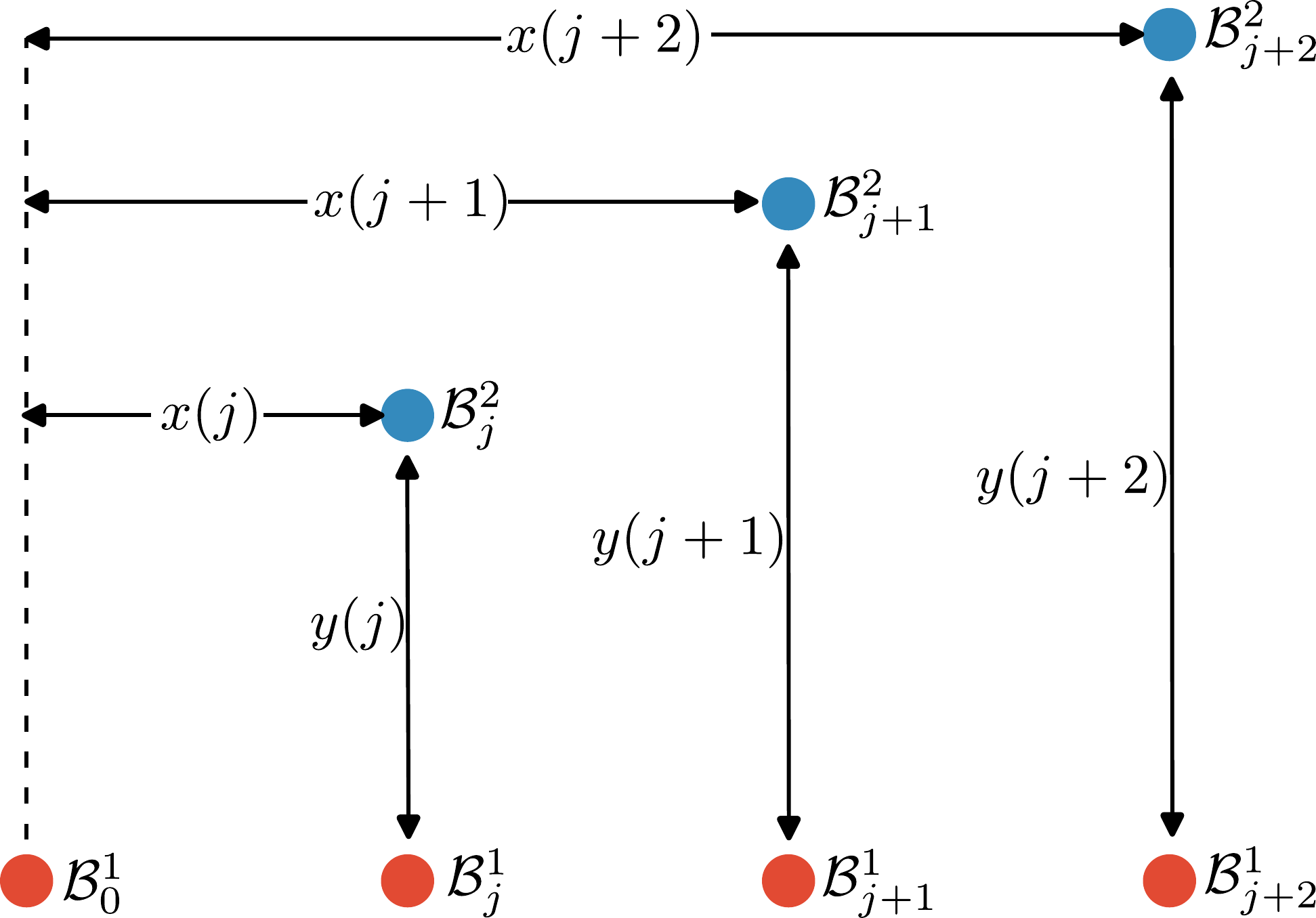}
	\caption{{\it Top:} Variation of the information distance \(d_z(0:j)\) (defined along the RG direction) under the scaling transformations. Obtained by plotting the expression for the distance, as a function of the number of transformations \(j\), for two values of \(z\). For \(z = 1\) (blue curve), the distance between two extreme points increases, showing a decrease in entanglement. For \(z=-1\) (red curve), the distance decreases, showing an increase in entanglement. {\it Bottom:} Geometry of subsystems used for calculating the emergent curvature. The red blobs represent the subsystems generated along the emergent direction, through the scaling transformations: \(\mathcal{A_2}_j \to \mathcal{B}_{j+1}\). The blue blobs represent subsystems generated within the \(1+1-\)dimensional field theory, transverse to the emergent direction.}
	\label{distance1}
\end{figure}
For \(z>0\), the distance increases as \(j\) increases (keeping \(i\) fixed), as the \(k-\)space points are made more coarse-grained under the RG transformations. On the other hand, for \(z<0\), the distance decreases as \(i\) is increased (keeping \(j\) fixed). The rate of increase or decrease depends on the value of \(z\). This scaling of the distance is shown in the \alert{top panel} of Fig.~\ref{distance1}.

At any given RG step \(j\), we define two subsystems \(\mathcal{B}_j^1\) and \(\mathcal{B}_j^2\).
The first subsystem \(\mathcal{B}_j^1\) is the entire set of states \(\mathcal{A}_{j,z}\) available at the \(j^\text{th}\) RG step, while the second subsystem \(\mathcal{B}_j^2\) is the set comprising those modes that remain to be considered at the next RG step if \(z>0\), and the set comprising only the modes that were present in the previous step for \(z<0\). This scheme is shown in the \alert{bottom panel} of Fig.~\ref{distance1}. The mutual information between these two sets defines a measure of distance \(y_j\) at the $j^{\text{th}}$ RG step:
\begin{equation}\begin{aligned}
	y_z(j) \equiv \ln \frac{I^2_\text{max}}{I^2_z(\mathcal{B}^1_j: \mathcal{B}^2_j)} = \begin{cases}
		\left(j+1\right)^z \ln 2~, \quad  z > 0~,\\
		\left(j-1\right)^z \ln 2, \quad  z < 0~.\\
	\end{cases}
	\label{info-dist3}
\end{aligned}
\end{equation}
The distance along the RG direction ($x_z(j)$) is, on the other hand, defined as the distance between the current RG step and the one with the maximum entanglement entropy:
\begin{equation}\label{info-dist4} \begin{aligned}
	x_z(j) \equiv \theta(z) d(0,j) + \theta(-z) d(j,\infty) = j^z \ln 2~,
\end{aligned}\end{equation}
where \(d(i,j)\) was defined in eq.~\eqref{info-dist}. 

\begin{figure}
	\centering
	\includegraphics[width=0.48\textwidth]{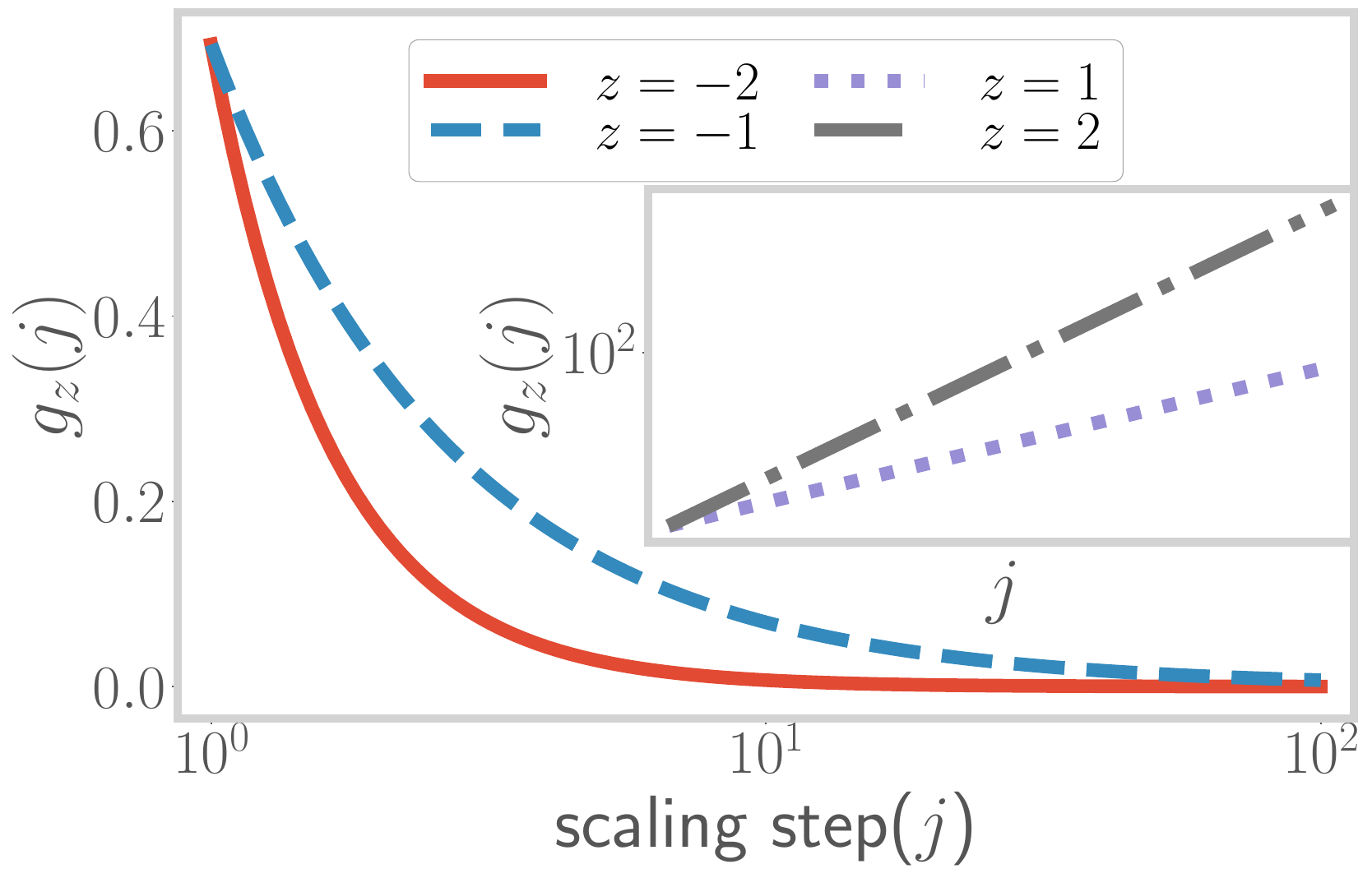}
	\includegraphics[width=0.48\textwidth]{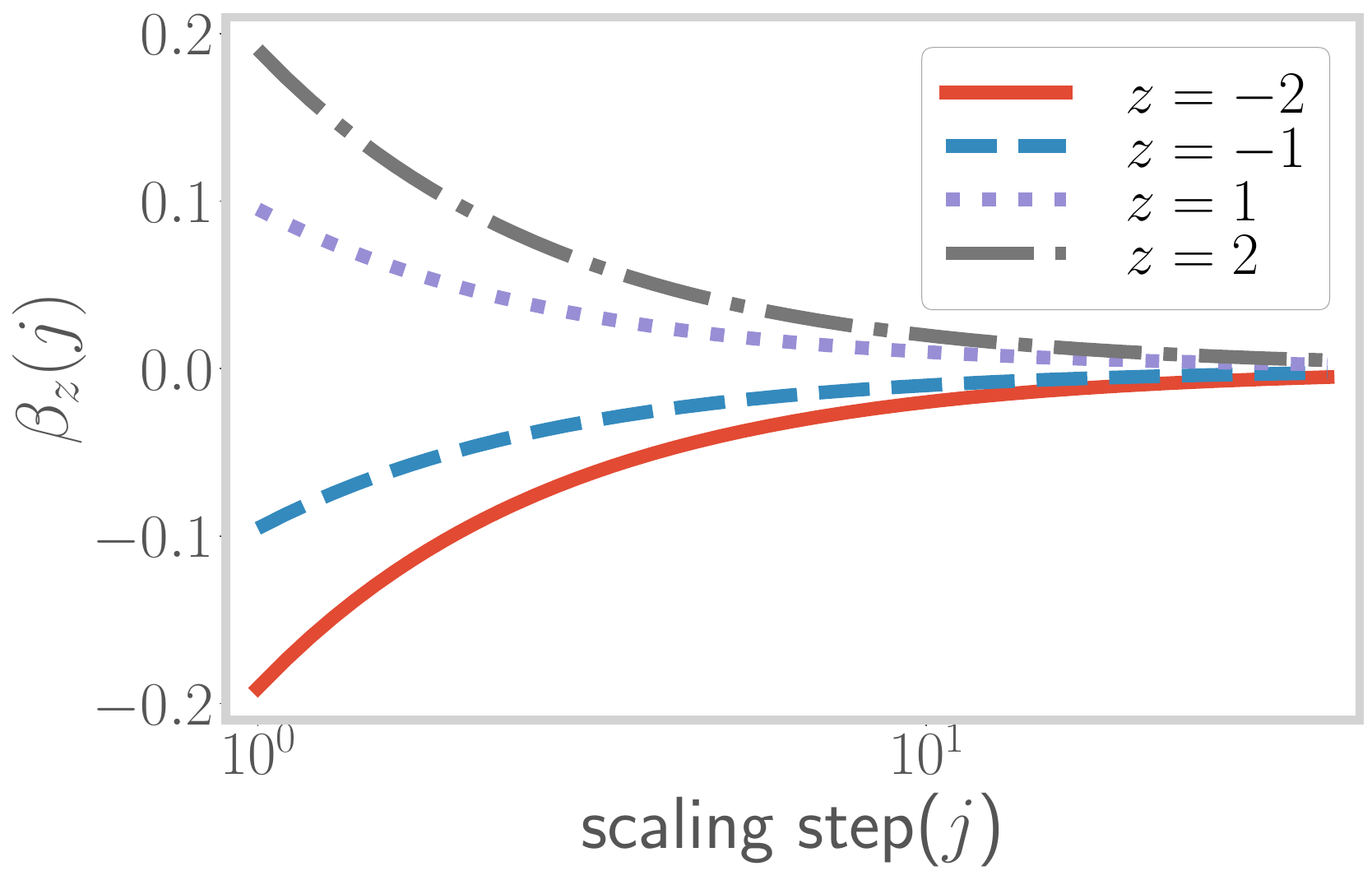}
	\caption{{\it Top:} Depiction of the RG flow of the coupling \(g_z(j)\). For \(z<0\) (blue and red curves), the coupling is RG-irrelevant and shows a decrease as \(j\) is increased. For \(z > 0\) (gray and violet curves), the coupling is RG-relevant and shows an increase as \(j\) is increased. {\it Bottom:} Variation of RG beta function \(\beta_z(j)\) for the coupling \(g_z(j)\), as a function of step index $j$. For all values of \(z\), the beta function converges to 0 as \(j \to \infty\), indicating a fixed point of the coupling \(g\).}
	\label{mass-beta}
\end{figure}

The connection of the \(x-\) and \(y-\)distances to the RG can be made manifest by relating them to the RG beta function. As we have seen above, even though we are working with massless Dirac fermions ($m=0$), the Aharonov-Bohm flux and the dimensional reduction into \(1+1-\)dimensional modes generate an effective mass \(M_{n,\phi} = \frac{2\pi}{L_x}(n+\phi)\) for \(n>0\). At a specific RG step \(j\), the \(n^\text{th}\) mode is given by \(t_{z}(j)\times n\) on account of the RG transformations in \(k_x-\)space. With these points in mind, we define a coupling \(g_z(j)\) that acts as a measure of the gap in the single-particle spectrum
\begin{equation}\begin{aligned}
	g_z(j) \equiv \ln \frac{M_{n+1,\phi}(j) - M_{n,\phi}(j)}{2\pi/L_x} = \ln t_{z}(j) = j^z \ln 2~.
\end{aligned}\end{equation}
The RG beta-function for the coupling \(g_z(j)\) can then be written as
\begin{equation}\begin{aligned}
	\beta_{z}(j) \equiv \frac{\Delta \log g_z}{\Delta j}= \log \frac{g_z(j+1)}{g_z(j)} = z\ln \left( 1 + \frac{1}{j} \right)~,
	\label{RGbetafunc}
\end{aligned}\end{equation}
where we have substituted \(\Delta j = 1\). The beta function is positive for \(z>0\), indicating that the coupling \(g_z(j)\) is RG-relevant in this regime (\alert{top panel} of Fig.~\ref{mass-beta}).
On the other hand, the beta function is negative for \(z<0\), leading to irrelevant RG flows in this regime. The RG beta-function \(\beta_z(j)\) itself decreases monotonically as \(j\) increases under the RG transformations (see \alert{bottom panel of} Fig.~\ref{mass-beta}): for a finite \(z\), a fixed point is reached at \(j \to \infty\) where the beta function goes to zero. 
 
As we will now see, the flow of the RG beta functions can be related to the extremisation of a potential defined in the space of Hamiltonians. This potential is a function of the mutual information \(I^2_z(0:j)\):
\begin{equation}\label{potential}\begin{aligned}
	\beta_z(j) = -\Delta \left[\ln \left(\ln \frac{I^2_z(0:j)}{I^2_\text{max}}\right)\right]~.
\end{aligned}\end{equation}
The potential given within the square brackets above corresponds to an RG monotone whose fixed point corresponds with the fixed point of the RG. For relevant flows, the above equation ensures that the function \(\ln \left(\ln \frac{I^2_z(0:j)}{I^2_\text{max}}\right)\) is always minimised under the RG flow, with the consequence that the trajectories ultimately lead to minimisation of the mutual information. This indicates that the relevance of the mass coupling leads in turn to the decrease of the correlation length and hence a lowering of entanglement between points that are separated in real space.
On the other hand, for irrelevant flows, the above-mentioned function is maximised, implying that the RG trajectories now maximise the entanglement. This reflects the fact that the decreasing mass leads to increased correlation lengths and an approach towards criticality.

Given eqs.\eqref{info-dist3}, \eqref{info-dist4} and \eqref{potential}, we can relate the distances $x$ and $y$ to the beta function $\beta_z$:
\begin{gather}
	\label{x-beta}
	x_z = \left( e^\frac{\beta_z}{z} - 1 \right)^{-z} \ln 2~,\\
	\label{y-beta}
	y_z = \begin{cases}
		x_z e^\beta\quad &z > 0~,\\
		x_z \left(2 - e^\frac{\beta}{z}\right)^z\quad &z < 0~.\\
	\end{cases}
\end{gather}

\begin{figure}
	\centering
	\includegraphics[width=0.48\textwidth]{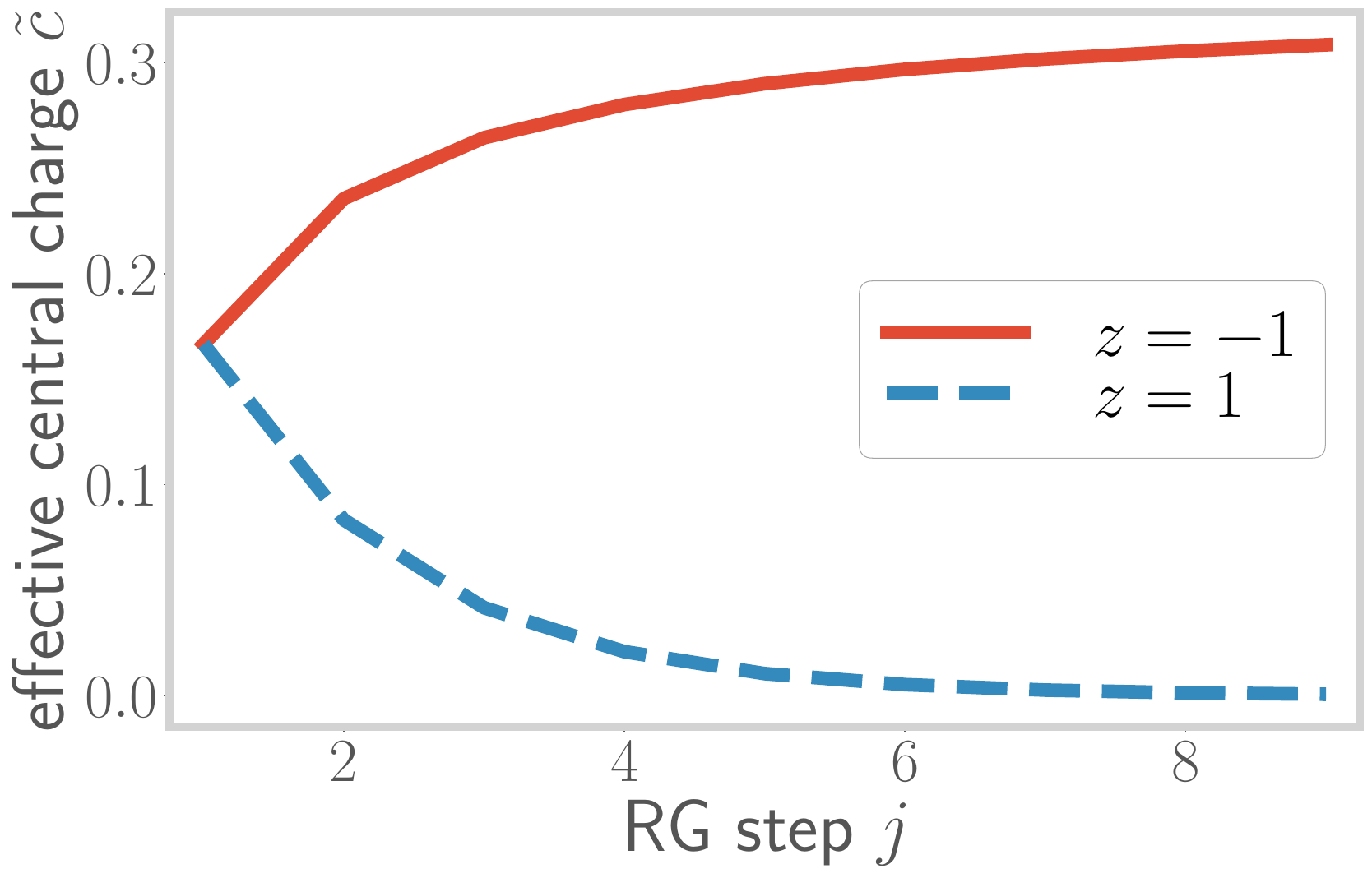}
	\includegraphics[width=0.48\textwidth]{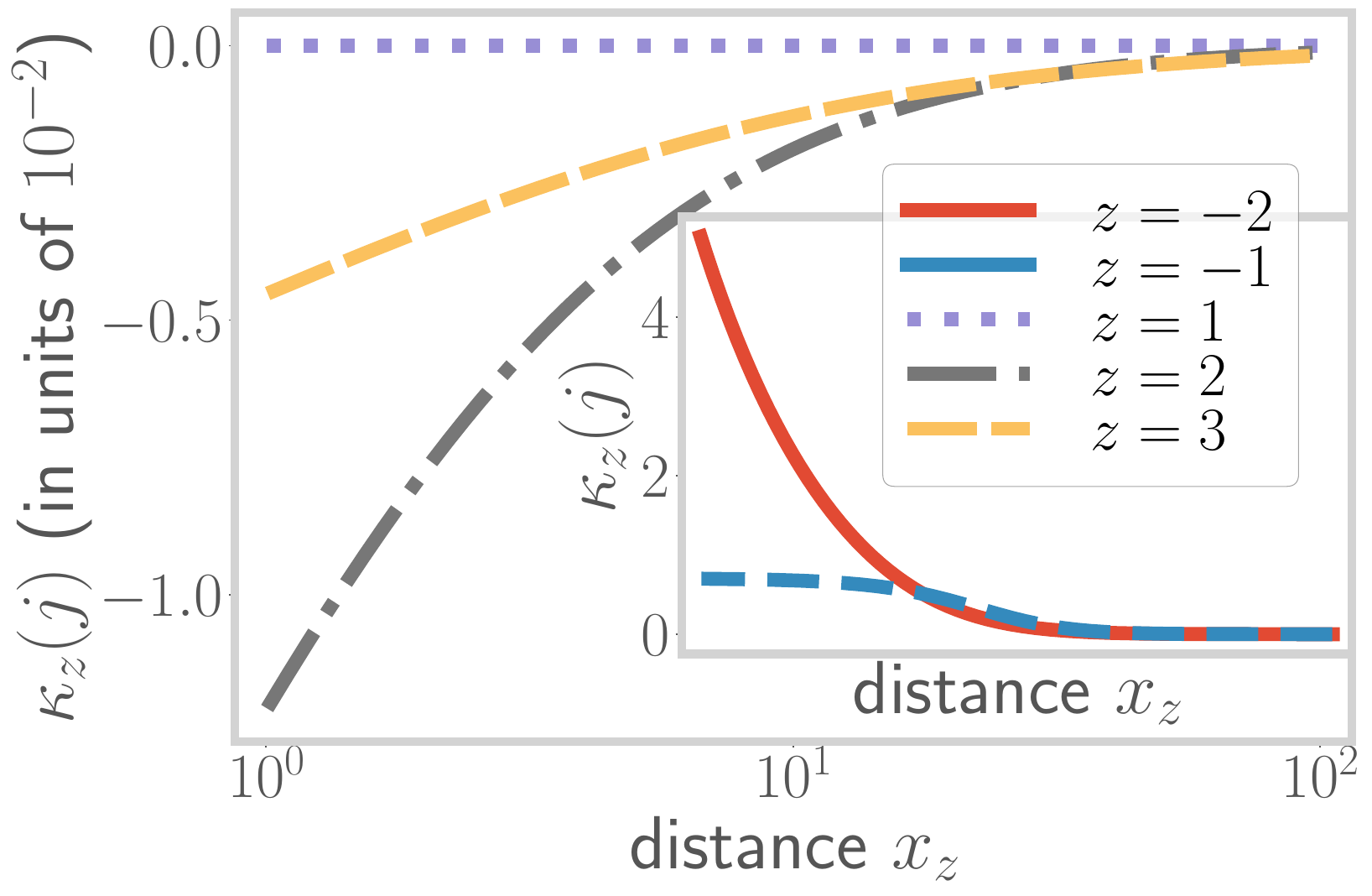}
	\caption{{\it Top:} Variation of the effective central charge defined in eq.~\eqref{c-charge-flow}, under the RG transformations for relevant ($z=1$, blue curve) and irrelevant ($z=-1$, red curve) flows. In the former case, the central charge starts from the fermionic gapless value of \(1/3\) and vanishes towards the fixed point, indicating the flow to a gapped massive phase. In the latter case, it starts from the gapped value of \(0\), and increases to \(1/3\) at the fixed point, indicating the flow to a gapless phase. {\it Bottom:} Variation of the curvature \(\kappa_z(j)\) along the RG for multiple values of \(z\). \(z=1\) leads to a flat space with zero curvature, while \(z>1\)  and $z<1$ (inset) lead to negative and positive curvature respectively.}
	\label{c-charge}
\end{figure}

The contrasting behaviour of the beta function depending on the sign of \(z\) indicates that it tracks the effective number of degrees of freedom that contribute to the entanglement (eq.~\eqref{EE-subsystem}) at any given RG step. Indeed, we find that the effective number of degrees of freedom is the effective \(c-\)function (or central charge) \(\tilde c_z(j)\) of the theory, taking the form 
\(\tilde c_z(j) = f_{z}(i) c\).
The beta function is thus related to the \(c-\)function as
\begin{equation}\begin{aligned}
\beta_z = \Delta\left[ \ln\left( \ln \frac{c}{\tilde c_z(j)}\right) \right]~.
\label{c-charge-flow}
\end{aligned}\end{equation}
For RG relevant flows, the \(c-\)function is found to diminish from the free fermion central charge \(c=1/3\), reaching the value of zero at the fixed point (where only the central \(n=0\) mode remains and the rest have been removed because of the large mass gap). The non-negative, non-increasing nature of the \(c-\)function, and its flow towards its minimum at the fixed point, leads to 
speculate that eq.\eqref{c-charge-flow} can be thought of as a holographic counterpart of Zamalodchikov's \(c-\)theorem~\cite{alvarez_1999,myers_2010,Myers2011,chu_2020}.
For RG irrelevant flows, the behaviour is reversed. The flow of the \(c-\)function is shown in the \alert{top panel} of Fig.~\ref{c-charge}.

\subsection{Embedding the emergent dimension on a metric space}
\label{metric_subsection}
We have shown above that the renormalisation group flow of the Hamiltonian leads to the emergence of an additional direction, quantified by the parameter \(x_j\). This will now be used to construct a metric, in the following way. We already know that the scaling transformations lead to a graph of theories that are connected via edges that represent the mutual information between any two theories. As a result, this ``information graph" can be mapped on to a new ``distance graph" where the edges now represent the distance between the theories at the nodes. Finally, these distances can be used to define a metric for the distance graph: the metric between any two nodes will be the distance along the shortest path between these two nodes. This approach as well as some of the terminology are inherited from reference~\cite{cao_2017}.

The RG flow can be represented as an ``information graph" \(G(V,E)\), where the systems \(\mathcal{B}^1_{j}\) (defined above eq.~\eqref{info-dist3}) constitute the set of vertices \(V = \left\{ V_j \right\} \) and the mutual information \(I_z^2(j:j^\prime)\) between the systems \(\mathcal{B}^1_j\) and \(\mathcal{B}^1_{j^\prime}\) constitute the edges \(E_{ij}\) between the vertices. This information graph is characterised by an adjacency matrix whose elements are defined by the mutual information edge weights \(I_z^2(j:j^\prime)\)~\cite{singharoy2020}. This graph structure is shown in Fig.~\eqref{distance1}, with the red spheres indicating the vertices. In order to define a metric space, we will map the graph \(G\) to another graph \(\mathcal{G}\) that has the same vertices as \(G\), but whose edges are weighted according to a well-defined notion of distance.

A path \(\mathcal{P}\) in \(\mathcal{G}\) is defined as a sequence of \(m\) vertices: \(\mathcal{P}(V_{1}, V_{2},\ldots V_{m})\). The distance along the path is defined as the sum of distances along every pair of consecutive vertices: \(l(\mathcal{P}) = \sum_{i=1}^{m-1} d(V_{i}, V_{i+1})\), where \(d(V_i, V_{i+1}) = d(i,i+1)\) (see eq.~\eqref{info-dist}). The {\it geodesic} between two points \(V_i\) and \(V_j\) is the path \(\mathcal{P}\) which minimises the distance \(l(P)\), and the distance along the geodesic then defines the {\it metric} \(\tilde d(V_i, V_j)\) between two vertices \(V_i\) and \(V_j\)~\cite{singharoy2020,cao_2017}:
\begin{equation}\begin{aligned}
	\tilde d(V_i, V_j) = {\text{min}\atop{\mathcal{P}}}\left\{l(\mathcal{P})\right\} = {\text{min}\atop{\mathcal{P}}}\left\{\sum_{i=1}^{m-1} d(V_{i}, V_{i+1})\right\}~,
\end{aligned}\end{equation}
where the minimisation is carried out over all possible connected paths \(\left\{ \mathcal{P} \right\} \) between \(V_i\) and \(V_j\), and \(m\) is the total number of vertices in a particular connected path \(\mathcal{P}_m\).
Following the expression for the distance \(d(i,j)\) obtained in eq.~\eqref{info-dist2}, the metric takes the form
\begin{equation}\begin{aligned}
	\tilde d(V_i, V_j) = \theta(z) j^z \ln 2 + \theta(-z) i^z \ln 2, \quad (i < j)~,
\end{aligned}\end{equation}
where we have assumed \(i < j\), without loss of generality. 

Finally, the metric dimension of the emergent space can be determined by finding the set with the lowest possible cardinality that can resolve the metric space~\cite{cao_2017,Bau2013}. In order for a set to resolve the metric space, we must have \(d(i,a) = d(j,a) \implies i = j\) for all points \(a\) in the set and for any pair of points \(i\) and \(j\) in the metric space. In our case, the singleton set \(\mathcal{R} = \left\{0\right\} \) can resolve the metric space for \(z > 0\), as \(d(i,0) = d(j,0) \implies i^z = j^z \implies i = j\). Similarly, the singleton set \(\mathcal{R} = \left\{\infty \right\} \) resolves the metric space for \(z < 0\). As a result, we conclude that our emergent metric space is one-dimensional in nature.

\section{Curvature of the emergent space}
\label{curvature}
Further insight on the spatial geometry of the RG flows can be obtained by looking at the curvature generated by the RG transformations. In the plane of the directions \(x\) and \(y\) generated by the RG flow, the curvature can be written using the ideas of analytic geometry as
\begin{equation}\begin{aligned}
	\label{kappa}
	\kappa_{z}(j) = \frac{v^\prime_z(j)}{\left[1 + v_z(j)^2\right]^\frac{3}{2}}~,
\end{aligned}\end{equation}
where
\begin{equation}\begin{aligned}
	v_z(j) \equiv \frac{\Delta y_z(j)}{\Delta x_z(j)} = \frac{y_z(j+1) - y_z(j)}{x_z(j+1) - x_z(j)} \\
	= \begin{cases}
		\frac{\left(j+2\right)^z - \left(j+1\right)^z}{\left(j+1\right)^z - j^z} \quad  z > 0~,\\
		\frac{\left(j\right)^z - \left(j-1\right)^z}{\left(j+1\right)^z - j^z} \quad  z < 0~,\\
	\end{cases}
\end{aligned}\end{equation}
and
\begin{equation}\begin{aligned}
	v^\prime_z(j) \equiv \frac{v_z(j+1) - v_z(j)}{x_z(j+1) - x_z(j)}~.
\end{aligned}\end{equation}

The curvature $\kappa_{z}(j)$ has been plotted against the RG step index \(j\) in \alert{bottom panel} of Fig.~\ref{c-charge}. For \(z=1\), the first derivative \(v_z(j)\) becomes unity, so the second derivative \(v^\prime_z(j)\) (and hence the curvature) vanishes, leading to a flat (Minkowski) space. For \(z>1\), the curvature becomes negative, while it is positive for \(z<0\). All the flows lead to an asymptotically flat space (i.e., the curvature vanishes) as \(j \to \infty\). 
 
The different signatures of the emergent curvatures can be attributed to the different ways in which the mass of the theory scales under the decimation procedure. For \(z>0\), the curvature defined in eq.~\eqref{kappa} can be written in terms of the beta function:
\begin{equation}\begin{aligned}
	\label{kappa-mass}
	\kappa_z(j) = \frac{\left[e^{\beta(j)} - 1\right]^\frac{1}{2} j^{-z}}{\left[e^{\beta(j) + \beta(j+1)} - 1\right]^{\frac{3}{2}}}\left[\frac{e^{\beta(j+2)} - 1}{1 - e^{-\beta(j+1)}} - \frac{e^{\beta(j+1)} - 1}{1 - e^{-\beta(j)}}\right]~.
\end{aligned}\end{equation}
This equation relates directly a geometric quantity (i.e., the curvature) of the bulk emergent space with the RG beta function for the field theory on the boundary. From Fig.~\ref{mass-beta}, as the beta functions flow towards zero, we see from eq.~\eqref{kappa-mass} that the first fraction inside the square brackets is the first to vanish, followed by the second fraction. The sequential vanishing of these two fractions prior to the vanishing of the denominator \(e^{\beta(j)} - 1\) in eq.\eqref{kappa-mass} leads to the asymptotic vanishing of the curvature towards the fixed point. Different choices of \(z >0\) can be thought to represent different relevant perturbations that drive the system towards {\it gapped fixed points}, leading to distinct curvatures of the bulk. By employing the relation \(v_z(j)=\Delta S_{j+1,z}/\Delta S_{j,z}\), eq.~\eqref{kappa} can also be cast in terms of the entanglement entropy \(S_{j,z}, S_{j+1,z}\) and \(S_{j+2,z}\). The resulting non-linear relationship between $\kappa_{z}(j)$ and $S_{j,z}$ obtained is, however, in contrast to the linear dependence obtained by Cao et al.~\cite{cao_2017} between entanglement perturbations on the Hilbert space and the curvature of the resultant holographic space.

\subsection{Topological nature of the sign of curvature}
We will now argue that the signature of the curvature $\kappa$ corresponds to a topological quantum number, and that a change in the signature as \(z\) is tuned through \(1\) amounts to a topological translation. For this, we first point out (following eq.~\eqref{kappa}) that the sign of the curvature is determined purely by the sign of \(\gamma_z(j) \equiv 1 - v_z(j+1)/v_z(j)\) and \(\alpha_z(j) = y_z(j+1) - y_z(j)\):
\begin{equation}\begin{aligned}
	\label{curvature-gamma-rel}
	\kappa_{z}(j) = \frac{v^\prime_z(j)}{\left[1 + v_z(j)^2\right]^\frac{3}{2}} = -\frac{\alpha_z(j)~\gamma_z(j)}{\left(\Delta x_z(j)\right)^2\left[1 + v_z(j)^2\right]^\frac{3}{2}} \\
	\implies \text{sign}\left[\kappa_z(j)\right] = -\text{sign}\left[\alpha_z(j)\right]\text{sign}\left[\gamma_z(j)\right]~.
\end{aligned}\end{equation}
We now promote \(j\) to a continuous complex variable with the real part being greater than or equal to 1: \(\text{Re}\left[j\right] \geq 1\). The quantities \(\alpha_z(j)\) and \(\gamma_z(j)\) can be written as
\begin{gather}
	\alpha_z(j) = \begin{cases}
		\left[\left( j+2 \right) ^z - \left( j+1 \right) ^z\right]\log 2, ~ z > 0\\
		\left[j^z - \left( j-1 \right) ^z\right]\log 2, ~ z < 0\\
	\end{cases},\\
	\gamma_z(j) = \begin{cases}
		1 - \frac{\left[\left( j+3 \right) ^z - \left( j+2 \right) ^z\right] \left[\left( j+1 \right) ^z - j^z\right]}{\left[\left( j+2 \right) ^z - \left( j+1 \right) ^z\right]^2}, ~ z >0\\
		1 - \frac{\left[\left( j+1 \right) ^z - j^z\right]^2}{ \left[j^z - \left(j-1\right)^z\right]\left[\left( j+2 \right) ^z - \left( j+1 \right) ^z\right]}, ~ z < 0~.\\
	\end{cases}
\end{gather}
\begin{figure}
	\centering
	\includegraphics[width=0.48\textwidth]{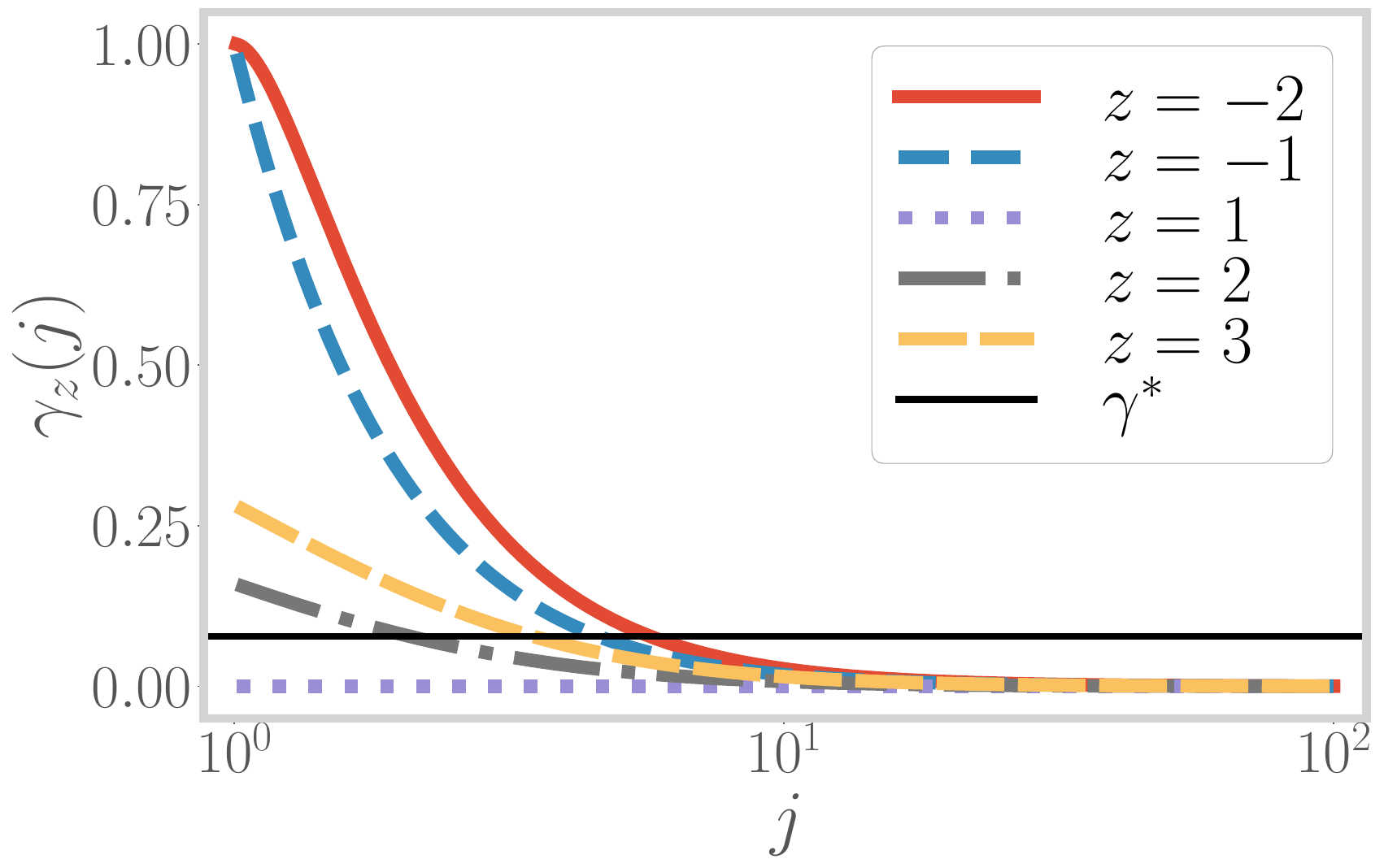}
	\includegraphics[width=0.48\textwidth]{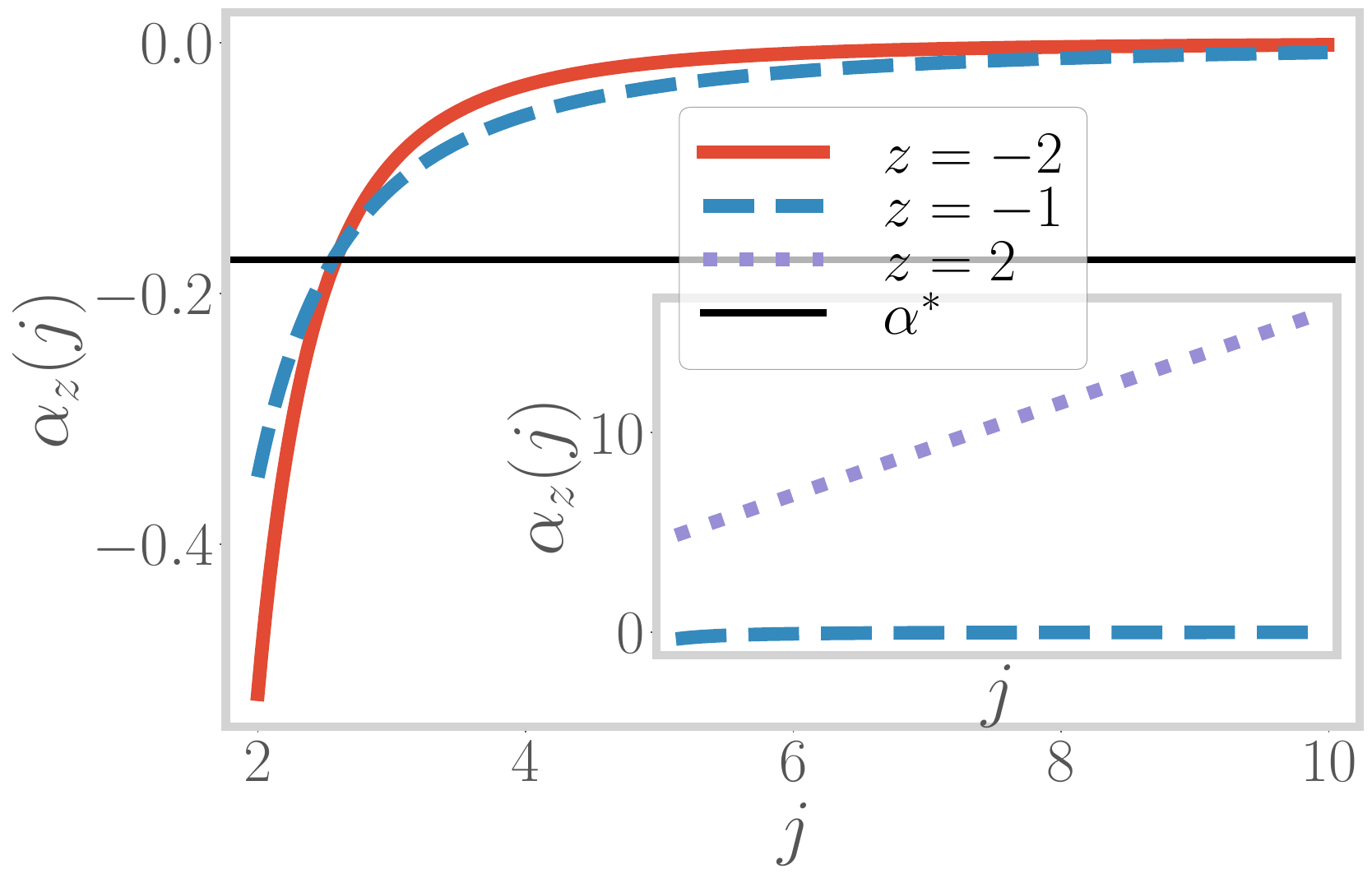}
	\caption{{\it Top:} Variation of \(\gamma_z\) along the RG for multiple values of \(z > 0\).
	All trajectories for \(z > 1\) intersect the line \(y = \gamma^*\).
{\it Bottom:} Variation of \(\alpha_z\) along the RG for multiple values of \(z < 0\).
All trajectories intersect the line \(y = \alpha^*\)}
	\label{gamma-fig}
\end{figure}

As shown in Fig.\ref{gamma-fig}(\alert{top panel}), the function \(\gamma_z\) is positive and monotonically decreasing for \(z > 1\) or \(z \leq -1\), and vanishes identically for \(z=1\). On the other hand, \(\alpha_z(j)\) is positive for \(z>1\), zero at \(z=1\) and negative for \(z \leq -1\) (see Fig.\ref{gamma-fig}(\alert{bottom panel})). Combining this, we find that the sign of the curvature can be written as:
\begin{equation}\begin{aligned}
	\text{sign}\left[\kappa_z\right] = \begin{cases}
		-1 , \quad  z \geq 1\\
		1 , \quad  z \leq -1\\
	\end{cases} = \begin{cases}
		-\text{sign}\left[\gamma_z(j)\right] , \quad  z \geq 1\\
		-\text{sign}\left[\alpha_z(j)\right] , \quad  z \leq -1~.\\
	\end{cases}
\end{aligned}\end{equation}

We start by considering the region \(z \geq 1\). Since \(z=0\) is excluded by definition, the denominator is always non-zero. This indicates that \(\gamma_z(j)\) is continuous and monotonically decreasing in this range: the maximum value of \(\left\{\gamma_z(j), j\in \left[1, \infty\right]\right\} \) occurs at \(j=1\). Further, the maximum value \(\gamma_z(1)\) increases with the value of \(z\), ranging between \(\gamma_2(2)\) at \(z=2\) and unity at \(z\to\infty\). These two features are demonstrated in the \alert{top panel} of Fig.~\ref{gamma-fig}, and imply that all curves \(\left\{ \gamma_z(j):j \right\} \) will take values both greater than and less than the value \(\gamma^* \equiv \gamma_2(2)\) (\alert{thin black line} in \alert{top panel} of Fig.~\ref{gamma-fig}).
Thus, the plot shows that there always exists a real value \(j_z^*\) for any given integer value \(z>1\) such that \(\gamma_z(j_z^*) = \gamma^* = \gamma_2(2)\). This is precisely what ceases to happen for \(z=1\), as that is the only curve in the \alert{top panel} of Fig.~\ref{gamma-fig} that does not pass through \(\gamma^*\).

\begin{figure}
	\centering
	\includegraphics[width=0.48\textwidth]{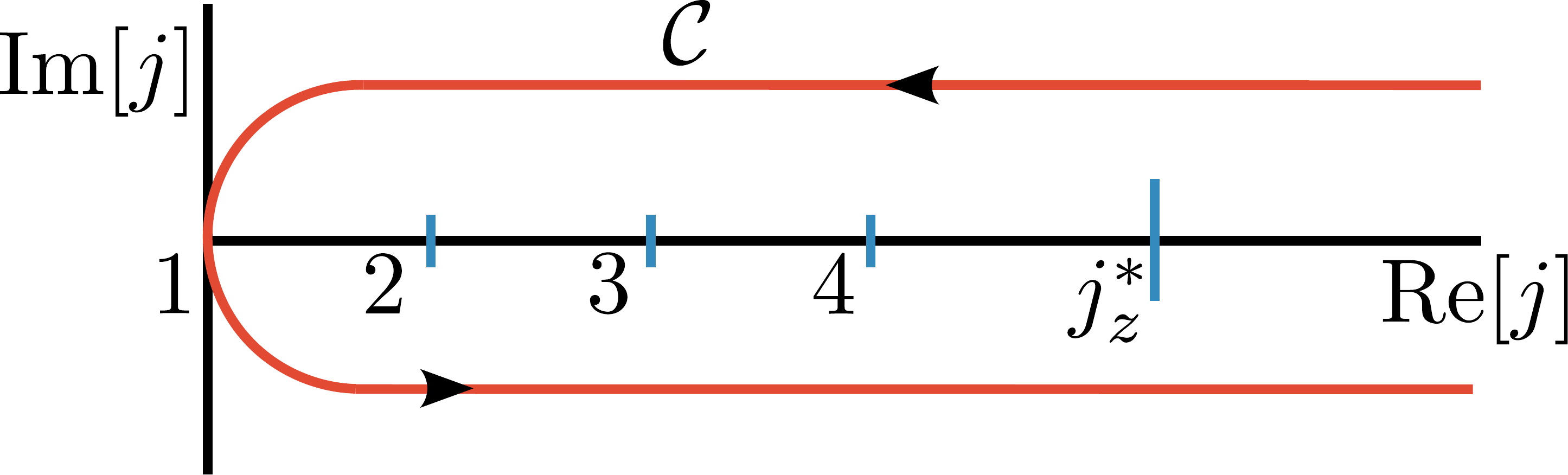}
	\includegraphics[width=0.48\textwidth]{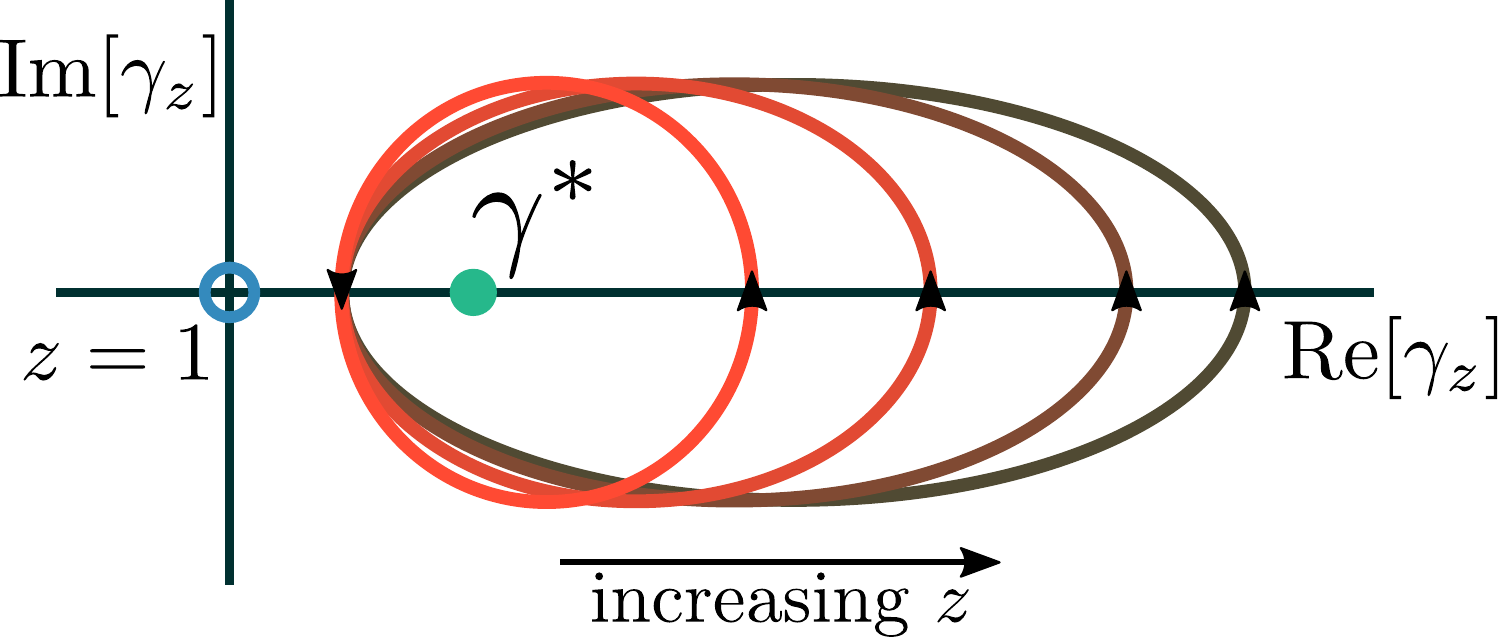}
	\includegraphics[width=0.48\textwidth]{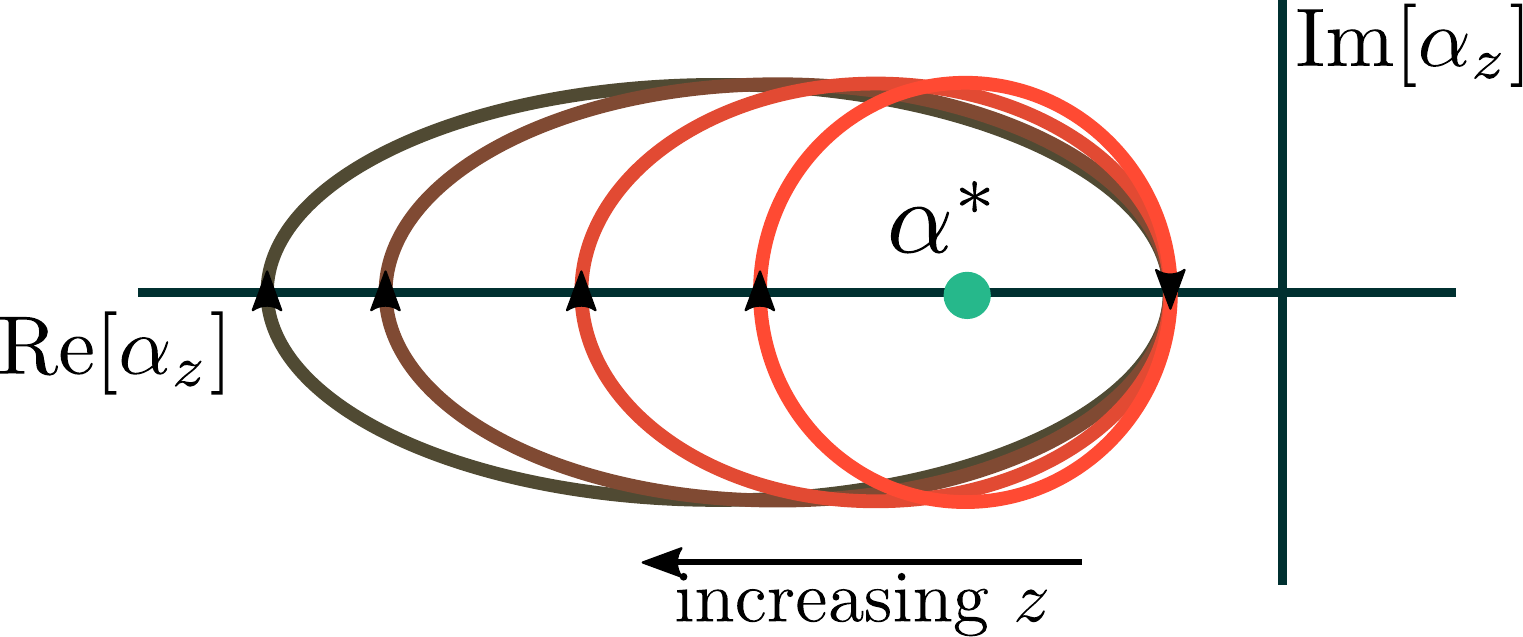}
	\caption{{\it Top:} The contour \(\mathcal{C}\) used in eq.~\eqref{integral-winding}. It extends across all values of \(j\), and the integral counts the number of poles of the function \(1/(\gamma - \gamma^*)\) encircled by the contour.
{\it Bottom:} {\it Middle:} Contours \(\gamma_z(\mathcal{C})\) for multiple values of \(z > 0\). As \(z\) increases, the contours extend in the positive \(\text{Re}\left[\gamma_z\right] \) direction. {\it Bottom:} Contours \(\alpha_z(\mathcal{C})\) for multiple values of \(z > 0\). As \(z\) increases, the contours extend in the negative \(\text{Re}\left[\gamma_z\right] \) direction. }
	\label{curvature-winding}
\end{figure}

Indeed, the difference between the curvature for the \(z=1\) and \(z>1\) classes can be captured by expressing the presence of the solution \(j_z^*\) in the form of a winding number. For this, we use the fact that the sign of \(\gamma_j(z)\) is given by the following integral:
\begin{equation}\begin{aligned}
	\label{integral-winding}
	\text{sign}\left[\gamma_z(j)\right]\bigg\vert_{z \geq 1} = \frac{1}{2\pi i}\oint_{\mathcal{C}} \mathrm{d}j \frac{\partial{}}{\partial{j}}\ln \left[\gamma_z(j) - \gamma^*\right]~,
\end{aligned}\end{equation}
where the contour \(\mathcal{C}\) extends from \(j=1\) to \(j=\infty\), and is shown in the \alert{top panel} of Fig.~\ref{curvature-winding}. As the value of \(z\) is tuned from 2 to 1, the pole at \(j_z^*\) disappears and the integral reduces to zero. As shown in the middle panel of Fig.~\ref{curvature-winding}, the integral can thus be expressed as the winding number of the curve \(\gamma_z(\mathcal{C}(j))\) that counts the number of times the contour \(\gamma_z(\mathcal{C})\) winds around the singularity at \(\gamma^*\):
\begin{equation}\label{contour1}\begin{aligned}
	\text{sign}\left[\gamma_z(j)\right] = \frac{1}{2\pi i}\oint_{\mathcal{C}} \mathrm{d}j \frac{\partial{\left(\gamma_z(j) - \gamma^*\right)}}{\gamma_z(j) - \gamma^*} = \frac{1}{2\pi i}\oint_{\gamma_z(\mathcal{C})}  \frac{\mathrm{d}\gamma_z}{\gamma_z - \gamma^*}~.
\end{aligned}\end{equation}

This is easily seen from the following argument. If we write a general complex variable \(y\) in the form \(y=re^{i\theta}\) and consider a general contour \(C\) that starts from \(r=r_0,\theta=0\) and ends at \(r=r_0,\theta=2 w \pi\) (such that the winding number of the curve is \(w\)), integrals of the form given above simplify and return the winding number itself:
\begin{equation}\begin{aligned}
	\frac{1}{2\pi i}\oint_{C} \frac{dy}{y} = \frac{1}{2\pi i}\int_{r_0}^{r_0} \frac{dr}{r} + \frac{1}{2\pi}\int_{0}^{2 w \pi} d\theta = w~,
\end{aligned}\end{equation}
where we used \(dy = y\left( \frac{dr}{r} + d\theta \right) \).

The winding numbers take integer values, and are topological in nature: they are robust against geometric deformations, and change only when the curve crosses the singularity at $\gamma=\gamma^{*}$. Thus, for \(z>1\), the contour \(\gamma_z(\mathcal{C})\) encloses the singularity at \(\gamma^*\) exactly once, leading to a winding number of unity. As \(z\) is lowered towards \(z=2\), the contours becomes smaller such that, in order to go from \(z=2\) to \(z=1\), the contour has to be moved across the singularity. This constituting a change in the topology of the contour: for \(z=1\), the contour no longer encloses the pole and the winding number is zero. In fact, at \(z=1\), the contour itself (depicted in blue) is of vanishing radius because \(\gamma_1(j)\) has a constant value of zero.
To complete the connection, we use eq.~\eqref{curvature-gamma-rel} to relate this topological winding number to the sign of the curvature:
\begin{equation}\begin{aligned}
	\label{sign-pos}
	\text{sign}\left[\kappa_z\right]\bigg\vert_{z \geq 1} = -\mathcal{W}_z(\gamma^*)~,
\end{aligned}\end{equation}
where \(\mathcal{W}_z(\gamma^*)\) is the winding number for \(\gamma_z(\mathcal{C})\) around \(\gamma^*\).

One can make a very similar argument for the case of \(z \leq -1\). Here, as shown in the \alert{bottom panel} of Fig.~\ref{gamma-fig}, the relevant quantity is \(\alpha_z(j)\). For each curve \(\alpha_z\), the minimum value is at \(j=1\). For all \(z\), this minimum starting value is largest for \(z=-1\), and all subsequent curves \(\alpha_z(j)\) start from even lower values. If we define a value \(\alpha^* \equiv \alpha_z(j=1)/2\), it is then clear that all curves \(\alpha_z(j)\) will pass through the value \(\alpha^*\) (thin black line in \alert{bottom panel} of Fig.~\ref{gamma-fig}) at some value of \(j\).
This is very similar to the situation discussed just above for the case of \(z \geq 1\) (with \(\gamma^*\) playing the role of \(\alpha^*\)); we, therefore, employ the same technique here.
Taking the same contour \(\mathcal{C}\) as defined above, we write the sign of \(\alpha_z\) as the winding number of the curve \(\alpha_z(\mathcal{C})\) around the point \(\alpha_z = \alpha^* \)
\begin{equation}\begin{aligned}
	\text{sign}\left[\alpha_z(j)\right] \bigg\vert_{z \leq -1} = -\mathcal{W}^\prime_z(\alpha^*) = \frac{1}{2\pi i}\oint_{\alpha_z(\mathcal{C})}  \frac{\mathrm{d}\alpha_z}{\alpha_z - \alpha^*}~.
\end{aligned}\end{equation}
For \(z>0\), \(\alpha_z\) is positive, and the curve \(\alpha_z(\mathcal{C})\) encloses points in the positive half of the real axis such that the winding number in the \alert{bottom panel} of Fig.~\ref{curvature-winding} is then zero. This indicates that the function \(1 - 2\mathcal{W}_z^\prime(\alpha^*)\) is unity for \(z > 0\), leaving any other function in product with it unchanged. On the other hand, it is \(-1\) for \(z < 0\), returning the correct sign of \(\alpha_z(j)\) in that range. Combining with eq.~\eqref{sign-pos}, the sign in the entire range can be expressed as
\begin{equation}\begin{aligned}
	\text{sign}\left[\kappa_z\right] = \mathcal{W}_z\left( \gamma^* \right) \times \left[2\mathcal{W}^\prime_z\left( \alpha^* \right) - 1\right]~. 
	\label{topo-kappa}
\end{aligned}\end{equation}

We can now summarise the topological nature of the sign of the curvature. For \(z>0\), the second winding number \(\mathcal{W}^\prime \) is 0. As \(z\) varies from 2 to 1, the other winding number \(\mathcal{W}\) changes from 1 to 0, constituting the first topological change. In contrast, upon varying \(z\) from 1 to \(-1\), both winding numbers change: \(\mathcal{W}\) from 0 to 1, and \(\mathcal{W}^\prime\) from 0 to 1. This constitutes the second topological change. Together, the two changes show that going from an open geometry (negative curvature) to a closed geometry (positive curvature) involves two topological transformations.

\begin{figure}
	\centering
	\includegraphics[width=0.45\textwidth]{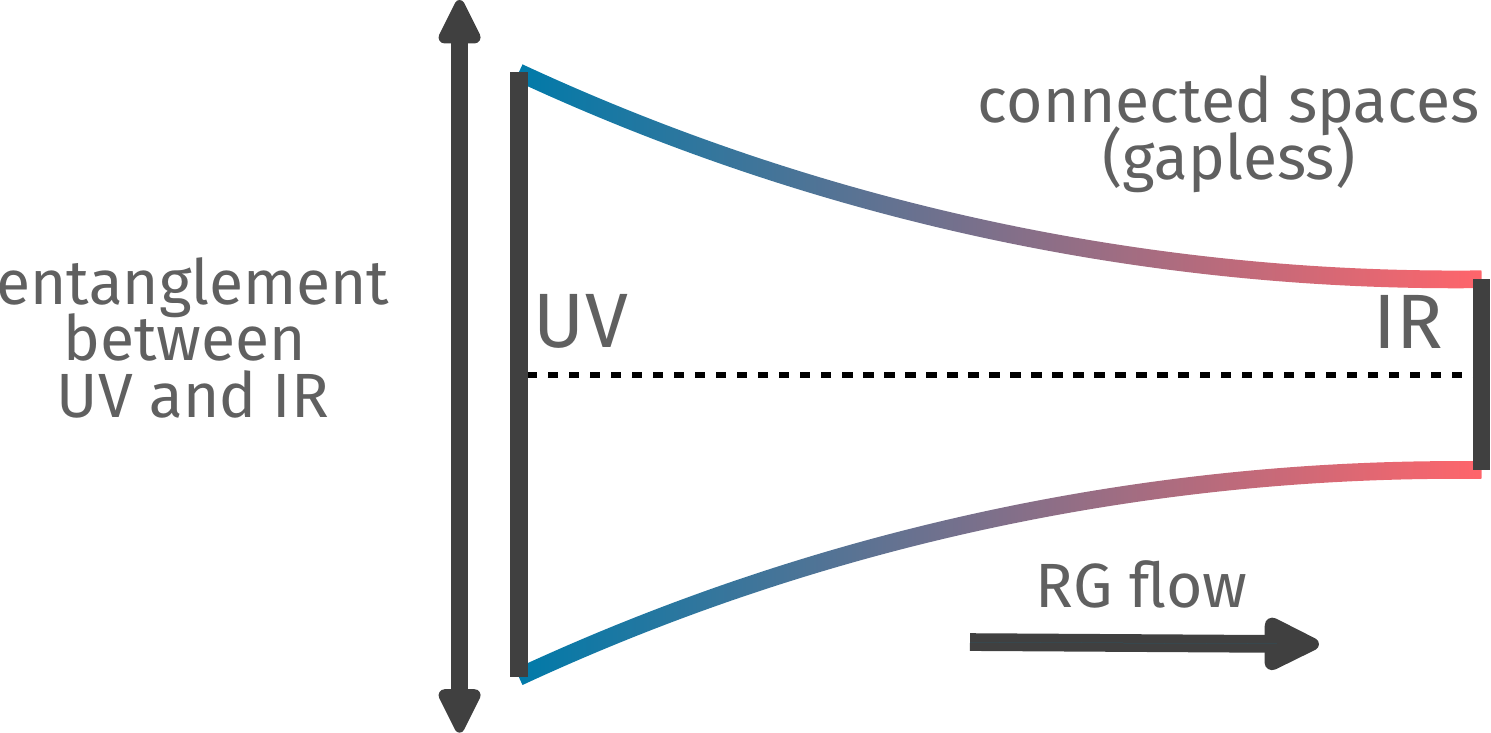}\\
	\vspace*{20pt}
	\includegraphics[width=0.45\textwidth]{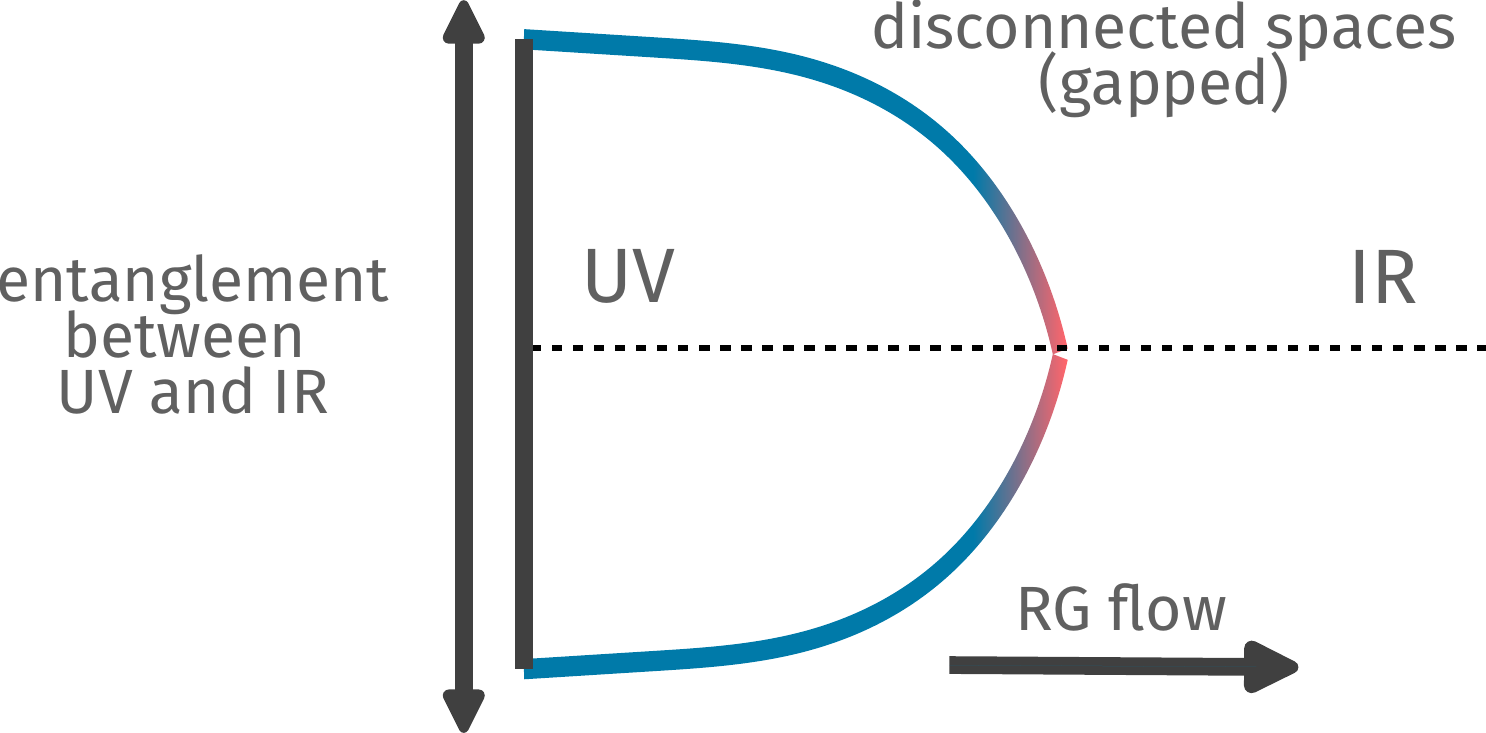}
	\caption{Schematic depiction of the emergent geometries in the cases of gapless (top) and gapped (bottom) RG flows. While the gapless RG flow involves a growth in the entanglement between the UV and IR spaces, the latter leads to the vanishing of UV-IR entanglement. This results in the emergence of connected and disconnected geometries, respectively.}
	\label{wormhole}
\end{figure}

\section{Entanglement Holography and Fermionic Criticality}
\label{fermi_critical}
\subsection{Equivalence between a critical Fermi surface and a holographic wormhole}
As mentioned previously under eq.~\eqref{mapping}, the quantity \(z\) corresponds to the anomalous dimension of the spectral gap \(g_z\) in the effective field theory: a change in the sign of \(z\) corresponds to a change in the nature of the RG flow from relevance to irrelevance and vice-versa. The holographic nature of the RG flow dictates that such a change arises by a variation in some quantity in the parent interacting theory from which the effective field theory of non-interacting Dirac fermions can be thought to have originated. Indeed, such a change in the RG relevance has drastic consequences: a relevant spectral distance \(g_z\) implies a gap in the excitations at the fixed point, while an irrelevant \(g_z\) indicates the presence of gapless excitations. The presence or absence of a gap in the low-energy excitations determines the topology of the electronic Fermi surface through the change of a topological invariant, the Luttinger volume of the system~\cite{luttinger1960fermi,oshikawa2000topological,seki2017topological,anirbanurg1,anirbanurg2}~(also see eq.~\eqref{LT-violation}). 

At the transition between the gapless (metallic) and gapped (insulating) states, the Fermi surface becomes critical, signalled by the vanishing lifetime of the gapless excitations close to it. Our formulation shows that such a gap-inducing transition and its associated quantum critical Fermi surface has consequences for the space generated along the RG flow. For the gapless case (\(z < 0\)), the degrees of freedom in the UV and the IR have finite entanglement, indicating that they are part of the same connected holographic space(see Figs.~\ref{distance1} and \ref{mass-beta}), while for the gapped case (\(z < 0\)), the entanglement vanishes into the IR, leading to a decoupling of these two regions and their separation into disconnected spaces~\cite{van2010building}. These two scenarios are shown schematically in Fig.~\ref{wormhole}. It is therefore tempting to conclude that at the transition between the gapped and gapless states, a {\it wormhole geometry} connects the UV and the IR via some minimal entanglement~\cite{van2010building,cao_2017}. Our results also suggest that the presence of such a wormhole would then coincide with two topological transitions, the first being the change in the sign of the curvature (as shown in the previous subsection), and the second being the change in the topological Luttinger volume of the underlying quantum field theory (as discussed in the previous paragraph). We end by noting that we have provided, in a later section of this work, additional support for the topological nature of Luttinger's volume and its connection to the holographically generated space by relating it to multipartite entanglement (see eq.~\eqref{winding-2}).

\subsection{Connecting modular energy with the metric}
The notion of an entanglement Hamiltonian was introduced in eq.~\eqref{ent_Ham}. At the transition between the gapless (metallic) and gapped (insulating) states, the system becomes critical, and the Dirac electrons are described by a CFT. It turns out that for such a CFT, the entanglement Hamiltonian takes a particularly simple form. In this subsection, we will relate the emergent metric to the entanglement Hamiltonian obtained at the IR fixed point, focusing on the theory at the transition.

For a 1D CFT, the entanglement Hamiltonian corresponding to a partition of length \(l\) is given by~\cite{bisognanoWichmann_scalar,bisognanoWichmann_quantum}
\begin{equation}\begin{aligned}
	\mathcal{H}_\text{1D} = \frac{2\pi L}{c}\int_0^l\mathrm{dx}~ \frac{L^2 - x^2}{2L^2}T_{00}(x)~,
\end{aligned}\end{equation}
where \(T_{\mu\nu}\) is the stress-energy tensor of the CFT and \(c\) is the speed of light. We have already shown in eq.~\eqref{ent_ham_2d} that the entanglement Hamiltonian for the 2D theory is simply a sum of the 1D counterparts, owing to the decoupling of 1D modes in the Lagrangian. Working in a region where \(T_{00}\) is approximately uniform, the 2D entanglement Hamiltonian can be expressed as
\begin{equation}\begin{aligned}
	\mathcal{H}_\text{2D} \sim \sum_n \mathcal{H}_\text{1D} = \frac{2\pi L^2}{3c}\sum_n T^n_{00}~,
\end{aligned}\end{equation}
where \(T^n_\text{00}\) is the average value of \(T_{00}\) within the partition for the CFT associated with the \(n^\text{th}\) mode.

In order to make the connection with the emergent geometry, we use the entanglement first law~\cite{Blanco2013}, which states that at first order, small variations \(\Delta S\) of the entanglement entropy can be equated to the associated change in the expectation value of the entanglement Hamiltonian:
\begin{equation}\begin{aligned}
	\label{entanglement_stresstensor}
	\Delta S(A) = \Delta I_2(A_0:A) = \left<\Delta \mathcal{H} \right> = \frac{2\pi L^2}{3c}\sum_n \left<T^n_{00}\right>,
\end{aligned}\end{equation}
where \(\left<\mathcal{H}\right>=\text{Tr}\left( \rho \mathcal{H} \right) \) is often referred to as the modular energy, and we have equated \(\Delta S(A)\) and \(\Delta I_2(A_0:A)\) following eq.~\eqref{mutinfo}. Further, since the entanglement at the critical point is expected to be large, we can approximate the expression for the metric between points \((i,j)\) obtained in subsection~\ref{metric_subsection} as \(\tilde d(i,j)  = -\ln I^2(i:j)/I^2_\text{max} \simeq |I^2(i:j)/I^2_\text{max} - 1|\). Combining this equation with eq.~\eqref{entanglement_stresstensor} leads to the following interesting relation:
\begin{equation}\begin{aligned}
	\Delta \tilde d(i,j) \simeq \frac{2\pi L^2}{3c}\sum_n \left<T^n_{00}\right>~.
\end{aligned}\end{equation}
The equation relates the change in a geometric measure (the emergent metric) with a mass/energy term. An analogous relation was obtained for states for which the entropy obeys an area law by Cao et al~\cite{cao_2017}, connecting the spatial curvature and the stress tensor, ultimately leading to a semi-classical Einsteins field equation for gravity. Our formulation does not however lead to a relation involving the curvature, and we speculate that this is due to the violation of the area law in entanglement of fermionic critical systems. Instead, we show in the next section that the curvature in our formulation is related to the rate of convergence of the RG flows.

\subsection{Kinematics of the RG flows}
\label{kinematics}
Further statements on the RG flow and the emergent holographic dimension can be made by studying quantities that can serve as candidates for the convergence parameter \(\theta\) for the RG flows of \(g_z\)~\cite{kar_2001,kar2007raychaudhuri}. We consider the following funcion as a candidate for the convergence parameter for the case of \(z>0\):
\begin{equation}\begin{aligned}
	\label{theta-def}
	\theta_z(j) = \frac{1}{\sqrt{v_z^{-2} + 1}} = \left[1 +\left\{ \frac{1 - e^{-\beta_z(j)}}{e^{\beta_z(j+1) - 1}} \right\}^2  \right]^{-\frac{1}{2}}~. 
\end{aligned}\end{equation}
\begin{figure}
	\includegraphics[width=0.48\textwidth]{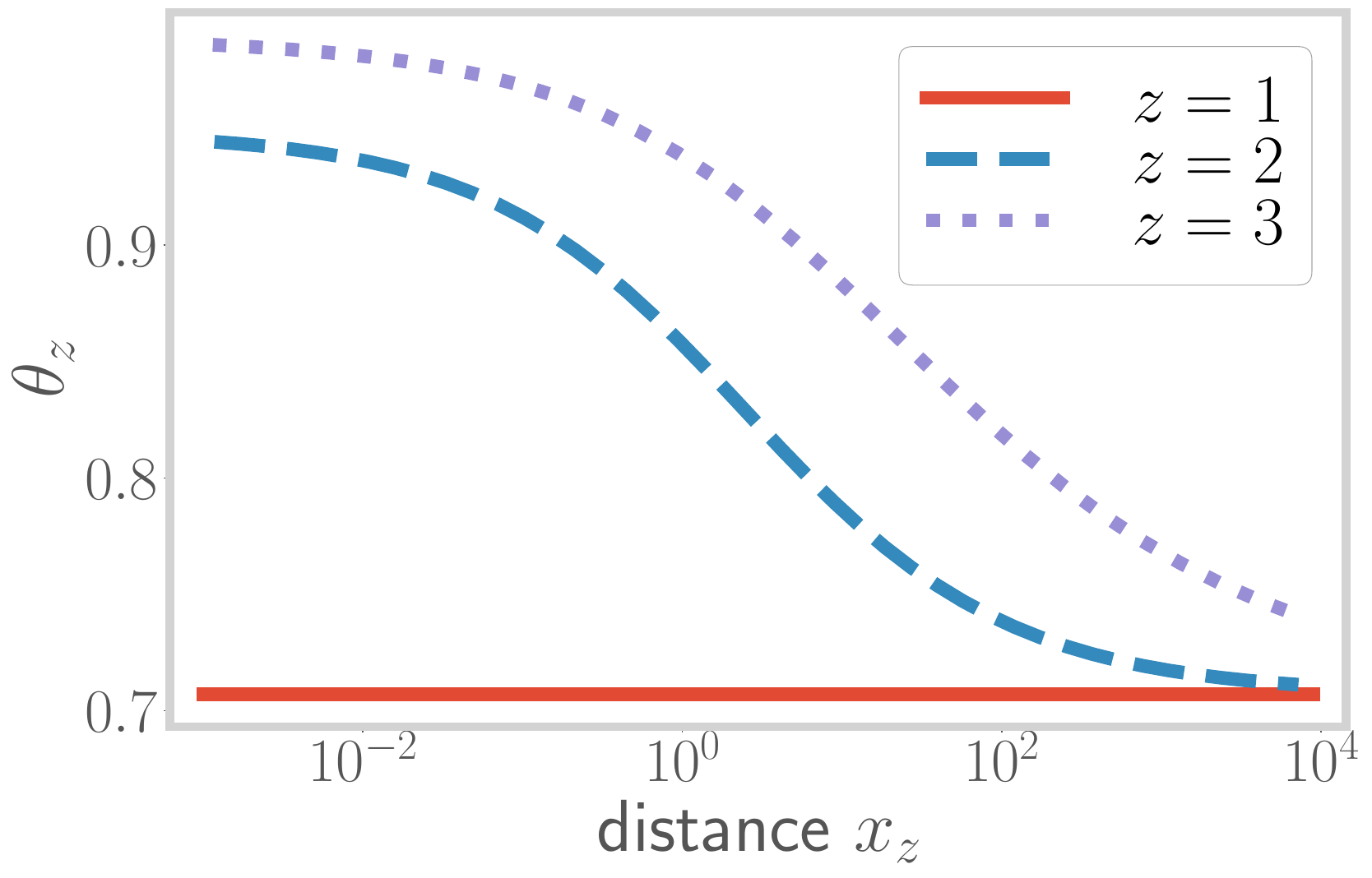}
	\caption{Evolution of expansion parameter \(\theta_z\) with distance $x_{z}$, leading to the fixed point at \(\theta_z^* = 1/\sqrt 2\).}
	\label{theta}
\end{figure}
The parameter \(\theta_z\) has been plotted in Fig.~\ref{theta}, and the behaviour is observed to be very similar to that of the beta function shown in Fig.~\ref{mass-beta}. Indeed, this similarity describes the holographic nature of the RG flow: the emergent geometry (as seen through \(\theta_z\)) is dictated by the RG flow described by \(\beta\).

The quantity $\theta$ can be interpreted as an estimate of the rate of change of area enclosed by a fixed number of RG trajectories. To see that this is the case, we note that by using \(v_j = \Delta y/\Delta x\), the expansion parameter can be written as \(\theta = \Delta y(j)/\sqrt{\Delta x^2 + \Delta y^2} \sim \Delta g(j+1)/\sqrt{\Delta x^2 + \Delta y^2}\). Considering RG flows for \(g_z(j)\) corresponding to multiple \(z\), the area enclosed by the curves is given by the maximum \(g_z\), and \(\Delta g_z\) is then the change in the area under the RG transformation. The denominator \(\sqrt{\Delta x^2 + \Delta y^2}\) is the distance covered in the emergent spacetime, rendering \(\theta\) as the rate of change of the area.

The equation of motion for this expansion parameter can then be obtained as:
\begin{equation}\begin{aligned}
\frac{\:\mathrm{d}\theta_z}{\:\mathrm{d}x} = \frac{v_z^{-3}}{\left(1 + v_z^{-2}\right)^\frac{3}{2}} \frac{\:\mathrm{d}v_z}{\:\mathrm{d}x_j} = \kappa_z~.
\label{RC}
\end{aligned}\end{equation}
The simple form of this relation makes it clear that for relevant flows (corresponding to a negative curvature \(\kappa_z<0\)), \(\theta_z\) flows to a minimum. Following Fig.~\ref{c-charge}, this is accompanied by the minimisation of the \(c-\)function to zero. It is, therefore, tempting to conclude that eq.~\eqref{RC} for the spatial evolution of the convergence parameter $\theta$ is dual to eq.~\eqref{c-charge-flow} that controls the flow of the $c$-function of the quantum theory along the RG flow.

\section{Extension to the massive case}
\label{massive}
We will now generalise several of the main results obtained above to the case of two-dimensional electrons with a single-particle mass-gap  \(M_{n,\phi} = \sqrt{m^2 + \frac{4\pi^2}{L_x^2}(n + \phi)^2}\). Given that they depend only on the geometric part of the entanglement (which remains unchanged on introducing the mass $M_{n,\phi}$), the following set of key qualitative results and observations discussed earlier continue to hold for the case of a dispersion with a mass-gap:
\begin{itemize}
	\item the hierarchical structure of the entanglement content (e.g., the area-law term in eqns.(\ref{EE-subsystem}) and (\ref{Ig_def})) and the fact that the entanglement satisfies the Ryu-Takayanagi bound (\alert{bottom panel} of Fig.\ref{area}),
	\item the dependence of the emergent distances \(x\) and \(y\) on the RG flow (Fig.\ref{distance1}, eqns.(\ref{info-dist2}) and (\ref{info-dist3})), and the fact that the RG beta function tracks the central charge (Fig.\ref{c-charge}),
	\item various qualitative and quantitative properties of the curvature (e.g., eq.\eqref{kappa}), including the topological nature of its sign (eq.\eqref{topo-kappa}), and
	\item the existence of a convergence parameter that is related to the RG flow, and the 
	equation that governs the evolution of the convergence parameter (eq.\eqref{RC}).
\end{itemize}

\begin{figure*}[!ht]
	\centering
	\includegraphics[width=0.9\textwidth]{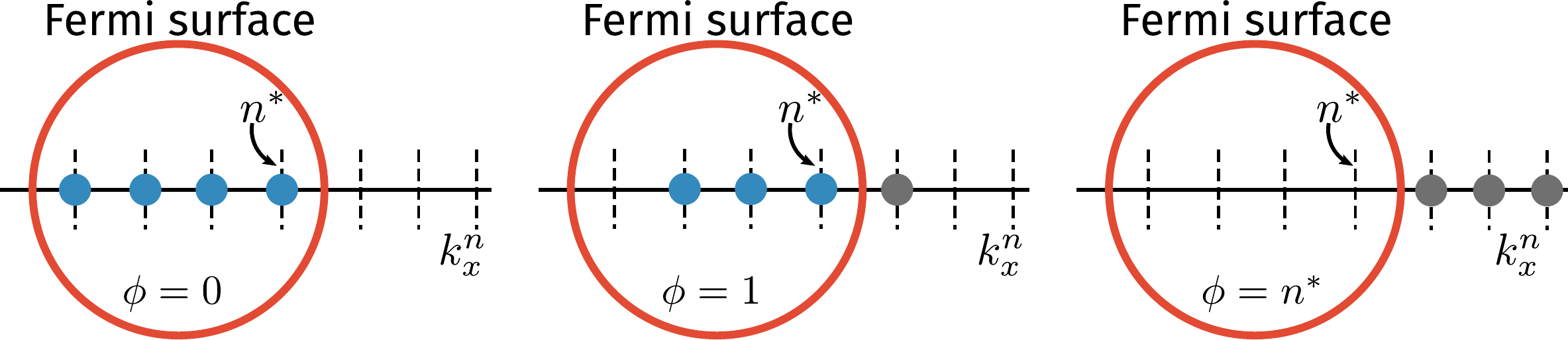}
	\caption{Spectral flow of the states within the Fermi voume. The red curve represents the Fermi surface. Upon tuning the flux \(\phi\), the \(k-\)space states (indicated in the form of dotted lines along the \(k_x-\)axis) are shifted towards the Fermi surface. The blue circles represent electrons with momenta below \(k_F\), and are part of the ground state. On the other hand, the grey circles represent particles with momenta above \(k_F\): they have moved out of the system, and into the reservoir connected to it (see discussion in main text). Each time \(\phi\) changes by 1, the state nearest to the Fermi surface moves out of the Fermi volume. Tuning \(\phi\) by \(n^*\) empties the entire Fermi volume.}
	\label{spectral-flow}
\end{figure*}

The functional forms of certain other results do, however, undergo changes. For instance, the coupling \(g_z(j)\) is given by the (more general) expression
\begin{equation}\begin{aligned}
	&g_z(n,j) \\
	&= \ln \frac{M_{n+1,\phi}(j) - M_{n,\phi}(j)}{2\pi/L_x} \\
	&= \ln \left[\sqrt{\tilde m^2 + \left(t_z(j)(n+1) + \phi\right)^2} -\sqrt{\tilde m^2 + \left(t_z(j)n + \phi\right)^2} \right]~, 
\end{aligned}\end{equation}
where \(\tilde m = m L_x/2\pi\) is a dimensionless parameter. We will work with the coupling for the mode \(n=0\):
\begin{equation}\begin{aligned}
	g_z(0,j) = \ln \left[\sqrt{\tilde m^2 + \left(t_z(j) + \phi\right)^2} -\sqrt{\tilde m^2 + \phi^2} \right]~. 
\end{aligned}\end{equation}
In order to simplify the analysis, we work in the large mass limit \(\frac{t_z + \phi}{\tilde m} \to 0\).
In this limit, upon expanding the square root and retaining only the lowest order term, we obtain
\begin{equation}\begin{aligned}
g_z(0,j) = \ln \frac{\left[t_z(j) + \phi\right] ^2 - \phi^2}{2\tilde m}~.\\
\end{aligned}\end{equation}
The RG beta function $\beta = \ln (g_{z}(j+1)/g_{z}(j))$ can, in the large mass limit, then be related to the emergent distances \(x\) and \(y\) in terms of the coupling \(g_z(j)\):
\begin{gather}
	x_z(j) = \ln t_{z}(j) = \ln \left[\sqrt{2\tilde m~ e^{g_{z}(0,j)} + \phi^2} - \phi\right]~, \\
	y_z(j) = \begin{cases}
		\ln \left[\sqrt{2\tilde m~ e^{g_{z}(0,j+1)} + \phi^2} - \phi\right],~ z > 0~, \\
		\\
		\ln \left[\sqrt{2\tilde m~ e^{g_{z}(0,j-1)} + \phi^2} - \phi\right],~ z < 0 ~.\\
	\end{cases}
\end{gather}
The quantitative dependence of the \(c-\)function on the RG beta function can also be expressed more generally as
\begin{equation}\begin{aligned}
	e^{\beta_z(j)} = \left[\ln \frac{\left(\frac{\tilde c(j+1)}{c} + \phi\right) ^2 - \phi^2}{2\tilde m}\right] \left[\ln \frac{\left(\frac{\tilde c(j)}{c} + \phi\right) ^2 - \phi^2}{2\tilde m}\right]^{-1}~.
\end{aligned}\end{equation}
Quantities such as the curvature $\kappa$ and the convergence parameter $\theta$ can similarly be expressed in terms of the coupling \(g_z(j)\).

\section{Topological content of the entanglement spectrum}
\label{topologicalContent}
\subsection{Luttinger's volume as a winding number: the massless case}
We will now show how to obtain Luttinger's volume (eq.\eqref{Luttvol}) from the $g$-partite information entanglement measure \(I^g\) of a system of massless 2D electrons. The flux-dependent part of the $g$-partite information \(I^g\) can be used to probe the \(k-\)space volume \(V_L\) within the Fermi surface. We define a quantity \(Q_g(\phi)\) that is periodic in the flux \(\phi\) and independent of the geometry:
\begin{equation}\begin{aligned}
	Q_g(\phi) = f_{g,z}\left[\frac{1}{\sqrt 2} - \frac{1}{2}e^{-\frac{1}{c}\left(I^g_{\left\{ \mathcal{A}_{z}(i) \right\}}(\phi) - I^g_{\left\{ \mathcal{A}_{z}(i) \right\}}(\phi^{(0)}f_{g,z})\right)}\right] \\
	= f_{g,z}\left[\frac{1}{\sqrt 2} - \big|\sin \left(\pi\phi f_{g,z}\right)\big|\right] ~,
\end{aligned}\end{equation}
where \(\phi^{(0)} = \frac{1}{6}\) is the value at which the flux-dependent part of \(S_0(\phi)\) vanishes. Further, from eq.\eqref{manyparty-info}, we have 
\begin{equation}
I^g_{\left\{ \mathcal{A}_{z}(i) \right\}}(\phi_g) - I^g_{\left\{ \mathcal{A}_{z}(i) \right\}}(\phi^{(0)}_g f_{g,z}) = -c\ln \left(2|\sin\left(\pi\phi f_{g,z}\right)|\right)~.
\end{equation}
Note that the non-differentiability of \(Q_g(\phi)\) at \(\phi f_{g,z} = 0, 1, 2, \ldots\) is not a concern as \(\phi\) is chosen to be non-integral, ensuring that \(\phi f_{g,z}\) is also non-integral.

In order to express the number \(N_L^x\) of positive \(k_x-\)states available inside the Fermi surface in the \(x-\)direction in terms of \(Q_g(\phi)\), we need to establish a relation between the flux \(\phi\) and the Fermi surface. The \(k_x-\)states of interest to us are given by \(k_x^n = \frac{2\pi n}{L_x}, n\in \mathbb{Z}, 0 \leq k_x^n \leq k^F_x\), where \(k_x^F\) is the \(x-\)component of the Fermi momentum \(\vec k_F\). Assuming the flux-free Hamiltonian has time-reversal symmetry, there should be an equal number of states with the opposite momenta. Let \(n^*\) be the integer for which \(k_x^{n^*} \leq k^F_x\) and \(k_x^{n^* + 1} > k^F_x\). Since the flux couples to the \(k_x-\)states as \(k_x^n \to k_x^n + eA = \frac{2\pi n + 2\pi \phi}{L_x}\), tuning \(\phi\) by 1 leads to the entire spectrum of \(k_x-\)states being shifted by one quantum number: \(k_x^n \to \frac{2\pi \left(n + 1\right)}{L_x} = k_x^{n+1}\).
This means that the state \(k_x^{n^*}\) moves to \(k_x^{n^*+1}\), and hence out of the Fermi volume by crossing \(k^F_x\). This flow of the spectrum is shown in Fig.~\ref{spectral-flow}.

\begin{figure}
	\centering
	\includegraphics[width=0.48\textwidth]{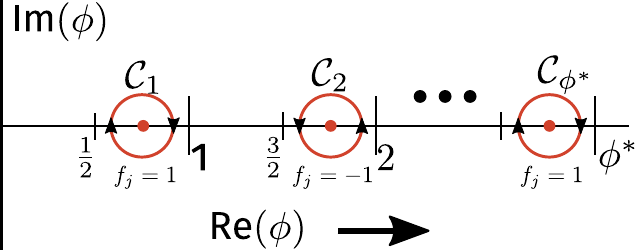}
	\includegraphics[width=0.48\textwidth]{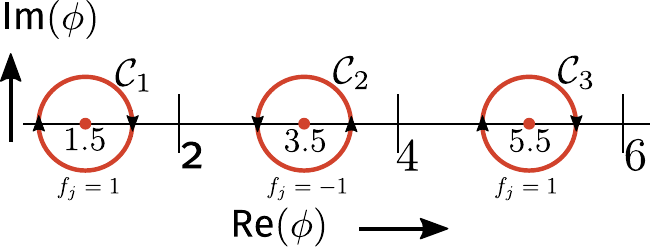}
	\caption{{\it Top:} Contours \(\mathcal{C}_j\) used to integrate over \(Q^{-1}_0\) in eq.~\eqref{Nlx_0}. Tall vertical lines represent integer values of \(\text{Re}(\phi)\). All contour start and end within two adjacent integer values. The final contour \(\mathcal{C}_{\phi^*}\) ends just short of \(\text{Re}(\phi) = \phi^*\). Note the alternating direction of the contours brought about by the alternation of \(f_j\) in going from \(j=1\) to \(j=2\), and so on.
{\it Bottom:} Contours \(\mathcal{C}_j^g\) for \(t_{g,z}=2\), centered at \(2\left( 1 - \frac{1}{4} \right),2\left( 2 - \frac{1}{4} \right)\) and \(2\left( 3 - \frac{1}{4} \right)\). Note the alternating directions of the contours. }
	\label{contour}
\end{figure}

If we now connect the system to a particle reservoir with the chemical potential set to zero, any \(k_x-\)state that moves out of the Fermi volume becomes unoccupied. We continue tuning the flux to a value \(\phi^*\), where the system becomes entirely devoid of any particles. The number of \(k_x-\)states that are below \(k_x^F\) can then be said to be given by the number of times \(\phi\) goes through an integer value while changing from \(0^+\) to \(\phi^* + 0^+\). Since we do not set the flux to an integer value, we can equivalently count the number of times \(\phi\) goes through, say, three-quarters of an integer - \(j - \frac{1}{4}: j = 1,2,\ldots\) - in changing from \(0^+\) to \(\phi^*-0^+\). We will now show that this count can be expressed in terms of \(Q_g\). We begin by considering \(Q_0(\phi) = \frac{1}{\sqrt 2} - |\sin \pi\phi|\)~. At every three-quarter value of the flux, the quantity \(\phi_j = j - \frac{1}{4}: j=1,2,\ldots\), \(|\sin\pi\phi_j|\) attains the value of \(\sin \frac{\pi}{4} = \frac{1}{\sqrt 2}\), such that \(Q^{-1}_0(\phi)\) goes through a simple pole. Each such pole \(\phi_j\) can be picked up via a complex integral around a contour \(\mathcal{C}_j\) that encloses the pole:
\begin{equation}\begin{aligned}
	\label{pole-1}
	-f_j \frac{1}{2\sqrt 2 i}\oint_{\mathcal{C}_j} \frac{d\phi}{Q_0(\phi)} = 1~,
\end{aligned}\end{equation}
where \(f_j\) is either \(+1\) or \(-1\) and is defined through the equation \(|\sin\left( \pi \left( j - \frac{1}{4} \right)  \right)| = f_j\sin\left( \pi \left( j - \frac{1}{4} \right)  \right)\), and \(\mathcal{C}_j\) is a circle of diameter \(1/4\) centered at \(\phi = j-\frac{1}{4}\) (shown in Fig.~\ref{contour-1}).
The evaluation of the integral eq.\eqref{pole-1} is shown in \alert{Appendix~\ref{contour-integral}}.
In order to drop the sign prefactor \(f_j\), we reorient those contours for which \(f_j\) is \(+1\), and such that they wind around in the clockwise direction. The contours are shown in the \alert{top panel} of Fig.~\ref{contour}, and defined as
\begin{equation}\begin{aligned}
	\label{curves0}
	C_j(\phi) = \left\{\phi: \phi = j - \frac{1}{4} + \frac{1}{8}e^{i\theta}, \theta \in \left[0, -2\pi f_j\right] \right\}~,
\end{aligned}\end{equation}
such that they host a singularity at their center. With this redefinition of the contours, we have
\begin{equation}\begin{aligned}
	\frac{1}{2 \sqrt 2 i}\oint_{\mathcal{C}_j} \frac{d\phi}{Q_0(\phi)} = 1~.
\end{aligned}\end{equation}

Counting the number of three-quarter values then reduces to counting the number of poles of \(Q^{-1}_0\) in the above-mentioned range \(\phi \in \left( 0, \phi^* \right) \). In this way, we can now express the count \(N_L^x\) in terms of \(Q_0\):
\begin{equation}\begin{aligned}
	\label{Nlx_0}
	N_L^x = \frac{1}{2} \sum_{j=1}^{\phi^*}\frac{1}{2\sqrt 2i}\oint_{\mathcal{C}_j(\phi)} \frac{d\phi}{Q_0(\phi)} = \frac{1}{2\sqrt 2}\sum_{j=1}^{\phi^*}\frac{1}{2 i}\oint_{\mathcal{C}_j(\phi)} \frac{d\phi}{Q_0(\phi)}~.
\end{aligned}\end{equation}
The factor of half in eq.\eqref{Nlx_0} is required as we want to count only the number of positive momentum states (and which is equal to the number of negative momentum states).

\begin{figure}[!t]
	\centering
	\includegraphics[width=0.48\textwidth]{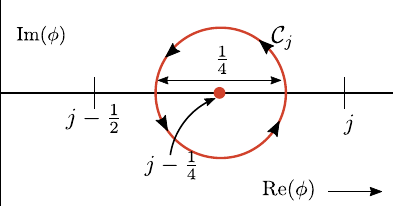}
	\caption{Contour \(\mathcal{C}_j\) used to integrate eq.~\eqref{pole-1}. The contour is chosen so as to encircle the simple pole of \(1/Q\) at \(\phi=j-1/4\).}
	\label{contour-1}
\end{figure}
In going from \(Q_0(\phi)\) to a general \(Q_g(\phi)\), the only change required is to redefine \(\phi \to \phi_g = f_{g,z}\phi\). This implies that while the states in \(k-\)space have become more coarse-grained, the flux \(\phi\) has become more fine-tuned by the same scaling factor.
This is a reflection of the overall conservation of particle number density: keeping the chemical potential fixed, the increased spacing between the \(k_x-\)states implies that a smaller number of particles are now occupying an equally smaller volume. As the number density is conserved, so is the Luttinger volume (eq.\eqref{Luttvol}), and this shows that our approach is consistent with the holographic Luttinger theorem~\cite{sachdev_2011}. Thus, given that \(\phi_g^* = \phi^*\), the expression for \(N_L^x\) remains practically unchanged:
\begin{equation}\begin{aligned}
	N_L^x = \frac{1}{2\sqrt 2}\sum_{j=1}^{\phi^*}\frac{1}{2 i}\oint_{\mathcal{C}_j(\phi_g)} \frac{d\phi_g}{\frac{1}{\sqrt 2} - |\sin \pi \phi_g|}~.
\end{aligned}\end{equation}
In order to rewrite $N_L^x$ in terms of the flux \(\phi\), we start by making the substitution \(\phi_g \to f_{g,z}\phi\):
\begin{equation}\label{NLx}\begin{aligned}
	N_L^x = \frac{1}{2\sqrt 2}\sum_{j=1}^{\phi^*}\frac{1}{2 i}\oint_{\mathcal{C}_j(f_{g,z}\phi)} \frac{f_{g,z} d\phi}{\frac{1}{\sqrt 2} - |\sin \left( \pi f_{g,z}\phi\right)| }~.
\end{aligned}\end{equation}
In the above equation, we can recognise  \(\frac{f_{g,z} }{\frac{1}{\sqrt 2} - |\sin \left( \pi f_{g,z}\phi\right)|}\) as \(Q_g^{-1}(\phi)\).
Also note that the contour can be written as
\begin{equation}\begin{aligned}
	\mathcal{C}_j(f_{g,z}\phi) \equiv \mathcal{C}^g_j(\phi) = \bigg\{\phi:\phi = t_{g,z}\left( j - \frac{1}{4} + \frac{e^{i\theta}}{8} \right),\\
\theta \in \left[0, -2\pi f_j\right]\bigg\}~.
\end{aligned}\end{equation}
These curves are again circles of radius \(1/8\), but centered at \(t_{g,z}\times \left( 1 - \frac{1}{4} \right) ,t_{g,z}\times \left( 2 - \frac{1}{4} \right) ,t_{g,z}\times \left( 3 - \frac{1}{4} \right) \) and so on. These are simply a generalisation of the curves defined in eq.~\eqref{curves0}; the curves for \(t_{g,z}=2\) are shown in the \alert{bottom panel} of Fig.~\ref{contour}.
Substituting this in eq.\eqref{NLx} gives
\begin{equation}\begin{aligned}
	\label{Lut_volume}
	N_L^x = \frac{1}{2\sqrt 2}\sum_{j=1}^{\phi^*}\frac{1}{2 i}\oint_{\mathcal{C}_j^g(\phi)} \frac{d\phi}{Q_g}~.
\end{aligned}\end{equation}

For a spatially isotropic system, the Fermi volume is given by
\begin{equation}\begin{aligned}
	\label{iso-LV}
	V_L = v_2 \times \pi \left(N_L^x\right)^2 = \frac{1}{2}\frac{\pi^3}{V}\left[\sum_{j=1}^{\phi^*}\frac{1}{2 i}\oint_{\mathcal{C}^g_j(\phi)} \frac{d\phi}{Q_g(\phi)}\right]^2~,
\end{aligned}\end{equation}
where \(V\) is the real space volume of the system, and \(v_d = \left( 2\pi \right)^d/V \) is the volume of the smallest unit in \(k-\)space. For a generally anisotropic Fermi volume, each direction in the \(d-\)dimensional \(k-\)space can be completely specified by an angular coordinate \(\theta\) with respect to some predefined coordinate system.
For each direction \(\hat \theta\), we can insert a gauge field \(\vec A = A \hat \theta\) and take the thin torus limit in this direction, obtaining thereby the \(g-\)partite information \(I^g(\phi_\theta)\) for length \(l\) in the direction perpendicular to \(\hat \theta\). With the \(\theta-\)dependent \(I^g\), one can then define the quantity \(Q_g(\phi_\theta)\) which counts the number of states \(N_L^\theta\) in the direction \(\hat \theta\) inside the Fermi volume. In this way, the total Fermi volume can be obtained by integrating over \(\theta\):
\begin{equation}\begin{aligned}
	V_L = \frac{\pi^2 }{4V} \int_0^{2\pi} d\theta \left[\sum_{j=1}^{\phi_\theta^*}\frac{1}{2 \pi i}\oint_{\mathcal{C}_j^g(\phi)} \frac{d\phi}{Q_g(\phi)}\right]^2~.\label{luttvolhier}
\end{aligned}\end{equation}
It is important to note that the quantity $V_{L}$ is independent of the RG index $g$, and is thus identical in value for all $g$ members of the hierarchy at the $g$th step of the coarse-graining process.

The topological nature of Luttinger volume can be made further manifest by expressing it in terms of winding numbers, following Ref.\cite{seki2017topological} for the case of interacting electrons. To do so, we point out that the integrand \(Q^{-1}_g(\phi)\) can be written as a derivative:
\begin{equation}\begin{aligned}
	\frac{1}{Q_g(\phi)} = \frac{\sqrt 2}{\pi} \frac{\partial{}}{\partial{\phi}} \ln Y^g_j = \frac{\sqrt 2}{\pi} \frac{1}{Y^g_j}\frac{\partial{Y^g_j}}{\partial{\phi}}~,
\end{aligned}\end{equation}
where 
\begin{equation}\begin{aligned}
	\label{y-def}
	Y^g_j(\phi) &\equiv \left[\frac{f_j \tan 3\pi/8 - \tan \left( \pi f_{g,z} \phi/2 \right)}{\tan \left(\pi f_{g,z} \phi/2 \right) - \tan \left(\pi/8 \right) }\right]^{f_j} \\
		    &= \begin{cases}
			    \frac{\tan 3\pi/8 - \tan \left( \pi f_{g,z} \phi/2 \right)}{\tan \left( \pi f_{g,z} \phi/2 \right) - \tan \left( \pi/8 \right) }, \text{ when }j=\text{odd},~\text{and}\\
			    \frac{\tan \left( \pi f_{g,z} \phi/2 \right) + \tan \left( \pi/8 \right)}{-\tan 3\pi/8 - \tan \left( \pi f_{g,z} \phi/2 \right)}, \text{ when }j=\text{even}~.
	\end{cases}
\end{aligned}\end{equation}
Substituting this into the expression for \(2-\)dimensional isotropic Luttinger volume in eq.~\eqref{iso-LV}, we obtain
\begin{equation}\begin{aligned}
	\label{winding}
	V_L = \frac{\pi^3}{V}\left[\sum_{j=1}^{\phi^*}\frac{1}{2\pi i}\oint_{Y^g_j(\mathcal{C}_j^g)} \frac{\:\mathrm{d} Y_j^g}{ Y_j^g}\right]^2~.
\end{aligned}\end{equation}
As before, the specific integral in the above expression is a winding number that counts the number of times the contour \(Y_j^g(\mathcal{C}_j^g)\) winds around the singularity at \(Y_j^g=0\). Note that these are the same singularities that exist at the centers of the contours \(\mathcal{C}_j^g\) and that are picked up in, for example, eq.~\eqref{pole-1}. To demonstrate this, we point out that the centers of the contours are given by \(\phi = t_{g,z}\left(j - \frac{1}{4}\right), j =1,2,\ldots,\phi^* \).
At these values, we have
\begin{equation}\begin{aligned}
	\tan \left( \frac{\pi f_{g,z} \phi}{2} \right) = \tan \left( \frac{\pi j}{2} - \frac{\pi}{8} \right) = \begin{cases}
	\tan \frac{3\pi}{8}, j=1,3,\ldots\\
	-\tan \frac{\pi}{8}, j=2,4,\ldots\\
\end{cases}
\end{aligned}\end{equation}
Comparing with eq.~\eqref{y-def}, we can see that the numerators of \(Y_g^j\) are zero at these values of \(\phi\) for both even and odd values of \(j\). This leads to \(Y_g^j=0\), and hence a singularity in \(\frac{d Y_g^j}{Y_g^j}\). These values of the flux are therefore the common singularities that are picked up by the poles of \(Q^{-1}_g(\phi)\) as well as the winding numbers.

We now define \(w_g^j\) as the winding number for the curve \(Y_g^j(\mathcal{C}_j^g)\) around \(Y_g^j=0\), and \(\mathcal{W}(\phi^*) = \sum_{j=1}^{\phi^*}w_j^g\) as the total winding number for all the curves \(\left\{Y_j^g\left( \mathcal{C}_j \right), j=1,2,\ldots \right\} \). Then, the isotropic Fermi volume takes the compact form
\begin{equation}
	\label{winding-2}
	V_L = \frac{\pi^3}{V}\mathcal{W}(\phi^*)^2~.
\end{equation}

\subsection{A relation between the quantum hall Chern numbers and Luttinger's volume}
\begin{figure}[!t]
	\centering
	\includegraphics[width=0.3\textwidth]{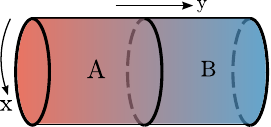}
	\caption{Cylindrical geometry for the integer quantum Hall system, with a magnetic field $\vec{B}$ pointing radially outwards through the surface of the cylinder. An Aharanov-Bohm vector potential acts in the \(x-\)direction (i.e., along the circumference). The entanglement is calculated between the subsystem \(A\) and its compliment \(B\).}
	\label{cylinder}
\end{figure}

Interestingly, the above method employed in using entanglement measures to extract the Luttinger volume can also be applied (with suitable modifications) to an integer quantum hall system (generated upon placing the system in a strong transverse magnetic field) to obtain the number of filled Landau levels (a Chern number). To begin, we imagine a system of free non-relativistic electrons in a cylinder geometry (\alert{Fig}.~\ref{cylinder}), and placed in a magnetic field \(\nabla\times \vec A = \vec B\) that points radially outwards. The field \(\vec B\) generates \(\Phi\) flux quanta through the surface of the cylinder. 
An Aharonov-Bohm flux \(\vec A^\prime = -2\pi\phi/L_x \hat x\) is also placed in the \(x-\)direction (i.e., along the circumference of the cylinder). We divide the cylinder into two subsystems \(A\) and \(B\) (\alert{Fig}.~\ref{cylinder}), such that tuning the dimensionless flux \(\phi\) by 1 transfers a charge equal to the first Chern number, from \(A\) to \(B\).
The translation invariance in the \(x-\)direction ensures that the momentum \(k_x\) are good quantum numbers; for each value of \(k_x\), we can define the matrix elements of the single-particle correlation matrix \(G(k_y)\) for the subsystem \(A\) as~\cite{Peschel_2003,Alexandradinata2011}
\begin{equation}\begin{aligned}
	G^A_{ij}(k_x) = \braket{\psi_\text{gs} | c^\dagger_{k_x,i} c_{k_x,j} | \psi_\text{gs}}~,
\end{aligned}\end{equation}
where \(\ket{\psi_\text{gs}}\) is the ground state and \(i,j\) represent the degrees of freedom other than \(k_x\). The total matrix \(G^A\) is constructed by joining the smaller blocks \(G^A(k_x)\).
As \(\phi\) is varied, the variation of the trace of the flux-dependent matrix \(G^A(\phi)\) is linear in the flux \(\phi\): \(\text{Tr}\left[G^A(\phi)\right] - \text{Tr}\left[G^A(0)\right] = -\frac{1}{e}\sigma_H\phi\).
This implies that as long as there are extended states at sufficiently small energies, we will have spectral flow that can be captured through the integral:
\begin{equation}\begin{aligned}
-\frac{\sigma_H}{2\pi i e}\int_{\mathcal{C}_H} \frac{d\phi}{\text{Tr}\left[G^A(\phi)\right] - \text{Tr}\left[G^A(0)\right]}~,
\end{aligned}\end{equation}
where \(\mathcal{C}_H\) is a circle of radius \(1/2\) in the complex plane centered at \(\phi=0\).
This integral is zero if there is no spectral flow, i.e., the effect of the gauge field can be removed by a gauge transformation and the trace is insensitive to the flux. In order to count the number of filled Landau levels (essentially, the filling factor \(\nu\)), we can now vary the flux quanta \(\Phi\) (due to \(\vec B\)) so as to change the filling from \(\nu\) to 0:
\begin{equation}\begin{aligned}
	\label{landau}
	\nu = -\frac{\sigma_H}{2\pi i e}\sum_{\Phi^\prime=\Phi}^{\Phi^\text{max}}\int_{\mathcal{C}_H} \frac{d\phi}{\text{Tr}\left[G^A_{\Phi^\prime}(\phi)\right] - \text{Tr}\left[G^A_{\Phi^\prime}(0)\right]}~.
\end{aligned}\end{equation}
The subscript \(\Phi^\prime\) on \(G^A\) indicates that the correlations depend on the variable magnetic field \(\vec B^\prime\). The upper limit \(\Phi^\text{max}\) is given by the lowest value of the magnetic field \(\vec B^\prime\) for which only one Landau level is filled. The magnetic field is varied such that exactly an integer number of Landau levels are filled at any given value. Importantly, the filling factor $\nu$ is known to be a Chern number $\mathcal{C}\left(\Phi\right)$~\cite{thouless1982quantized,niu1985quantized}.

We can also relate eq.~\eqref{landau} to the entanglement of the system by introducing the entanglement spectrum~\cite{li2008entanglement,laflorencie2016quantum}, which refers to the set of eigenvalues \(\left\{ \lambda_\chi^A(k_x) \right\} \) of the reduced density matrix \(\rho^A(k_x)\) for subsystem \(A\) corresponding to the momentum mode \(k_x\) in the \(x-\)direction and \(\chi\) is a particular configuration of the degrees of freedom \(\left\{ i \right\} \) in the \(y-\)direction.
We have the relation~\cite{Alexandradinata2011}
\begin{equation}\begin{aligned}
	\lambda^A_\chi(k_x) = \text{Det}\left[P_\chi G^A(k_x) \left( 1 - P_\chi \right) G^A(k_x)\right] ~,
\end{aligned}\end{equation}
where \(P_\chi\) is an operator that projects on to the subset of states \(\left\{ j \right\} \) in \(\left\{ i \right\} \) that are occupied in the configuration \(\chi\).
This relation between \(G^A\) and the eigenvalues \(\lambda\) then leads to the same conclusion as in the case without the magnetic field: the pole structure of the entanglement spectrum encodes the topological information of the system revealed in eq.\eqref{landau}. In the case of zero magnetic field, the topological quantity revealed from the pole structure was a winding number ($\mathcal{W}$), while in the present case it is a Chern number.

Finally, we reveal the connection between the winding numbers $\mathcal{W}$ in the metallic case and the Chern numbers in the integer quantum Hall system. The degeneracy of each Landau level is equal to the number of flux quanta \(\Phi\) generated by the magnetic field \(B\).
Since the total number of particles in the system is fixed, the electronic states of the filled Fermi sea that form the Luttinger volume (in the absence of the external B-field) are redistributed into the Landau levels with the appropriate degeneracy as a strong transverse magnetic field is turned on. Thus, we have the constraint that the total number of occupied states \(\mathcal{C} \times \Phi\) in the quantum hall system (\(\mathcal{C}\) being the Chern number and equal to the number of filled Landau levels) must be equal to \(\pi \left(N_L^x\right)^2 \), the total number of occupied states in the metallic system. Combining eqs.~\ref{Lut_volume} and \ref{winding-2} with this discussion then allows us to relate $\mathcal{C}$ to the winding number $\mathcal{W}(\phi^*)$:
\begin{equation}\begin{aligned}
	\mathcal{C} \times \Phi = \bigg\lfloor \frac{\pi}{4}\mathcal{W}(\phi^*)^2\bigg\rfloor = \frac{V_L}{v_2}~,
\end{aligned}\end{equation}
where \(v_2 = 4\pi^2/V\) is the volume of the \(k-\)space unit cell (defined below eq.~\eqref{Lut_volume}) and \(\lfloor \cdot \rfloor\) is the floor function and returns the largest integer less than or equal to the argument. 

\section{Conclusions and Outlook}
\label{conclusions}

In summary, we have shown that considering the renormalisation group flows of the spectral gap coupling of a system of non-interacting electrons in \(2+1-\)dimensional theory leads to hierarchical arrangement of the entanglement with scale, and that the geometrical part of the entanglement can be described as the emergence of an additional spatial dimension, leading to holographic expressions for geometric parameters like distance and curvature. We show that changing the boundedness of the holographic space involves a topological transition with the Fermi surface of the boundary theory turning critical, and involves the generation of a geometric wormhole between the UV and IR degrees of freedom.
For the massless case, the presence of an Aharonov-Bohm flux generates a geometry-independent term in the entanglement entropy which can be linked to the Luttinger volume of the gapless fermionic system. By placing the system of $2+1$ dimensional electrons in a strong magnetic field, we argue that the Chern number topological invariant for the ground states of the integer quantum Hall ground system are constrained by the value of the Luttinger volume in the underlying metallic system (i.e., prior to the insertion of the field).

How robust are our conclusions towards the inclusion of electronic correlations? Given the adiabatic continuity that links Fermi liquids to the non-interacting Fermi gas, i.e., the preservation of the $k$-space quantum numbers under adiabatic inclusion of (screened) inter-particle interactions, we expect our results to be equally valid for the many-particle entanglement of the excitations described by the fixed point effective theory for the Fermi liquid. This also suggests that our results for the wormhole geometry and the stress-energy tensor relation at the critical Fermi surface apply very generally to phase transitions that destabilise and gap out the Fermi surface. Specifically, this has profound implications for metal-insulator transitions driven by strong correlations.
Non-Fermi liquids are, however, not adiabatically continuous to the Fermi gas. Thus, it should be interesting to adapt our approach to the study of interacting models of electrons that lead to gapless non-Fermi liquids (including those that are known to appear at quantum critical points related to the breakdown of a Fermi liquid). For instance, Ref.\cite{Flynn_2023} reports that the second Renyi entropy of a thin $k$-space shell proximate to the Fermi surface is related to the quasiparticle residue; how would this change in a non-Fermi liquid?  It also appears interesting to identify the effects of electronic correlations on various aspects of the emergent space. 
The renormalisation group techniques applied in Refs.\cite{balasubramanian2012momentum,mcdermott2013,costa2022,Flynn_2023,PatraEntanglement2023} could also be worth investigating towards answering these questions. 

It is interesting to note that our results point towards a broader similarity between the gapless fermionic systems considered here and gapped topologically ordered quantum liquids. Recall that the latter are known to possess a ground state degeneracy that is dependent on the topological nature of the interacting electronic system, leading to a subsystem entanglement entropy that has a geometry-independent subleading piece (\(S_\text{topo}\)) referred to as the topological entanglement entropy. It was shown by some of us recently~\cite{siddhartha_TEE} that for a collection of \(N\) subsystems (CSS) that form a closed annular structure, the \(n-\)partite quantum entanglement information measures \(I^n\) \((3 \leq n \leq N)\) are a topological invariant equal to the product of \(S_\text{topo}\) and the Euler characteristic of the planar manifold on which the CSS is embedded. This appears to be analogous to our finding of the hierarchical sequence of quantities \(\left\{Q_g\right\}\) defined at a particular RG step (and arising out of the topological flux-dependent piece of the entanglement entropy), all of whom encode the same Luttinger volume for the gapless systems we have studied (eq.~\eqref{luttvolhier}). Further, the robustness of the $n$-partite information \(I^n\) against small deformations of the CSS in a topologically ordered system can be compared to the invariance of \(N^x_L\) to small deformations of the contours \(\mathcal{C}_1, \mathcal{C}_2, \ldots\) (see \alert{bottom panel} of Fig.~\ref{contour}). It appears tempting to speculate on a possible universal origin underlying these similarities (see also \cite{PatraEntanglement2023} for a recent investigation of the entanglement properties of various quantum liquids).

Our results lead to a number of interesting questions. 
Firstly, the convergence parameter \(\theta\) is reminiscent of the expansion rate of congruence flows on Riemannian manifolds in the context of the Raychaudhuri equation~\cite{kar_2001}.
While the evolution equation for \(\theta\) (eq.~\eqref{RC}) obtained by us is not of the same form (as there is no time variable, neither is there a \(\theta^2\) term in the expansion rate), we do expect such an equation to hold for an appropriately defined expansion parameter~\cite{kar2007raychaudhuri,balasubramanian1999}. This likely requires the introduction of temporal dynamics of the emergent spatial dimension, and lies well beyond the purview of the present work.
Secondly, it remains to be seen how the geometry-independent part of the entanglement behaves in the presence of a mass (\(m > 0\)) for the Dirac fermions: a spectral gap is expected to have consequences for Luttinger's theorem~\cite{anirbanurg1} (see also eq.\eqref{LT-violation}), as well as for the many-particle entanglement. For instance, Ref.\cite{Flynn_2023} reports that the second Renyi entropy shows the signatures of the condensation of Cooper pairs in the BCS problem.

Finally, we would like to point out an interesting conclusion that emerges from our analysis. As observed by us, the system considered here shows a hierarchy of multipartite entanglement measures beyond just the mutual information.
Further, we have shown that the mutual information leads to the emergence of the holographic dimension. However, whether the higher-party measures also have a holographic interpretation remains an open question.
We note that studies of the time-variation of tripartite information during an equilibrium process after the injection of energy suggest that it remains negative throughout the process~\cite{balasubramanian2011,asadi2018,aliakbari_2019}, indicating that care must be taken in defining geometric measures using it. Nevertheless, our study raises the possibility of obtaining a generalisation of the gauge-gravity duality from the perspective of many-particle entanglement. We speculate that the higher order information measures will lead to a modification of the universal entanglement Hamiltonian obtained at criticality. Further, that these modifications can be visualised as a {\it hypergraph}: just as a graph contains edges that connect any two nodes, a hypergraph contains {\it hyperedges} that can connect more than two nodes. The eigenvalues \(\epsilon^{(n)}\) (with \(n>2\)) of the entanglement Hamiltonian would act as hyperedges connecting the degrees of freedom \(\{i,j,\ldots\}\). We leave an investigation of these aspects to future works.

\acknowledgments
AM thanks IISER Kolkata for funding through a junior and a senior research fellowship. SP thanks the CSIR, Govt. of India for funding through a senior research fellowship. SL thanks the SERB, Govt. of India for funding through MATRICS grant MTR/2021/000141 and Core Research Grant CRG/2021/000852. The authors thank N. Banerjee and S. G. Choudhury for fruitful discussions and feedback. The authors are also deeply grateful to two anonymous referees for their helpful suggestions and comments.

\bibliography{metals-EE-manuscript}
\appendix
\section{Reduction of the \(2+1-\)D system to a tower of \(1+1-\)D massive modes}
\label{app-red}

In the absence of the gauge field, the Lagrangian for \(2+1-\)dimensional massive non-interacting relativistic electrons (in natural units) is
\begin{equation}\begin{aligned}
	\mathcal{L} = \int \mathrm{d}x~\mathrm{d}y~\overline\psi(x,y)\left[i\gamma^\mu\partial_\mu - m \right]\psi(x,y)~.
\end{aligned}\end{equation}
The integral ranges over the area of the torus on which the electrons reside, and the time dependence of the fields has been suppressed. Each component \(\psi^a,a=0,1,2,3\), of the Dirac spinor \(\psi = \left( \psi^0\quad\psi^1\quad\psi^2\quad\psi^3 \right) \) satisfies the Dirac equation
\begin{equation}\begin{aligned}
	\left[i\gamma^\mu\partial_\mu  -m \right]\psi^a(x,y) = 0~,
\end{aligned}\end{equation}
which is equivalent to the form
\begin{equation}\begin{aligned}
\int dx~ \overline{\psi^a}(x,y)\left[i\partial^\nu\gamma_\nu  + m \right]\left[i\gamma^\mu\partial_\mu  - m \right]\psi^a(x,y) = 0~.
\end{aligned}\end{equation}
Due to the periodic boundary conditions (PBCs) in the \(x-\)direction, the momenta \(k_x\) are quantised:~\( k_x^n = \frac{2\pi n}{L_x}, n \in \mathbb{Z}\). We expand the fields \(\psi^a(x,y)\) in these momenta: 
\begin{equation}\begin{aligned}
\psi^a(x,y) = \sum_{n=-\infty}^{\infty} e^{ik_x^n x} \psi^a(k_x^n,y)~.
\end{aligned}\end{equation}
Writing the Dirac equation in terms of the dual fields \(\psi^a(k_x^n,y)\) gives
\begin{equation}\label{DEdual}\begin{aligned}
	\sum_{m,n}\left(\overline{\psi^a}(k_x^m,y) e^{-ik_x^m x}\right) \left[i\partial^\nu\gamma_\nu  + m \right]\left[i\gamma^\mu\partial_\mu  - m \right]\times\\
	\left(e^{ik_x^n x}\psi^a(k_x^n,y)\right) = 0~,
\end{aligned}\end{equation}
where
\begin{equation}\begin{aligned}
	\left[i\gamma^\mu\partial_\mu  - m \right]\left(e^{ik_x^n x}\psi^a\right) = e^{ik_x^n x}\left[i\gamma^\mu\partial_\mu - \left( \gamma^x k_x^n + m \right) \right]\psi^a~.
\end{aligned}\end{equation}
Similarly, 
\begin{equation}\begin{aligned}
	\left(\overline{\psi^a}(k_x^m,y) e^{-ik_x^m x}\right) &\left[i\partial^\nu\gamma_\nu  + m \right] \\
							      &= \left[\left[i\gamma^\nu\partial_\nu + m\right]\left(e^{ik_x^m x}\gamma_0\psi^a\right)\right]^\dagger \\
							      &= \overline{\psi^a} e^{-ik_x^m x}\left[i\partial^\nu\gamma_\nu - \left( \gamma^x k_x^n - m \right) \right]~.
\end{aligned}\end{equation}
Substituting these into the equation eq.\eqref{DEdual} gives
\begin{equation}\begin{aligned}
	\sum_{n}\int\mathrm{d}y ~ \overline{\psi^a}(k_x^n,y) \left[\partial_\mu \partial^\mu + {k_x^n}^2 + m^2\right]\psi^a(k_x^n,y)= 0~,
\end{aligned}\end{equation}
where we have used \(\int \mathrm{d}x ~e^{ix\left( k_x^m - k_x^n \right) } = \delta_{m,n}\).

\section{Evaluation of the multi-partite information}
\label{g-partite info}

We define the multipartite information among \(g\) sets \(\left\{ \mathcal{A}_{z}(i)\right\} \) as~\cite{siddhartha_TEE}
\begin{equation}\begin{aligned}
	I^g_{\left\{ \mathcal{A}_{z}(i) \right\}} = \sum_{r=1}^g \left( -1 \right)^{r-1} \sum_{\beta \in \mathcal{P}\left(\left\{ \mathcal{A}_{z}(i) \right\}\right)}^{|\beta|=r} S_\beta~,
\end{aligned}\end{equation}
where \(\mathcal{P}\left(\left\{ \mathcal{A}_{z}(i) \right\}\right)\) is the power set of \(\left\{ \mathcal{A}_{z}(i) \right\}\), and \(|\beta|=r\) implies that we sum only over those sets \(\beta\) that have \(r\) number of elements in them. We will first evaluate the internal sum \(\sigma_r\) over the set \(\beta\) for a general \(r\): 
\begin{equation}\begin{aligned}
\sigma_r = \sum_{\beta \in \mathcal{P}\left(\left\{ \mathcal{A}_{z}(i) \right\}\right)}^{|\beta|=r} S_\beta~.
\end{aligned}\end{equation}
For this, we first note that following eq.~\eqref{ent-union}, the entanglement entropy (EE) of the union of a given sequence of sets \(\left\{ i \right\} \) is given by the EE of the set with the highest (lowest) index for a positive (negative) \(z\). As a result, the EE \(S_\beta\) for the set \(\beta\) is simply \(S_\beta = \theta(z)S_{\text{min}(\beta)} + \theta(-z)S_{\text{max}(\beta)}\). If \(n^{(r,+)}_j\) (\(n^{(r,-)}_j\)) is the number of sets of cardinality \(r\) that have the element \(\mathcal{A}_{j,z}\) as the element with the smallest (largest) index \(j\), the internal sum can be written as
\begin{equation}\begin{aligned}\label{internal-sum}
	\sigma_r = \sum_{j=1}^{g-(r-1)} \theta(z) n_j^{(r,+)} S_{j,z} + \sum_{j=r}^{g}\theta(-z) n_j^{(r,-)} S_{j,z}~.
\end{aligned}\end{equation}
The number \(n_j^{(r,+)}\) corresponds to the number of ways one can pick \(r-1\) elements out of a set of \(g-j\) elements: 
\(n_j^{(r,+)} = {{g-j}\choose{r-1}}\). Similarly, \(n_j^{(r,-)}\) corresponds to the number of ways one can pick \(r-1\) elements out of a set of \(j\) elements: \(n^{(r,-)}_j = {{j}\choose{r-1}}\). We can also extend the limits on the two summations by defining \({n\choose m} = 0 \) for \( n < m\). With these considerations, eq.~\eqref{internal-sum} takes the form
\begin{equation}\begin{aligned}
	\sigma_r = \sum_{j=1}^{g} \left[\theta(z) {{g-j}\choose{r-1}} + \theta(-z) {{j}\choose{r-1}}\right] S_{j,z}~.
\end{aligned}\end{equation}
This can now be substituted into the full expression for the multipartite equation, giving
%\begin{widetext}
\begin{equation}\begin{aligned}
	&I^g_{\left\{ \mathcal{A}_{z}(i) \right\}} \\
	&= \sum_{r=1}^g \left( -1 \right)^{r-1}\sum_{j=1}^{g} \left[\theta(z) {{g-j}\choose{r-1}} + \theta(-z) {{j}\choose{r-1}}\right] S_{j,z} \\
	&=\sum_{j=1}^{g} S_{j,z} \left[\theta(z) \sum_{r=0}^{g-j} \left( -1 \right)^{r}{{g-j}\choose{r}} + \theta(-z) \sum_{r=0}^{j} \left( -1 \right)^{r}{{j}\choose{r}}\right]~.
\end{aligned}\end{equation}
%\end{widetext}
\begin{figure*}[!t]
	\includegraphics[width=0.9\textwidth]{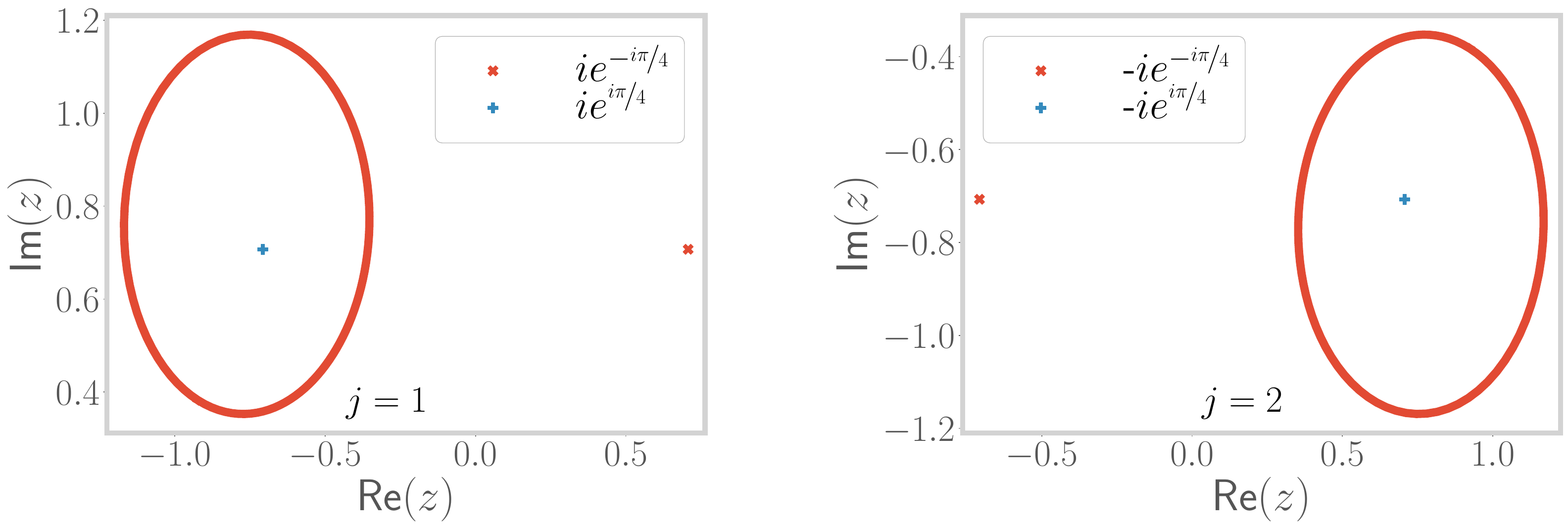}
	\caption{Contours \(z(\mathcal{C}_j)\) for two values of \(j\) employed for evaluating the integral eq.\eqref{appcont}, and the locations of the poles \(z^\pm_\pm\). For each \(j\), only one pole (shown in blue) contributes, as the other (shown in red) is not enclosed by the contour.}
	\label{pole-contour}
\end{figure*}
From the binomial theorem, we can now recognise each internal sum as over \(r\) as a polynomial expansion. That is, upon comparing with \((x-1)^n = \sum_{m=0}^n {n\choose m} (-1)^m x^{n-m}\), we obtain \(\sum_{r=0}^{g-j} \left( -1 \right)^{r}{{g-j}\choose{r}} = \lim_{x \to 1}(x-1)^{g-j} = \delta_{g,j}\). Similarly, the other sum evaluates to \(\delta_{j,0}\). Substituting these into the expression for $I^g_{\left\{ \mathcal{A}_{z}(i) \right\}}$ gives
\begin{equation}\begin{aligned}
	I^g_{\left\{ \mathcal{A}_{z}(i) \right\}} &=\sum_{j=1}^{g} S_{j,z} \left[\theta(z) \delta_{g,j} + \theta(-z) \delta_{j,0}\right] \\
						  &= \theta(z)S_g + \theta(-z)S_0~.
\end{aligned}\end{equation}
By noting that \(S_g\) and \(S_0\) are, respectively, the EE of the intersection set of \(\left\{\mathcal{S}_{z}(i)\right\}\), one can combine the two terms in the previous equation into
\begin{equation}\begin{aligned}
	I^g_{\left\{ \mathcal{A}_{z}(i) \right\}} = \theta(z)S_g + \theta(-z)S_0 = S_{\cap \left\{ \mathcal{A}_{z}(i) \right\}}~.
\end{aligned}\end{equation}

\section{Evaluation of the contour integral that counts poles}
\label{contour-integral}
We prove eq.~\eqref{pole-1} in this appendix. The integral in question is given by
\begin{equation}\label{appcont}\begin{aligned}
	\mathcal{I} = \oint_{\mathcal{C}_j} \frac{d\phi}{Q_0(\phi)} = \oint_{\mathcal{C}_j} \frac{d\phi}{\frac{1}{\sqrt 2} - |\sin\pi\phi|}~,
\end{aligned}\end{equation}
where \(\mathcal{C}_j\) is the contour shown in Fig.~\ref{contour-1} and the variable $\phi$ is given by \(\phi = \left\{\phi_j + \frac{1}{8}e^{i\theta}: \theta \in\left[0,2\pi\right] \right\} \), where \(\phi_j = j - \frac{1}{4}\) is the center of the contour. In order to evaluate the integral, we make a variable change: \(z = e^{i\pi\phi}\), giving \(dz = i\pi z~d\phi\) and \(\sin\pi\phi = \frac{1}{2i}\left( z - \frac{1}{z} \right)\). Depending on the value of \(j\), \(\sin\pi\phi\) can be either positive or negative. We denote either case by a subscript \(\mathcal{I}_\pm\) corresponding to the contour \(\mathcal{C}_\pm\), giving \(|\sin \left(\pi\phi\right)| = \pm \sin \left(\pi\phi\right)\).
Casting the integral in terms of \(z\), we obtain
\begin{equation}\begin{aligned}
	\mathcal{I}_\pm = \frac{1}{i \pi}\oint_{z\left(\mathcal{C}_\pm\right)} \frac{d z/z}{\frac{1}{\sqrt 2} \mp \frac{1}{2i}\left( z - \frac{1}{z} \right) } = \oint_{z\left(\mathcal{C}_\pm\right)} \frac{\mp \frac{2}{\pi}~d z}{\left( z\mp \frac{i}{\sqrt 2} \right) ^2 - \frac{1}{2}}~.
\end{aligned}\end{equation}
The integrand in \(\mathcal{I}_+\) has two simple poles at \(z_+^\pm = +ie^{\mp i \pi/4} \).
Similarly, the integrand in \(\mathcal{I}_-\) has two simple poles at \(z_-^\pm = -z_+^\pm\).
We plot the contour and the two poles for two contours of opposite signature in Fig.~\ref{pole-contour}; it is clear that only the pole at \(z^+_\pm\) contributes to the integral. Using the residue theorem, the integral evaluates to
\begin{equation}\begin{aligned}
	\mathcal{I}_\pm &= \oint_{z\left(\mathcal{C}_\pm\right)} \frac{\mp \frac{2}{\pi}~d z}{\left(z - z^+_\pm\right)\left(z - z^-_\pm\right)} \\
			&= \mp \frac{2}{\pi}2\pi i \lim_{z \to z^+_\pm} \frac{1}{z - z^-_\pm} \\
			&= \mp 2\sqrt 2 i~.
\end{aligned}\end{equation}
This proves eq.~\eqref{pole-1}.
\end{document}